\documentclass[12pt,a4paper]{article}
\pdfoutput=1
\usepackage[utf8]{inputenc}
\usepackage{simpler-wick}
\usepackage{jheppub}
\usepackage{amsmath}
\usepackage{tcolorbox}
 \usepackage{dsfont}
\usepackage{float}
\usepackage{nicematrix,tikz}
\usepackage{subcaption}
 
\newcommand*\bell{\ensuremath{\boldsymbol\ell}}

\newcommand{\dd}{\mathrm{d}}
\newcommand{\hd}{\hat{\dd}}
\newcommand{\hdelta}{\hat{\delta}}

\usepackage{amsmath}
\usepackage{amssymb}
\usepackage{braket}
\usepackage{bm}
\renewcommand{\[}{\begin{equation}\begin{aligned}}
\renewcommand{\]}{\end{aligned}\end{equation}}

\usepackage{xcolor}
\usepackage{graphicx}
\usepackage{subcaption}

\usepackage{tikz}
\def\lexp{\biggl\langle\!\!\!\biggl\langle}
\def\rexp{\biggr\rangle\!\!\!\biggr\rangle}

\usetikzlibrary{external}
\usepackage[compat=1.1.0]{tikz-feynman}
\definecolor{airforceblue}{rgb}{0.36, 0.54, 0.66}
\definecolor{blue(ncs)}{rgb}{0.0, 0.53, 0.74}
\definecolor{caribbeangreen}{rgb}{0.0, 0.8, 0.6}

\usepackage{relsize}

\renewcommand{\[}{\begin{equation}\begin{aligned}}
\renewcommand{\]}{\end{aligned}\end{equation}}

\renewcommand{\Im}{\operatorname{Im}}
\renewcommand{\Re}{\operatorname{Re}}

\renewcommand{\vec}[1]{\mathbf{#1}}

\newcommand{\rma}{{\rm a}}
\newcommand{\rmb}{{\rm b}}
\newcommand{\rmc}{{\rm c}}

\author[1]{Rafael Aoude,}
\emailAdd{rafael.aoude@ed.ac.uk}
\author[2]{Andrea Cristofoli,}
\emailAdd{cristofoli@yukawa.kyoto-u.ac.jp}
\author[1]{Asaad Elkhidir,}
\emailAdd{A.E.H.Elkhidir@sms.ed.ac.uk}
\author[3]{Matteo Sergola}
\emailAdd{matteo.sergola@ipht.fr}

\affiliation[1]{Higgs Centre for Theoretical Physics,
School of Physics and Astronomy, \\
The University of Edinburgh, Edinburgh EH9 3JZ, Scotland, UK}
\affiliation[2]{Center for Gravitational Physics and Quantum Information,
Yukawa Institute for Theoretical Physics, Kyoto University, 606-8502, Kyoto, Japan}
\affiliation[3]{
Institut de Physique Théorique, CEA, CNRS, \\Université Paris-Saclay, F–91191 Gif-sur-Yvette cedex, France}

\title{Inelastic Coupled-Channel Eikonal Scattering}

\abstract{
Emitted radiation and absorption effects in black hole dynamics lead to inelastic scattering amplitudes. In this paper, we study how these effects introduce an inelasticity function to the  $2\rightarrow2$ eikonalised $S$-matrix and how they can be described using unequal mass and spin on-shell amplitudes. To achieve this, we formulate the inelastic coupled-channel eikonal (ICCE) using the KMOC formalism and the language of quantum channels, where off-diagonal channels involve mass and spin changes. This formulation allows us to re-use usual eikonal results but also suggests a different resummation of inelastic effects. We then apply this formulation to calculate classical inelastic processes, such as the mass change in binary dynamics due to the presence of an event horizon. Additionally, we provide a complementary analysis for the case of wave scattering on a black hole, considering absorption effects. In both scenarios, we derive unitarity relations accounting for inelastic effects.
}

\begin{document}
{\baselineskip0pt
\rightline{\baselineskip16pt\rm\vbox to-20pt{
           \hbox{YITP-24-142}
\vss}}%
}
\maketitle
\newpage

\section{Introduction}

Ever since the eikonal equation was introduced in its modern form  by Hamilton  \cite{hamilton1828theory} and popularised by the lectures of Glauber \cite{glauber1959lectures}, eikonal-inspired methods have been applied to a wide range of problems in physics. Most notably, the eikonal was employed in high-energy physics, where its field theoretic features were developed in the sixties through different seminal works \cite{PhysRevLett.23.53,Torgerson:1966zz,PhysRevLett.22.666,PhysRev.186.1611,PhysRev.186.1656,PhysRevD.16.3565,PhysRevD.23.1411}. Later on, systematic corrections to the leading approximation were  pioneered by Wallace~\cite{Wallace:1973iu,PhysRevD.8.1846}, pushing the framework to higher orders of precision. In recent years the eikonal has  gained reinvigorated  attention in the context of the gravitational two-body problem; widening its range of applicability from the scale of  spacetime curvatures  all the way down to the  hadronic scale.

%%%
In fact, lately the eikonal approximation has been applied to the scattering of compact bodies and the emission of gravitational waves \cite{Bern:2020gjj,DiVecchia:2022nna,DiVecchia:2022piu, Cristofoli:2021jas,Damgaard:2021ipf} (see also \cite{DiVecchia:2023frv} for a recent review and references therein). These methods rely on a QFT-rooted scattering amplitudes approach to model the classical limit of the two-body problem~\cite{Buonanno:2022pgc, Arkani-Hamed:2017jhn,Bjerrum-Bohr:2018xdl,Bellazzini:2022wzv,Guevara:2018wpp,Kosower:2018adc,Cheung:2018wkq,Bautista:2019tdr,Guevara:2019fsj,Bern:2019nnu,Bjerrum-Bohr:2019kec,Bern:2020buy,Cristofoli:2020uzm,Bjerrum-Bohr:2021vuf,Herrmann:2021tct,Bern:2021dqo,DiVecchia:2021bdo,Bern:2021yeh,Bautista:2021wfy,Cristofoli:2021vyo,Herrmann:2021lqe,Travaglini:2022uwo,Monteiro:2020plf,Bjerrum-Bohr:2022blt,Bern:2023ccb,Kosower:2022yvp,Bern:2024adl,Driesse:2024xad,Gambino:2024uge,Luna:2023uwd,Herderschee:2023fxh,Brandhuber:2023hhy,Georgoudis:2024pdz,Brunello:2024ibk,Bini:2024rsy,Alaverdian:2024spu}. Many advances have also been made in both the conservative  and radiative sectors in the Post-Minkowskian (PM) expansion through effective field theory and worldline QFT methods \cite{Levi:2015msa,Porto:2016pyg,Levi:2018nxp,Damour:2014jta,Cristofoli:2019neg,Kalin:2019rwq,Kalin:2020fhe,Kalin:2020mvi,Dlapa:2021npj,Dlapa:2021vgp,Cho:2022syn,Dlapa:2022lmu,Dlapa:2024cje}. Recent works have also tackled the spinning 
eikonal~\cite{Haddad:2021znf,Adamo:2021rfq,Gatica:2023iws,Luna:2023uwd} and, more generally, the inclusion of high-spin effects in conservative classical amplitudes \cite{Damgaard:2019lfh,Bautista:2019tdr,Aoude:2020onz,Bern:2022kto,Aoude:2022trd,Moynihan:2019bor,Damgaard:2022jem,Aoude:2022thd,Haddad:2023ylx,Akpinar:2024meg,Chiodaroli:2021eug,Cangemi:2022bew,Cangemi:2023bpe,Bianchi:2023lrg,Ben-Shahar:2023djm,Scheopner:2023rzp,Cangemi:2022abk,Azevedo:2024rrf,Bautista:2022wjf,Bautista:2021wfy,Bautista:2023sdf, Bautista:2024agp,FebresCordero:2022jts}. 

%%%
Extensively applied to the conservative sector, the inelastic eikonal was recently endowed with radiative dynamics,  in the important context of emitted gravitational radiation~\cite{Cristofoli:2021vyo,Elkhidir:2023dco,Georgoudis:2023lgf,Georgoudis:2023eke,Georgoudis:2023ozp} (See also \cite{Ivanov:2024sds}).
In this case, one has to consider inelastic scattering since the number of particles changes in the final state, \textit{e.g.} $2 \rightarrow 3$. This type of inelasticity allows one to incorporate graviton emission -- and produce the corresponding observable waveform \cite{Cristofoli:2021vyo} -- however, it does not entail the fact that the black hole might change its mass (or the total angular momentum magnitude) during the process. We will argue that this naturally leads to the concept of  ``coupled channels", which will be the central topic of this paper.
In nuclear and hadronic physics, a coupled
channel describes the system's internal degrees of freedom (d.o.f.) -- or internal excitations -- throughout the scattering. In this work, we will use the same terminology to indicate that the BH's induced d.o.f. will be taken into account similarly, in a manner that will soon be clear.
 
%%%
The study of inelastic and couple-channel effects however is a well-investigated area in the field of scattering amplitudes and it is popular in hadronic~\cite{Rudin:1970jv}, nuclear~\cite{Hagino:2022jkz}, atomic~\cite{TTGien_1984}, and black-hole perturbation theory~\cite{Pound:2021qin}. For instance, in Hadronic physics, a channel is usually associated to possible resonances or changing particle states (See~\cite{Oller:2019opk} for a review). In the case of $\pi\pi$ scattering, the system evolves either through an elastic channel
$\pi\pi \rightarrow \pi\pi$  or via  rescattering $\pi\pi \rightarrow K\bar{K} \rightarrow \pi \pi$. At  low enough energies, one can even perform a resummation of the all-loop bubble diagrams  to obtain the $\pi\pi$ scattering length~\cite{Oller:1997ti}. Even more, with the aid of the $K$-matrix formalism~\cite{AITCHISON1972417}, one can include coupled channels in the resummation analysis.  The advantage of using the coupled-channel approach is to include the $K\bar{K}$ transitions without spoiling the previous pion resummation. In this paper, we are going to suggest a resummation with coupled channels, leading to a scattering-equivalent of the inelasticity parameter. We will show how typical eikonal results can be recycled and inelastic effects can be easily incorporated.

%%%
Recently, absorption effects were included in the language of on-shell scattering amplitude by means of mass-changing three-point vertices~\cite{Aoude:2023fdm} and  black hole spectral density functions (See also~\cite{Chen:2023qzo,Ivanov:2024sds,Jones:2023ugm,Bautista:2022wjf}). These effects furnish a natural microscopic model for the coupled-channel inelasticity parameter of a  scattering theory. As we will see, this is mainly due to different Hilbert spaces induced by changing mass and spin. Indeed, the eigenvalues of Casimir invariants defines a particle state and -- borrowing language from  quantum information theory -- when the system changes set of little-group labels, one  says that a channel transition took place. (See also~\cite{Bern:2023ity,Alaverdian:2024spu,Akpinar:2024meg} for effects of casimir change in scattering amplitudes). The language of diagonal and off-diagonal Casimir channels will act as a guideline for us to incorporate traditional couple-channel literature in the context of scattering amplitudes for black holes, which will be the goal of this paper. Interpreting these interactions as channels will help us organize and re-use previous eikonal computations.

%%%
In Section~\ref{sec:Channel_Analysis} we introduce our notation for using the language of quantum channels to consider off-diagonal (in channel space) to the usual scattering. This also allows us to split the resolution of the identity in the diagonal/off-diagonal pieces. In Section~\ref{sec:ICCE_QFT} we formulate the ICCE in the QFT with channels being different mass/spin transitions. This induces an inelasticity parameter which we call $\eta$. We give it an interpretation in terms of QFT and Feynman diagrams. The concept of inelasticity due to a coupled-channel is then put into practice in Section~\ref{sec:FinalStatesObs} by providing an ansatz for the final state and by studying observables. Using the unitarity of the full state, we also obtain all-order relations between the inelasticity parameters. Then, we conclude in
Section~\ref{sec:Conclusion}. Furthermore, we also give an explicit example of a  multi-channel  resummation by studying a particular toy model in  Appendix \ref{MCExp}.

\subsection*{Miscellaneous conventions}
We will use mostly minus signature throughout the paper: $\eta_{\mu\nu}=\text{diag}(+1,-1,-1,-1)$ and consistently hide powers of $(2\pi)$ in measures and delta function as~\cite{Kosower:2018adc} 
$$
\hat{\dd}^n p\equiv\frac{\dd^n p}{(2\pi)^n},\,\,\,\,\,\, \hat{\delta}^{n}(p)\equiv(2\pi)^n\delta^n(p).
$$
We will also write covariant phase space measures and deltas as
$$
\dd\Phi(p)\equiv\hat{\dd}^4 p\,\hat{\delta}(p^2-M^2)\Theta(p_0), \,\,\,\, \delta_{\Phi}(p-p')\equiv 2E_\vec{p}\hat{\delta}^3(\vec{p}-\vec{p}'),
$$
and write products as
$$
\dd\Phi(p_1, \cdots, p_n)\equiv \dd\Phi(p_1) \cdots \dd\Phi( p_n).
$$
When considering products of integrals we use the notation:
$$
\int_{\{x\}_{N}} \equiv \int \prod_{i=1}^N \dd x_i, \quad \int_{\{\ell\}_{N}} \equiv \int \prod_{i=1}^N \hd \ell_i,
$$
in position and momentum space respectively.
Finally, we use an amplitude notation inspired by the fragments of \cite{Cristofoli:2021jas}, namely that:
$$
    \mathcal{A}^{(L)}_{n, \rma \rmb}(p_1, p_2, \cdots \to p_1', p_2', \cdots),
$$
is the leading $\hbar$ piece of an $L$-loop $n$-point amplitude, with coupled channels $\rma\to \rmb$. Another choice of convention
we adopt is to take incoming momenta of initial states to be ingoing and final momenta to be outgoing. We also set $\hbar=1=c.$

\section{KMOC with Casimir channels}
\label{sec:Channel_Analysis}

In quantum information language, the concept of channel is associated~\cite{Aolita_2015} to a map (or transformations) acting on density matrices in the associated Hilbert space. 
The concept of channels allows one to introduce the formalism of transitions which were not included in the Hilbert space of the system. In this section, we will try to bring the language of hadronic physics and quantum information to the inelastic scattering in QFT, with the end goal of black hole and neutron star scattering as well as black-hole-wave interaction in Sec~\ref{sec:FinalStatesObs}. In QFT, states are labelled with eigenvalues of  Casimir operators~\footnote{As well as the sign of $P^0$.}
\[
\mathds{P}^2 |p,M,s\rangle = M^2|p,M,s\rangle \quad , \quad 
\mathds{W}^2 |p,M,s\rangle = M^2 s(s+1)|p,M,s\rangle \, ,
\]
where $\mathds{W}_\mu \equiv \frac{1}{2} \epsilon_{\mu\nu\rho\sigma}\mathds{P}^\nu\mathds{J}^{\rho\sigma}$ is the Pauli-Lubanski
operator. These states are then labelled with the eigenvalues of these operators $(M,s)$, which can be naturally extended to multiple particles. Therefore, it seems natural to define a channel as a map from a set of Casimir operator eigenvalues to another set. For example, considering the usual eikonal scattering off a potential, the channel is kept through the multiple interactions

\[
\begin{tikzpicture}[x=0.75pt,y=0.75pt,yscale=-1,xscale=1]
\tikzset{every picture/.style={line width=0.75pt}}  
  
% Text Node
\draw (144.15,16278.26) node    {$$};
% Text Node
\draw (487.61,16277.28) node    {$$};
% Text Node
\draw (155,16276) node    {$( M_{\rma},s)$};
% Text Node
\draw (241,16276) node    {$( M_{\rma},s)$};
% Text Node
\draw (321,16276) node    {$( M_{\rma},s)$};
% Text Node
\draw (403,16276) node    {$( M_{\rma},s)$};
% Text Node
\draw (483,16276) node    {$( M_{\rma},s)$};
% Text Node
\draw (155.82,16278.26) node    {$$};
% Text Node
\draw (321.61,16278.28) node    {$$};
% Text Node
\draw (350,16271) node [anchor=north west][inner sep=0.75pt]    {$\cdots $};
% Text Node
\draw (140.95,16278.28) node    {$$};
% Text Node
\draw (84,16266.35) node [anchor=north west][inner sep=0.75pt]    {$| \text{in} \rangle =$};
% Text Node
\draw (506.67,16267.35) node [anchor=north west][inner sep=0.75pt]    {$=| \text{out} \rangle. $};
% Connection
\draw    (180,16276) -- (213,16276) ;
\draw [shift={(216,16276)}, rotate = 180] [fill={rgb, 255:red, 0; green, 0; blue, 0 }  ][line width=0.08]  [draw opacity=0] (8.93,-4.29) -- (0,0) -- (8.93,4.29) -- (5.93,0) -- cycle    ;
% Connection
\draw    (266,16276) -- (293,16276) ;
\draw [shift={(296,16276)}, rotate = 180] [fill={rgb, 255:red, 0; green, 0; blue, 0 }  ][line width=0.08]  [draw opacity=0] (8.04,-3.86) -- (0,0) -- (8.04,3.86) -- (5.34,0) -- cycle    ;
% Connection
\draw    (428,16276) -- (455,16276) ;
\draw [shift={(458,16276)}, rotate = 180] [fill={rgb, 255:red, 0; green, 0; blue, 0 }  ][line width=0.08]  [draw opacity=0] (8.04,-3.86) -- (0,0) -- (8.04,3.86) -- (5.34,0) -- cycle    ;

\end{tikzpicture}
\]
\hspace{0cm}Where from now on we call the initial mass $M_\rma\equiv M$.

However, in this work we aim to consider the case where the set of Casimir operators can change. In other words, starting with $(M_\rma, s)$ which could end in the same or in a different set 
\[
\begin{tikzpicture}[x=0.75pt,y=0.75pt,yscale=-1,xscale=1]
\tikzset{every picture/.style={line width=0.75pt}}

% Text Node
\draw (131.15,16135.26) node    {$$};
% Text Node
\draw (392.15,16192.26) node    {$$};
% Text Node
\draw (474.61,16134.28) node    {$$};
% Text Node
\draw (142,16133) node    {$( M_{\rma},s)$};
% Text Node
\draw (228,16133) node    {$( M_{\rma},s)$};
% Text Node
\draw (308,16133) node    {$( M_{\rma},s)$};
% Text Node
\draw (227,16196.5) node    {$( M_{\rmb},s')$};
% Text Node
\draw (307,16196) node    {$( M_{\rmb},s')$};
% Text Node
\draw (390,16133) node    {$( M_{\rma},s)$};
% Text Node
\draw (470,16133) node    {$( M_{\rma},s)$};
% Text Node
\draw (391,16197) node    {$( M_{\rmb},s')$};
% Text Node
\draw (471,16195) node    {$( M_{\rmb},s')$};
% Text Node
\draw (142.82,16135.26) node    {$$};
% Text Node
\draw (228.61,16192.28) node    {$$};
% Text Node
\draw (228.15,16193.26) node    {$$};
% Text Node
\draw (308.61,16135.28) node    {$$};
% Text Node
\draw (337,16128) node [anchor=north west][inner sep=0.75pt]    {$\cdots $};
% Text Node
\draw (339,16191) node [anchor=north west][inner sep=0.75pt]    {$\cdots $};
% Text Node
\draw (57.15,16185.26) node    {$$};
% Text Node
\draw (118.28,16200.61) node    {$$};
% Text Node
\draw (57.15,16183.26) node    {$$};
% Text Node
\draw (127.95,16135.28) node    {$$};
% Text Node
\draw (71,16123.35) node [anchor=north west][inner sep=0.75pt]    {$| \text{in} \rangle =$};
% Text Node
\draw (493.67,16124.35) node [anchor=north west][inner sep=0.75pt]    {$=| \text{out} \rangle $};
% Text Node
\draw (499.67,16186.68) node [anchor=north west][inner sep=0.75pt]    {$=| \text{out} '\rangle .$};
% Connection
\draw    (401.15,16185.93) -- (463.16,16142.33) ;
\draw [shift={(465.61,16140.61)}, rotate = 144.89] [fill={rgb, 255:red, 0; green, 0; blue, 0 }  ][line width=0.08]  [draw opacity=0] (8.04,-3.86) -- (0,0) -- (8.04,3.86) -- (5.34,0) -- cycle    ;
% Connection
\draw    (151.82,16141.24) -- (217.12,16184.64) ;
\draw [shift={(219.61,16186.3)}, rotate = 213.61] [fill={rgb, 255:red, 0; green, 0; blue, 0 }  ][line width=0.08]  [draw opacity=0] (8.04,-3.86) -- (0,0) -- (8.04,3.86) -- (5.34,0) -- cycle    ;
% Connection
\draw    (167,16133) -- (200,16133) ;
\draw [shift={(203,16133)}, rotate = 180] [fill={rgb, 255:red, 0; green, 0; blue, 0 }  ][line width=0.08]  [draw opacity=0] (8.93,-4.29) -- (0,0) -- (8.93,4.29) -- (5.93,0) -- cycle    ;
% Connection
\draw    (253,16133) -- (280,16133) ;
\draw [shift={(283,16133)}, rotate = 180] [fill={rgb, 255:red, 0; green, 0; blue, 0 }  ][line width=0.08]  [draw opacity=0] (8.04,-3.86) -- (0,0) -- (8.04,3.86) -- (5.34,0) -- cycle    ;
% Connection
\draw    (256.5,16196.32) -- (274.5,16196.2) ;
\draw [shift={(277.5,16196.18)}, rotate = 179.64] [fill={rgb, 255:red, 0; green, 0; blue, 0 }  ][line width=0.08]  [draw opacity=0] (8.04,-3.86) -- (0,0) -- (8.04,3.86) -- (5.34,0) -- cycle    ;
% Connection
\draw    (415,16133) -- (442,16133) ;
\draw [shift={(445,16133)}, rotate = 180] [fill={rgb, 255:red, 0; green, 0; blue, 0 }  ][line width=0.08]  [draw opacity=0] (8.04,-3.86) -- (0,0) -- (8.04,3.86) -- (5.34,0) -- cycle    ;
% Connection
\draw    (420.5,16196.26) -- (438.5,16195.81) ;
\draw [shift={(441.5,16195.74)}, rotate = 178.57] [fill={rgb, 255:red, 0; green, 0; blue, 0 }  ][line width=0.08]  [draw opacity=0] (8.04,-3.86) -- (0,0) -- (8.04,3.86) -- (5.34,0) -- cycle    ;
% Connection
\draw    (237.15,16186.78) -- (297.18,16143.52) ;
\draw [shift={(299.61,16141.77)}, rotate = 144.22] [fill={rgb, 255:red, 0; green, 0; blue, 0 }  ][line width=0.08]  [draw opacity=0] (8.04,-3.86) -- (0,0) -- (8.04,3.86) -- (5.34,0) -- cycle    ;

\end{tikzpicture} 
\]  

Let us focus in the simple case in which we only have one extra channel. The initial one we will call it $(\rma)$ while the other possibility $(\rmb)$. 
Thus, we divide the Hilbert space into $\mathcal{H} = \mathcal{H}_\rma \otimes \mathcal{H}_\rmb$. Since we are focusing on a single particle spaces, the $(\rma)$ space would be for the massive spinless while the $(\rmb)$ space would be a massive spin-$s$ with a possible different mass.  

These Hilbert spaces are defined w.r.t the Casimir operators, changing the set of Casimir eigenvalues, changes the channel.
{This is our channel definition for this analysis}.
The full Fock space can be constructed out of this but in this paper we are going to stick to one particle changing channels. 
Taking the simplest initial state (product and pure) in composite space as 
\begin{align} 
|\Psi\rangle = 
\begin{bmatrix}
|\varphi_\rma\rangle \\
|\varphi_\rmb\rangle  
\end{bmatrix}
\qquad \text{such that}\qquad
|\varphi_\rma\rangle \in \mathcal{H}_{\rma}, \qquad
|\varphi_\rmb\rangle \in \mathcal{H}_{\rmb}.
\end{align}
The simplest non-spinning state will be written down as~\cite{Kosower:2018adc}
\begin{align}
\label{KMOCalpha}
|\varphi_\rma\rangle = 
\int \dd\Phi(p_{\rma})\, \varphi_{\rma}(p)\, e^{ib\cdot p}|p_{\rma}\rangle \, ,
\end{align}
where we have introduced the label $(\rma)$ in the phase-space integral and in the momentum states, representing
\begin{align}
       \dd\Phi(p_{\rma}) = \hat{\dd}^4p_\rma\, \hat{\delta}^{(+)}(p_\rma^2 - M_\rma^2)
       \qquad
       \text{and}
       \qquad
       \mathds{P}^2 |p_{\rma}\rangle = M_{\rma}^2|p_{\rma}\rangle.
\end{align}
Similarly, we can define a state with a different mass (representing the different channel) and different spin as~\cite{Maybee:2019jus} 
\begin{align}\label{KMOCbeta}
|\varphi_\rmb\rangle = 
\int \dd\Phi(p_{\rmb}) \varphi_{\rmb}(p)\, \xi^{\{c\}} e^{ib\cdot p}|p_{\rmb},\{c\}\rangle \, ,
\end{align}
where the indices $\{c\}$ represent the SU(2) massive little-group indices\footnote{Spinning particles could also be represented by coherent-spin states~\cite{Aoude:2021oqj}, which are more suitable for the classical limit.}.
We now evolve this in the full system as
$S|\Psi\rangle$ in the \emph{channel-space}:
\begin{align}
\label{eq:ChannelSpaceMatrix}
S|\Psi\rangle = 
\begin{bmatrix}
S_{\rma\rma} & S_{\rma\rmb} \\
S_{\rmb\rma} & S_{\rmb\rmb} 
\end{bmatrix}
\begin{bmatrix}
|\varphi_\rma\rangle\\
|\varphi_\rmb\rangle\\ 
\end{bmatrix} \, .
\end{align}
One important point is that we will be mostly interested in the sub-space evolution in the same Hilbert space, from $\mathcal{H}_{\rma} \rightarrow \mathcal{H}_{\rma}$, given by $\langle \varphi_{\rma} | S_{\rma\rma} | \varphi_\rma\rangle$.
In Sec.~\ref{sec:FinalStatesObs} we are going to obtain relations derived from the unitarity of the $S$-matrix in channel space. As a consequence, it is already possible to see here that the usual unitarity relation to the purely diagonal submatrix $S_{\rma\rma}$ does not hold anymore. In general, we now have that $S_{\rma\rma}S^\dagger_{\rma\rma} \leq 1$ since the full channel-space $S$-matrix should be the one in which unitarity holds.
All of these operators are still operators in their own subspaces of momenta and spin. We now want to understand the eikonal properties of such $S$ matrix in terms of mass-changing amplitudes.
Choosing the available channels implicitly assumes a mass spectra of the theory, so that we have the following resolution of the identity~\cite{Aoude:2023fdm}
\begin{equation}\begin{aligned}	
\mathds{1} = \sum_{s=0}^\infty \sum_{\{c\}}\int_{0}^\infty \dd M_\rmb^2\, \rho_{\rm full}(M_\rmb^2)\int \dd\Phi(p_{\rmb}) |p_{\rmb},s,\{c\}\rangle\langle p_{\rmb},s,\{c\}| \, .
\end{aligned}\end{equation}	
We could have added a spin dependence on the spectral density function $\rho_{\rm full}$, but let us keep it simpler for the moment. We then assume that we for sure have the initial state in the full spectrum, i.e. $\rho_{\rm full}(M_\rmb^2) \supset \delta(M_{\rmb}^2-M_{\rma}^2) $. Thus, we can split this function into 
\begin{equation}
\begin{aligned}	
    \rho_{\rm full}(M_{\rmb}^2) = \delta(M_{\rmb}^2-M_{\rma}^2) + \Theta(M_\rmb^2 - M_\rma^2) \rho(M_{\rmb}^2) \, ,
\end{aligned}\end{equation}	
where the second term contains masses strictly greater than $M_\rma^2$. Hence, we split the identity into two terms
\begin{equation}
\begin{aligned}	
    \mathds{1} 
    &= \sum_{s=0}^\infty \sum_{\{c\}}\int \dd\Phi(p_{\rma}) |p_{\rma},s,\{c\}\rangle\langle p_{\rmc},s,\{c\}| \\
    &+\sum_{s=0}^\infty \sum_{\{c\}}\int_{M_\rma^2}^\infty\, \dd M^2_\rmb \rho(M_\rmb^2)\int \dd\Phi(p_{\rmb}) |p_{\rmb},s,\{c\}\rangle\langle p_{\rmb},s,\{c\}| \, .
\end{aligned}\end{equation}	
We also want to split the spin summation into a non-spinning and a spinning contribution $\sum_{s=0}^\infty  = \delta_s^0 + \sum_{s>0}^\infty$, which leaves us with four terms, relatively to the original initial state: 
\textit{(i)} preserves the mass and spin; 
\textit{(ii)} preserves the spin and changes the mass;
\textit{(iii)} preserves the mass and changes the spin; 
\textit{(iv)} changes both.
To make our notation less cumbersome, we name each contribution as follows
\begin{equation}\label{eq:identity}
\begin{aligned}	
    \mathds{1}_{\text{el.}} &=\int \dd\Phi(p_{\rma}) |p_{\rma}\rangle\langle p_{\rma}| \ , \\
    \mathds{1}_{\Delta \rm M} &= \int_{M_\rma^2}^\infty \dd M_\rmb^2 \rho(M_\rmb^2)\int \dd\Phi(p_{\rmb}) |p_{\rmb}\rangle\langle p_{\rmb}| \ ,\\
    \mathds{1}_{\Delta \rm S} &=  \sum_{s>0}^\infty \sum_{\{c\}}\int \dd\Phi(p_{\rma}) |p_{\rma},s,\{c\}\rangle\langle p_{\rma},s,\{c\}| \, ,\\
    \mathds{1}_{\Delta \rm MS} &= \sum_{s>0}^\infty \sum_{\{c\}}\int_{M_\rma^2}^\infty \dd M_\rmb^2 \rho(M_\rmb^2)\int \dd\Phi(p_{\rmb}) |p_{\rmb},s,\{c\}\rangle\langle p_{\rmb},s,\{c\}| \, ,
\end{aligned}\end{equation}	 
where ``el." is for elastic. The splitting of the identity is given by~\footnote{This equation should be read as the left hand side being an identity in the full Hilbert space while the right hand side being identities in the subspaces with zeros in the other entries.}
\begin{equation}
\begin{aligned}	
    \mathds{1}  &=	 \mathds{1}_{\rm el.} +	\mathds{1}_{\Delta \rm M}+\mathds{1}_{\Delta \rm S}+	\mathds{1}_{\Delta \rm MS} \, ,
\end{aligned}\end{equation}	
which is a good separation \emph{given}  the initial state being a spinless state of mass $M_{\rma}$. 
Of course, if we start with a spinning particle, we would split accordingly with its spin. Coming back to our simple example, the $\varphi_\rma$ is related to the first part of the identity while the $\varphi_\rmb $ to the last one. 
Thus, the diagonal element of the $S$-matrix\footnote{As we will explain later we find it convenient to discuss probe limit amplitudes. } will be given by
\begin{align}
 S_{\rma\rma} 
 = \mathds{1}_{\rm el.}+i
 &\int \dd\Phi(p_{\rma},p'_{\rma})
 \mathcal{A}(p_\rma \to p_\rma')a^\dagger(p_\rma')a(p_\rma) +\cdots ,
\end{align}
where $p_{\rma}^{'2} = p_{\rma}^2 =M_{\rma}^2$ and dots represent higher orders in the coupling and multiplicities.
Instead,  the off-diagonal elements are given by
\begin{align}
    S_{\rmb\rma} 
    =i\sum_{s,\{c\}}
    \int_{M_\rma^2}^\infty \dd M_\rmb^2 \,\,\rho(M_\rmb^2)
    &\int \dd\Phi(p_{\rma},p'_{\rmb})
     \mathcal{A}(p_\rma \to p_\rmb')a^{\dagger}(p_\rmb';\{c\},s)a(p_\rma) +\cdots ,
\end{align}
where $p_{\rma}^{2} = M_{\rma}^2 \neq p_{\rmb}^{'2}  = M_\rmb^2$ and we extended to the off-diagonal formula by summing over the newly-generated Casimirs.  Note that since $S_{\rmb\rma}$ is an operator its ``inverse" labels indicate a process where $\rma\to\rmb.$

Since we are interested in applying the formalism of quantum field theory to describe classical physics, let us briefly describe how the classical limit emerges from quantum amplitudes. In doing so, we shall adopt the KMOC formalism \cite{Kosower:2018adc}. In a nutshell, this amounts to prescribing powers of $
\hbar$ to various quantities, and demanding that classical observables are finite in the $\hbar \rightarrow 0$ limit. For our purposes, it is important to note the distinction between momenta of massive particles $p^\mu$ and messenger (photon or graviton) momenta $q^\mu$. Messenger momenta are $\hbar$-suppressed relative to the classical wavenumber $\bar{q}^\mu$
\[
q^\mu = \hbar \bar{q}^\mu,
\]
while massive momenta (in eg. channel $\rma$) are related to the classical four-velocity $u^\mu$ by 
\[\label{ClassicalMomentum}
p_\rma^\mu  = M_\rma u^\mu. 
\]
This hierarchy ensures that the messenger particles cannot probe scales of order of the Compton wavelength $\ell_c \sim M_\rma/\hbar$ associated with particle $p_\rma^\mu$. The wavepacket $\varphi_\rma(p)$ in eq.\eqref{KMOCalpha} is sharply peaked around the value $p_\rma^\mu = M_\rma u^\mu$ in the classical limit. Where appropriate, we will denote this wavepacket integral using the double-bracket notation
\[
\int \dd \Phi(p_\rma, \cdots ) |\varphi_\rma(p_\rma)|^2 f(p_\rma, \cdots )  = \lexp f(p_\rma, \cdots ) \rexp,
\]
where it is understood that the RHS is evaluated on the value $p_\rma^\mu=M_\rma u^\mu$ in the classical limit. The above discussion applies equally to channel $\rmb$.  However, in allowing transitions between channels of different masses $M_\rma$ and $M_\rmb$, the classical limit further imposes the restriction \cite{Aoude:2023fdm,Jones:2023ugm}
 \[\label{MassChangeHbar}
 w \equiv \frac{M_\rmb^2 - M_\rma^2}{2M_\rma} \sim \mathcal{O} (\hbar).
 \]
 This is because the momentum difference $p^\mu_\rmb - p^\mu_\rma \sim \mathcal{O}(q^\mu)$, implying that the momentum $p_\rmb$ cannot go on-shell unless \eqref{MassChangeHbar} is satisfied. Also note $\dd M_\rmb^2=2M_\rma \dd w$. Finally, for clarity of notation we will proceed by setting $\hbar=1$.

\section{Inelastic coupled-channel eikonal in QFT}
\label{sec:ICCE_QFT}
In this part of the paper we will describe a way to source inelastic effects with scattering  amplitudes. For simplicity in this section we will mostly consider a simple    model  where one particle is much heavier than the other. This is usually referred to as the probe limit and we  begin by introducing it in relation to channels in the next section. 

\subsection{Channel space amplitudes and their probe limit }
It is convenient to explore multi-channel scattering in the probe limit of quantum field theory. In this limit, the QFT amplitudes coincide with those of potential scattering theory. This allows us to make contact with several results in the literature and to exploit results in eikonal and potential theory.  We start by reviewing how the probe limit emerges from QFT scattering amplitudes in the eikonal limit. Here we will closely follow Weinberg's discussion in \cite{Weinberg:1995mt}, generalised to account for multiple channels. 

It is useful to start by briefly reviewing the standard eikonal approximation\footnote{See eg. \cite{DiVecchia:2023frv} for a detailed review.}. To this end, consider a generic $2\rightarrow 2$ amplitude containing massive particles with initial momenta $p_i$ and final momenta $p'_i = p_i + q_i$. Focusing on the diagram in figure \ref{figure1} with $N$ messenger exchanges, we find the following contribution
 \[
 \bra{p_1',p_2'}S\ket{p_1,p_2}_N =  \sum_{\text{topologies}}\int \left(\prod_{i=1}^N  \hd^4 \ell_i \, G(\ell_i) \right) I_N(\ell_{1}, \cdots , \ell_{n})\, \hdelta^4(\ell_i -q),
 \]
where $ G(\ell_i)$ are the $N$ messenger propagators and $I_N$ is determined by the Feynman rules.

\tikzset{every picture/.style={line width=0.75pt}} %set default line width to 0.75pt       
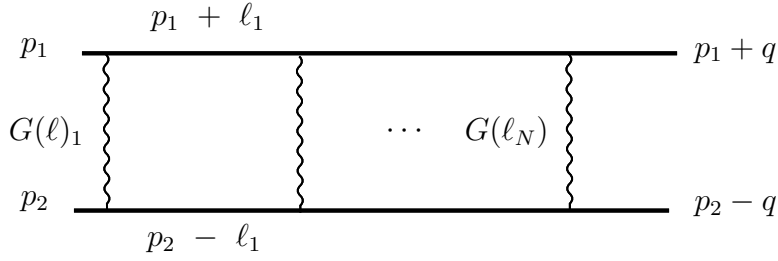
\begin{figure}
    \centering
    \label{fig:enter-label}

% Pattern Info
 
\tikzset{
pattern size/.store in=\mcSize, 
pattern size = 5pt,
pattern thickness/.store in=\mcThickness, 
pattern thickness = 0.3pt,
pattern radius/.store in=\mcRadius, 
pattern radius = 1pt}
\makeatletter
\pgfutil@ifundefined{pgf@pattern@name@_m49efpkn3}{
\pgfdeclarepatternformonly[\mcThickness,\mcSize]{_m49efpkn3}
{\pgfqpoint{0pt}{0pt}}
{\pgfpoint{\mcSize+\mcThickness}{\mcSize+\mcThickness}}
{\pgfpoint{\mcSize}{\mcSize}}
{
\pgfsetcolor{\tikz@pattern@color}
\pgfsetlinewidth{\mcThickness}
\pgfpathmoveto{\pgfqpoint{0pt}{0pt}}
\pgfpathlineto{\pgfpoint{\mcSize+\mcThickness}{\mcSize+\mcThickness}}
\pgfusepath{stroke}
}}
\makeatother

% Pattern Info
 
\tikzset{
pattern size/.store in=\mcSize, 
pattern size = 5pt,
pattern thickness/.store in=\mcThickness, 
pattern thickness = 0.3pt,
pattern radius/.store in=\mcRadius, 
pattern radius = 1pt}
\makeatletter
\pgfutil@ifundefined{pgf@pattern@name@_420zsw13h}{
\pgfdeclarepatternformonly[\mcThickness,\mcSize]{_420zsw13h}
{\pgfqpoint{0pt}{0pt}}
{\pgfpoint{\mcSize+\mcThickness}{\mcSize+\mcThickness}}
{\pgfpoint{\mcSize}{\mcSize}}
{
\pgfsetcolor{\tikz@pattern@color}
\pgfsetlinewidth{\mcThickness}
\pgfpathmoveto{\pgfqpoint{0pt}{0pt}}
\pgfpathlineto{\pgfpoint{\mcSize+\mcThickness}{\mcSize+\mcThickness}}
\pgfusepath{stroke}
}}
\makeatother

% Pattern Info
 
\tikzset{
pattern size/.store in=\mcSize, 
pattern size = 5pt,
pattern thickness/.store in=\mcThickness, 
pattern thickness = 0.3pt,
pattern radius/.store in=\mcRadius, 
pattern radius = 1pt}
\makeatletter
\pgfutil@ifundefined{pgf@pattern@name@_5ome1be7s}{
\pgfdeclarepatternformonly[\mcThickness,\mcSize]{_5ome1be7s}
{\pgfqpoint{0pt}{0pt}}
{\pgfpoint{\mcSize+\mcThickness}{\mcSize+\mcThickness}}
{\pgfpoint{\mcSize}{\mcSize}}
{
\pgfsetcolor{\tikz@pattern@color}
\pgfsetlinewidth{\mcThickness}
\pgfpathmoveto{\pgfqpoint{0pt}{0pt}}
\pgfpathlineto{\pgfpoint{\mcSize+\mcThickness}{\mcSize+\mcThickness}}
\pgfusepath{stroke}
}}
\makeatother
\tikzset{every picture/.style={line width=0.75pt}} %set default line width to 0.75pt        

\begin{tikzpicture}[x=0.75pt,y=0.75pt,yscale=-1,xscale=1]
%uncomment if require: \path (0,300); %set diagram left start at 0, and has height of 300

%Straight Lines [id:da15893471741940324] 
\draw [color={rgb, 255:red, 0; green, 0; blue, 0 }  ,draw opacity=1 ][line width=1.5]    (132.17,135.67) -- (381.56,135.63) ;
%Straight Lines [id:da06078962567816615] 
\draw    (155.61,134.69) .. controls (157.28,136.36) and (157.28,138.02) .. (155.62,139.69) .. controls (153.95,141.36) and (153.95,143.02) .. (155.62,144.69) .. controls (157.29,146.36) and (157.29,148.02) .. (155.63,149.69) .. controls (153.97,151.36) and (153.97,153.02) .. (155.64,154.69) .. controls (157.31,156.36) and (157.31,158.02) .. (155.64,159.69) .. controls (153.98,161.36) and (153.98,163.02) .. (155.65,164.69) .. controls (157.32,166.36) and (157.32,168.02) .. (155.66,169.69) .. controls (154,171.36) and (154,173.02) .. (155.67,174.69) .. controls (157.34,176.36) and (157.34,178.02) .. (155.67,179.69) .. controls (154.01,181.36) and (154.01,183.02) .. (155.68,184.69) .. controls (157.35,186.36) and (157.35,188.02) .. (155.69,189.69) .. controls (154.02,191.36) and (154.02,193.02) .. (155.69,194.69) .. controls (157.36,196.36) and (157.36,198.02) .. (155.7,199.69) .. controls (154.04,201.36) and (154.04,203.02) .. (155.71,204.69) .. controls (157.38,206.36) and (157.38,208.02) .. (155.72,209.69) -- (155.72,211.65) -- (155.72,211.65) ;
%Shape: Ellipse [id:dp4815430360779086] 
\draw  [pattern=_m49efpkn3,pattern size=3.2249999999999996pt,pattern thickness=0.75pt,pattern radius=0pt, pattern color={rgb, 255:red, 0; green, 0; blue, 0}] (155.72,211.65) .. controls (160.98,211.68) and (165.23,215.83) .. (165.2,220.93) .. controls (165.18,226.02) and (160.88,230.13) .. (155.62,230.1) .. controls (150.35,230.07) and (146.11,225.91) .. (146.13,220.82) .. controls (146.16,215.72) and (150.45,211.62) .. (155.72,211.65) -- cycle ;
%Straight Lines [id:da8006767780153381] 
\draw    (232.87,134.69) .. controls (234.56,136.34) and (234.57,138.01) .. (232.92,139.69) .. controls (231.27,141.37) and (231.29,143.04) .. (232.97,144.69) .. controls (234.65,146.34) and (234.67,148.01) .. (233.02,149.69) .. controls (231.37,151.37) and (231.39,153.04) .. (233.07,154.69) .. controls (234.75,156.34) and (234.77,158.01) .. (233.12,159.69) .. controls (231.47,161.37) and (231.49,163.04) .. (233.17,164.69) .. controls (234.85,166.34) and (234.86,168.01) .. (233.21,169.69) .. controls (231.56,171.37) and (231.58,173.04) .. (233.26,174.69) .. controls (234.94,176.34) and (234.96,178.01) .. (233.31,179.69) .. controls (231.66,181.37) and (231.68,183.04) .. (233.36,184.69) .. controls (235.04,186.34) and (235.06,188.01) .. (233.41,189.69) .. controls (231.76,191.37) and (231.78,193.04) .. (233.46,194.69) .. controls (235.14,196.34) and (235.16,198.01) .. (233.51,199.69) .. controls (231.86,201.37) and (231.88,203.04) .. (233.56,204.69) .. controls (235.24,206.34) and (235.26,208.01) .. (233.61,209.69) -- (233.63,211.65) -- (233.63,211.65) ;
%Shape: Ellipse [id:dp8466154453982724] 
\draw  [pattern=_420zsw13h,pattern size=3.2249999999999996pt,pattern thickness=0.75pt,pattern radius=0pt, pattern color={rgb, 255:red, 0; green, 0; blue, 0}] (233.63,211.65) .. controls (238.89,211.68) and (243.14,215.83) .. (243.11,220.93) .. controls (243.08,226.02) and (238.79,230.13) .. (233.53,230.1) .. controls (228.26,230.07) and (224.02,225.91) .. (224.04,220.82) .. controls (224.07,215.72) and (228.36,211.62) .. (233.63,211.65) -- cycle ;
%Straight Lines [id:da7183944955605346] 
\draw    (355.15,135.36) .. controls (356.82,137.03) and (356.82,138.69) .. (355.15,140.36) .. controls (353.48,142.03) and (353.48,143.69) .. (355.15,145.36) .. controls (356.82,147.03) and (356.82,148.69) .. (355.15,150.36) .. controls (353.48,152.03) and (353.48,153.69) .. (355.15,155.36) .. controls (356.82,157.03) and (356.82,158.69) .. (355.15,160.36) .. controls (353.48,162.03) and (353.48,163.69) .. (355.15,165.36) .. controls (356.82,167.03) and (356.82,168.69) .. (355.16,170.36) .. controls (353.49,172.03) and (353.49,173.69) .. (355.16,175.36) .. controls (356.83,177.03) and (356.83,178.69) .. (355.16,180.36) .. controls (353.49,182.03) and (353.49,183.69) .. (355.16,185.36) .. controls (356.83,187.03) and (356.83,188.69) .. (355.16,190.36) .. controls (353.49,192.03) and (353.49,193.69) .. (355.16,195.36) .. controls (356.83,197.03) and (356.83,198.69) .. (355.16,200.36) .. controls (353.5,202.03) and (353.5,203.69) .. (355.17,205.36) .. controls (356.84,207.03) and (356.84,208.69) .. (355.17,210.36) -- (355.17,211.65) -- (355.17,211.65) ;
%Shape: Ellipse [id:dp778517553410219] 
\draw  [pattern=_5ome1be7s,pattern size=3.2249999999999996pt,pattern thickness=0.75pt,pattern radius=0pt, pattern color={rgb, 255:red, 0; green, 0; blue, 0}] (355.17,211.65) .. controls (360.43,211.68) and (364.68,215.83) .. (364.65,220.93) .. controls (364.62,226.02) and (360.33,230.13) .. (355.07,230.1) .. controls (349.8,230.07) and (345.56,225.91) .. (345.58,220.82) .. controls (345.61,215.72) and (349.9,211.62) .. (355.17,211.65) -- cycle ;
%Straight Lines [id:da9742814454484854] 
\draw [color={rgb, 255:red, 0; green, 0; blue, 0 }  ,draw opacity=1 ][line width=1.5]    (165.2,220.93) -- (224.04,220.82) ;
%Straight Lines [id:da6913388692556836] 
\draw [color={rgb, 255:red, 0; green, 0; blue, 0 }  ,draw opacity=1 ][line width=1.5]    (243.11,220.93) -- (345.11,221.06) ;
%Straight Lines [id:da7855681019943294] 
\draw [color={rgb, 255:red, 0; green, 0; blue, 0 }  ,draw opacity=1 ][line width=1.5]    (364.65,220.93) -- (381.5,221) ;
%Straight Lines [id:da8016094619190814] 
\draw [color={rgb, 255:red, 0; green, 0; blue, 0 }  ,draw opacity=1 ][line width=1.5]    (131.81,220.93) -- (137.44,220.89) -- (146.13,220.82) ;

% Text Node
\draw (110.19,126.74) node [anchor=north west][inner sep=0.75pt]    {$p_{1}$};
% Text Node
\draw (266.69,175.00) node [anchor=north west][inner sep=0.75pt]    {$\cdots $};
% Text Node
\draw (110.19,210.33) node [anchor=north west][inner sep=0.75pt]    {$p_{2}$};
% Text Node
\draw (165.72,112.69) node [anchor=north west][inner sep=0.75pt]    {$p_{1} \ +\ \ell _{1}$};
% Text Node
\draw (165.72,227.19) node [anchor=north west][inner sep=0.75pt]    {$p_{2} \ -\ \ell _{1}$};
% Text Node
\draw (110.15,170.00) node [anchor=north west][inner sep=0.75pt]    {$G( \ell )_{1}$};
% Text Node
\draw (389.00,126.74) node [anchor=north west][inner sep=0.75pt]    {$p_{1} +q$};
% Text Node
\draw (389.00,210.33) node [anchor=north west][inner sep=0.75pt]    {$p_{2} -q$};
% Text Node
\draw (185.59,170.00) node [anchor=north west][inner sep=0.75pt]    {$G( \ell )_{2}$};
% Text Node
\draw (300.04,170.00) node [anchor=north west][inner sep=0.75pt]    {$G( \ell )_{N}$};
% Text Node
% \draw (-56.15,170.00) node [anchor=north west][inner sep=0.75pt]    {$\bra{p_1',p_2'}S\ket{p_1,p_2}_N=\ $};

\end{tikzpicture}
 \caption{Ladder diagram of the type contributing to $\bra{p_1',p_2'}S\ket{p_1,p_2}_N$ in the eikonal approximation. The leading eikonal phase is obtained by summing over permutations of diagrams of this topology. }
 \label{figure1}
\end{figure}
In the eikonal regime, we take the messenger momenta $\ell_i$ to be negligible compared to $p_i$. In terms of the mandelstam variables, this implies that $s \gg t$. In this regime, the dominant contribution to $I_N$ is of the form 

\[
I_N = \frac{N_1(\ell_1,\cdots \ell_N)}{2p_1 \cdot \ell_1 + i \epsilon \cdots (2p_1\cdot \ell_{1 \cdots N} + i \epsilon)}  \frac{N_2(\ell_1,\cdots \ell_N)}{2p_2 \cdot \ell_1 - i \epsilon \cdots (2p_2\cdot \ell_{1 \cdots N} - i \epsilon)},
\]
where $\ell_{1 \cdots N} \equiv \ell_1 + \cdots \ell_N$. Now, we can obtain all distinct Feynman diagrams of this type by fixing the momentum labels on one leg (say $p_1$) and summing over permutations $\sigma(i)$ of the labels $i = \{1,\cdots,N \}$ on the other. To express this in a more symmetric manner, we further \emph{average} over the momentum labels attached to line $1$. Performing these two operations, we end up with
\[
\sum_{\text{topologies}} I_N = \frac{1}{N!}\sum_{\sigma , \sigma'} &\frac{N_1(\ell_{\sigma(1)},\cdots \ell_{\sigma(N)})}{2p_1 \cdot \ell_{\sigma(1)} + i \epsilon \cdots (2p_1\cdot \ell_{\sigma(1) \cdots \sigma(N)} + i \epsilon)} \\
& \times \frac{N_2(\ell_{\sigma'(1)},\cdots \ell_{\sigma'(N)})}{2p_2 \cdot \ell_{\sigma'(1)} - i \epsilon \cdots (2p_2\cdot \ell_{\sigma'(1) \cdots \sigma'(N)} - i \epsilon)}. 
\]
A key simplification occurs when the numerators $N_1$ and $N_2$ are invariant under permutations of the momenta $\ell_i$. This is guaranteed to leading order in the eikonal approximation where these numerators are independent of the momenta $\ell_i$. In this case we invoke the eikonal identity~\cite{Akhoury:2013yua}
\[\label{eq:EikonalIdentity}
\sum_\sigma \frac{\hdelta(p_i \cdot \ell_{1\cdots N})}{(p \cdot \ell_{\sigma(1)} +i \epsilon) \cdots ( p \cdot \ell_{\sigma(1) \cdots \sigma(N-1)}+ i\epsilon)} = i^N \prod_{i=1}^N \hdelta(p \cdot \ell_i),
\]
to perform the sum over permutations. Further defining
\[
N_1 ( \ell_1, \cdots, \ell_N) N_2 ( \ell_1, \cdots, \ell_N) = (n(p_1,p_2))^N + \mathcal{O}(\ell_i),
\]
we find at leading order in the eikonal approximation:
\[
\bra{p_1',p_2'}S\ket{p_1,p_2}_N = \frac{1}{N!} \int \left( i^{2N} \prod_{i=1}^N  \hd^4 \ell_i \, G(\ell_i) \,  n(p_1,p_2) \, \hdelta(p_1 \cdot \ell_i) \hdelta(p_2 \cdot \ell_i) \right) \, \hdelta^4(\ell_i -q).
\]
At this point, a straightforward calculation reveals that taking the Fourier transform to impact parameter space and summing over $N$ leads to the exponentiation 
\[
1 + i  \mathcal{A}(s,b) = e^{i \chi(s,b)}, 
\]
where 
\[\label{eq:on-shell-FT}
\mathcal{A}(s,b)  =  \int \hd^4 q \, \hdelta(2 p_1 \cdot q + q^2) \hdelta(2p_2 \cdot q - q^2) \, e^{i q \cdot b} \, \mathcal{A}(p_1,p_2 \rightarrow p_1 + q, p_2 - q). 
\]
Let us emphasise that the standard eikonal approximation is obtained by applying the eikonal identity \eqref{eq:EikonalIdentity} to both matter lines. As we will see below, this eikonalisation also features in the probe limit. In that case however, the eikonal approximation is applied only to the matter line associated with the heavy particle.

Turning to the multi-channel case, we allow channel transitions in both internal and external lines for particle $p_1$ only, while particle $p_2$ is restricted to remain in the same channel.
With the aim of approaching the probe limit, we take particle $p_2$ to be centered at the origin such that $p_2^\mu=  (M_2,\vec{0})$. In the probe limit we assume that $M_2$ is much larger than the masses $M_\rma$ and $M_\rmb$ associated with particle 1. With this setup, we define the initial state
\[
\ket{\varphi_{\rma}} = \int \dd \Phi(p_{1\rma}, p_2) \varphi_\rma(p_1)\varphi(p_2) e^{i b \cdot p_1} \ket{p_{1\rma},p_2},
\]
where the subscript $\rma$ indicates the initial channel of particle $p_1$. Defining the conjugate state in a similar manner, we denote the amplitude
\[
\bra{\varphi_\rmb} S_{\rmb \rma} \ket{\varphi_\rma} = \int \dd \Phi(p_{1\rma}, p_{1\rmb}',p_2,p_2') \varphi_\rma(p_1)\varphi(p_2)\varphi^*_\rmb(p_1')\varphi^*(p_2') e^{i b \cdot (p_1 - p_1')}  \\
\times \bra{p_{1\rmb}',p_2'}S_{\rmb \rma}\ket{p_{1\rma},p_2}.
\]
Now, consider diagrams consisting of $N$ photon exchanges between two matter lines. Gauge invariance requires that the photon field $A_\mu(x)$ couples to some conserved current $J_{\mu}(x)$. Noting this, it is convenient to represent the amplitude in terms of the off-shell matrix elements:
\[\label{eq:GreensF}
&G^{(1) \mu_1 \cdots \mu_N}(\ell_1, \cdots \ell_N) = \int_{\{x\}_{N}}\; e^{ i \ell_1 \cdot x_1} \cdots e^{ i \ell_N \cdot x_N}    \bra{p_{1\rmb}'} T\{ J_{\rmb c_1}^{\mu_1}(x_1)  \cdots  J_{c_{N-1} \rma}^{\mu_N}(x_N)  \} \ket{p_{1\rma}}, \\
&G^{(2) \nu_1 \cdots \nu_N}(\ell_1, \cdots \ell_N) = \int_{\{y\}_{N}} \; e^{-i \ell_1 \cdot y_1} \cdots e^{ -i \ell_N \cdot y_N}    \bra{p_2'} T\{ J^{\nu_1}(y_1)  \cdots  J^{\nu_N}(y_N)  \} \ket{p_2}.
\]
Diagrammatically, these matrix elements represent the matter lines for particles $p_1$ and $p_2$ with $N$ vertex insertions, and with photon lines truncated. For particle $p_1$, we have also defined the current matrix $J_{c_1 c_2}^\mu$ as the conserved current which couples a photon to particles in channels $c_1$ and $c_2$. More specifically, this current is related to the three-point vertex $N^\mu$ in the limit of zero photon-momentum as follows
\[\label{eq:EikonalVertex}
N^\mu_{c_2 c_1}(p_1) \equiv \lim_{k \rightarrow 0} N^\mu(p_{1,c_1} \rightarrow p_{1,c_2}, k)  = \frac{\bra{p_{1,c_2}} J_{c_2 c_1}^{\mu}(0)\ket{p_{1,c_1}}}{(p_{1,c_1})_0 (2\pi)^3}.
\]
We can now construct the scattering amplitude with $N$ photon exchanges by sewing the two Green's functions in \eqref{eq:GreensF} with $N$ photon propagators and summing over all distinct permutations to get 
\[\label{eq: FactorizedAmplitude}
\bra{p_{1\rmb}',p_2'}S_{\rmb \rma} \ket{p_{1\rma},p_2}_N = \sum_{\sigma} \int_{ \{\ell\}_N }  \left(\frac{-i \eta^{\mu_1 \nu_{\sigma(1)}}}{\ell_1^2 + i \varepsilon} \cdots \frac{-i \eta^{\mu_N \nu_{\sigma(N)}}}{\ell_N^2 + i \varepsilon} \right)\\
(2p_{1\rma})_0 (2\pi)^3  G^{(1)}_{\mu_1 \cdots \mu_N}(\ell_1, \cdots \ell_N)  (2p_2)_0 (2\pi)^3  G^{(2)}_{\nu_1 \cdots \nu_N}(\ell_{\sigma(1)}, \cdots \ell_{\sigma(N)}).
\]
Here, we have generated all the relevant diagrams by fixing the vertices in line $p_1$ and summing over permutations of vertices in line $p_2$. Now, we apply the eikonal approximation by expressing all quantities to leading order in the photon momenta $\ell_i$. Focusing on line $p_2$, and working to leading order in the eikonal approximation we have 
\[
G^{(2)}_{\nu_1 \cdots \nu_N}(\ell_1, \cdots \ell_N) \approx \frac{(-i)^{N-1}}{2 (p_2)_0 (2\pi)^3} \hdelta^4 (\ell_{1\cdots N} + p_2'- p_2) \frac{ N_{\nu_1}(p_2) \cdots N_{\nu_N}(p_2) }{(2 p_2 \cdot \ell_1 + i\epsilon) \cdots(2p_2\cdot \ell_{1\cdots N-1} +  i\epsilon)},
\]
where we have dropped terms subleading in $\ell$ and defined $N^\nu(p_2)$ as in \eqref{eq:EikonalVertex}. In this limit we can further make the following approximation 
\[
\hdelta^4 (\ell_{1\cdots N} +p_2' - p_2) \approx (p_2)_0 \;\hdelta(p_2\cdot \ell_{1 \cdots N}) \hdelta^3(\bell_{1 \cdots N}+ \Vec{p}'_2-\Vec{p}_2).
\]
Putting everything together, and performing the sum over permutations in Eq. \eqref{eq: FactorizedAmplitude} using the identity \eqref{eq:EikonalIdentity}
we obtain 
\[\label{eq:ProbeEikonal}
\sum_\sigma G^{(2)}_{\nu_1 \cdots \nu_N}(\ell_{\sigma(1)}, \cdots \ell_{\sigma(N)})=   \frac{i}{(2\pi)^3}\; \hdelta^3(\bell_{1 \cdots N}+ \Vec{p}'_2-\Vec{p}_2) \left(\prod_{i=1}^N  \hdelta(2p_2\cdot \ell_i) N_{\nu_i}(p_2) \right).
\]
Using this result in Eq.\eqref{eq: FactorizedAmplitude} and reinstating the KMOC wavepackets yields the following expression   
\[\label{eq:AmplitudePotential}
\bra{\varphi_\rmb} S_{ \rmb \rma}  \ket{\varphi_\rma}_N =  i \int \dd \Phi(p_{1\rma}, p_{\rmb}',p_2,p_2') \varphi_\rma(p_1)\varphi(p_2)\varphi^*_\rmb(p_1')\varphi^*(p_2) e^{i b \cdot (p_1 - p_1')} \\
\times   2(p_{1\rma})_0 2 (p_2)_0 (2\pi)^3  \int_{\{x\}_{N}} \; \bra{p_{1\rmb}'} T\{ J_{\rmb c_1}^{\nu_1}(x_1)  \cdots  J_{c_{N-1} \rma}^{\nu_N}(x_N)  \} \ket{p_{1\rma}} \\
\times \left( \prod_{i=1}^N \int \hd^4 \ell_i \;e^{i \ell_i \cdot x_i}  \frac{ \hdelta(2p_2\cdot \ell_i) N_{\nu_i}(p_2)   }{\ell^2_i + i \epsilon}\right)  \hdelta^3(\bell_{1 \cdots N}+ \Vec{p}'_2-\Vec{p}_2)  .
\]
To see how this expression relates to the probe limit, we will find it convenient to isolate the following combination 
\[
V_{\nu_1 \cdots \nu_N}(\vec{x}_1 \cdots \vec{x}_N) &= \int \dd \Phi(p_2,p_2') \varphi(p_2)\varphi^*(p_2') \; 2 (p_2)_0  \\
&\times \left( \prod_{i=1}^N \int \hd^4 \ell_i \;e^{i \ell_i \cdot x_i}  \frac{ \hdelta(2p_2\cdot \ell_i) N_{\nu_i}(p_2)   }{\ell^2_i + i \epsilon}\right) \hdelta^3(\bell_{1 \cdots N}+ \Vec{p}'_2-\Vec{p}_2).
\]
Working in the rest frame of particle 2, defining $q = p_2'-p_2$, and using the delta functions we get
\[\label{eq:Potential}
V_{\nu_1 \cdots \nu_n}(\vec{x}_1 \cdots \vec{x}_N) =  \prod_{i=1}^N V_{\nu_i}(\vec{x}_i)\equiv \prod_{i=1}^N \left(\frac{1}{2(p_2)_0}\int \hd^{3} \ell_i \;   \frac{N_{\nu_i}(p_2)   }{\ell^2_i + i \epsilon} e^{-i \ell_i \cdot \vec{x}_i } \right),
\]
where we have left the wavefunction dependence implicit. In this notation, equation \eqref{eq:AmplitudePotential} reads:
\[\label{eq:ProbeAmplitude}
\bra{\varphi_\rmb} S_{\rmb \rma} \ket{\varphi_\rma}_N &= i \int \dd \Phi(p_{1\rma}, p_{\rmb}') \varphi_\rma(p_1)\varphi^*_\rmb(p_1') e^{i b \cdot (p - p')} \int_{\{x\}_{N}} \prod_{i=1}^N V_{\nu_i}(\vec{x}_i)  \\
& \times   (2p_{1\rma})_0 (2\pi)^3 \bra{p_{1\rmb}'} T\{ J_{\rmb c_1}^{\nu_1}(x_1)  \cdots  J_{c_{N-1} \rma}^{\nu_N}(x_N)  \} \ket{p_{1\rma}} \, .
\]
We can now identify this as the amplitude for particle $p_1$ in a background potential $V_{\nu_i}(\vec{x}_i)$. Notice that we have arrived at this result by applying the eikonal approximation to the matter line carrying particle $p_2$. Interactions with this heavy particle can therefore be captured by the background potential $V_\mu(\vec{x})$ in \eqref{eq:Potential}. We have not applied any approximations to the matter line $p_1$, which can be evaluated at any order in the eikonal approximation. 

Having seen how the eikonal approximation in QFT corresponds to the probe limit, we will now proceed by showing how to describe on-shell inelastic effects in a generic background looking at a specific model of spin and mass changing interactions. 

\subsection{Inelasticity effects from amplitudes}

In this part of the paper we will see a way to incorporate inelasticity effects through on-shell amplitudes that involve mass and spin-changing bodies, without ever referring to an action. To remain general we will begin by considering a generic vertex with changing degrees of freedom which we write as~\footnote{Similar inelastic amplitudes were studied in the context of Supergravity theories, where a dilaton is transformed into a Ramond-Ramond field~\cite{KoemansCollado:2018hss}. (See also~\cite{DiVecchia:2023frv} for off-diagonal transitions in the context of string theory). We thank Paolo di Vecchia for the discussion on this topic.} 

\[\label{vertexino}
\begin{tikzpicture}[x=0.8pt,y=0.8pt,yscale=-1,xscale=1]
\tikzset{
pattern size/.store in=\mcSize, 
pattern size = 5pt,
pattern thickness/.store in=\mcThickness, 
pattern thickness = 0.3pt,
pattern radius/.store in=\mcRadius, 
pattern radius = 1pt}
\makeatletter
\pgfutil@ifundefined{pgf@pattern@name@_r86dr5nuv}{
\pgfdeclarepatternformonly[\mcThickness,\mcSize]{_r86dr5nuv}
{\pgfqpoint{0pt}{0pt}}
{\pgfpoint{\mcSize+\mcThickness}{\mcSize+\mcThickness}}
{\pgfpoint{\mcSize}{\mcSize}}
{
\pgfsetcolor{\tikz@pattern@color}
\pgfsetlinewidth{\mcThickness}
\pgfpathmoveto{\pgfqpoint{0pt}{0pt}}
\pgfpathlineto{\pgfpoint{\mcSize+\mcThickness}{\mcSize+\mcThickness}}
\pgfusepath{stroke}
}}
\makeatother
\tikzset{every picture/.style={line width=0.75pt}} %set default line width to 0.75pt        

%uncomment if require: \path (0,16989); %set diagram left start at 0, and has height of 16989

%Straight Lines [id:da6034819781721428] 
\draw    (421.33,13777.83) -- (377.16,13778.16) ;
\draw [shift={(399.25,13778)}, rotate = 179.57] [fill={rgb, 255:red, 0; green, 0; blue, 0 }  ][line width=0.08]  [draw opacity=0] (5.36,-2.57) -- (0,0) -- (5.36,2.57) -- cycle    ;
%Straight Lines [id:da48803251948054593] 
\draw    (421.33,13777.83) .. controls (423.01,13779.48) and (423.02,13781.15) .. (421.37,13782.83) .. controls (419.72,13784.51) and (419.73,13786.18) .. (421.41,13787.83) .. controls (423.09,13789.48) and (423.1,13791.15) .. (421.45,13792.83) .. controls (419.8,13794.51) and (419.81,13796.18) .. (421.49,13797.83) .. controls (423.17,13799.48) and (423.19,13801.15) .. (421.54,13802.83) .. controls (419.89,13804.51) and (419.9,13806.18) .. (421.58,13807.83) .. controls (423.26,13809.48) and (423.27,13811.15) .. (421.62,13812.83) .. controls (419.97,13814.51) and (419.98,13816.18) .. (421.66,13817.83) -- (421.66,13818.88) -- (421.66,13818.88) ;
%Straight Lines [id:da21373913443961667] 
\draw    (463.99,13779.5) -- (421.33,13779.33)(464.01,13776.5) -- (421.34,13776.33) ;
\draw [shift={(442.67,13777.92)}, rotate = 180.23] [fill={rgb, 255:red, 0; green, 0; blue, 0 }  ][line width=0.08]  [draw opacity=0] (7.14,-3.43) -- (0,0) -- (7.14,3.43) -- cycle    ;
%Shape: Ellipse [id:dp7792992060279529] 
\draw  [fill={rgb, 255:red, 255; green, 255; blue, 255 }  ,fill opacity=1 ] (421.34,13815.17) .. controls (417.48,13815.17) and (414.34,13818.27) .. (414.34,13822.09) .. controls (414.34,13825.9) and (417.48,13829) .. (421.34,13829) .. controls (425.21,13829) and (428.34,13825.9) .. (428.34,13822.09) .. controls (428.34,13818.27) and (425.21,13815.17) .. (421.34,13815.17) -- cycle ;
%Shape: Ellipse [id:dp7448844087911832] 
\draw  [pattern=_r86dr5nuv,pattern size=3.2249999999999996pt,pattern thickness=0.75pt,pattern radius=0pt, pattern color={rgb, 255:red, 0; green, 0; blue, 0}] (421.19,13814.77) .. controls (425.04,13814.79) and (428.15,13817.98) .. (428.12,13821.9) .. controls (428.1,13825.82) and (424.96,13828.98) .. (421.1,13828.96) .. controls (417.25,13828.94) and (414.15,13825.75) .. (414.17,13821.83) .. controls (414.19,13817.91) and (417.33,13814.75) .. (421.19,13814.77) -- cycle ;
%Shape: Ellipse [id:dp5799512775975793] 
\draw  [color={rgb, 255:red, 0; green, 0; blue, 0 }  ,draw opacity=1 ][fill={rgb, 255:red, 74; green, 144; blue, 226 }  ,fill opacity=1 ] (421.36,13782.33) .. controls (423.83,13782.32) and (425.83,13780.29) .. (425.82,13777.81) .. controls (425.8,13775.32) and (423.78,13773.32) .. (421.31,13773.33) .. controls (418.83,13773.35) and (416.84,13775.37) .. (416.85,13777.86) .. controls (416.87,13780.34) and (418.88,13782.34) .. (421.36,13782.33) -- cycle ;

% Text Node
\draw (426,13792) node [anchor=north west][inner sep=0.75pt]    {$q$};
% Text Node
\draw (360.16,13773.16) node [anchor=north west][inner sep=0.75pt]    {$p$};
% Text Node
\draw (470.16,13768.16) node [anchor=north west][inner sep=0.75pt]    {$p+q, s,\{c\}$};
\draw (545.16,13775.16) node [anchor=north west][inner sep=0.75pt]    {$=i \mathcal{A}_{2,\rma\rmb}^{(0)} (p\to p+q;s, \{ c\}) =i\displaystyle{g_{\rma\rmb}(w)\frac{ \mathcal{D}_{\{ c\}}^{\mu_1\cdots \mu_s}\tilde J_{\mu_1\cdots \mu_s} }{q^2}}.$};

\end{tikzpicture}
\]
Let us explain this expression. First of all we note that the off-diagonal effective coupling $g_{\rma\rmb}(w)$  (represented as a non-fundamental interaction by the blue blob in the figure) is in principle  energy dependent and will have to be matched. However, we will see that for the class of diagrams we are discussing here only  the combination $|g_{\rma \rmb}(w)|^2 \rho(w)$ appears, and this can be matched to the results of \cite{Aoude:2023fdm}. 
Next, above we introduced  the symbol
\begin{equation}
\mathcal{D}_{\{ c\}}^{\mu_1\cdots \mu_s}\equiv\mathcal{D}^{\mu_1\cdots \mu_s}(p_\rma, p_\rmb, q; \varepsilon^s_{\{ c\}}),
\end{equation}
which is a generic Lorentz-invariant tensor which gathers all kinematic variables describing the mass/spin changing  state. This is also an $SU(2)$ tensor w.r.t. the little group index $c$. Instead, 
\begin{equation}
\tilde J_{\mu_1\cdots \mu_s} \equiv\tilde J_{\mu_1\cdots \mu_s} (q; V),
\end{equation}
is a gauge invariant  current/tensor that describes the background, above $V^\mu$ would be the time-like four-velocity of the source: $p^\mu_2= M_2 V^\mu$. We observe that since  the numerator is unspecified the background can be either gravitational or electromagnetic by a straightforward generalisation of the index structures. This will also give us the chance to discuss possible double copy relations in a future work~\footnote{Note that the minimal coupling of these amplitudes do indeed double-copy~\cite{Aoude:2023fdm} following the prescription of~\cite{Johansson:2019dnu}.}.

One can also express the tensors given above in terms of the massive-spinor helicity variables of~\cite{Arkani-Hamed:2017jhn}. These were already interpreted in terms of mass-changing effects by one of the authors and by his collaborator in~\cite{Aoude:2023fdm}; in our language this naturally becomes the off-diagonal three-point amplitude. Note also that in generic mass-changing amplitudes  higher dimensional operators can lead to multiple coefficients and Lorentz structures for the currents. However, as in the mass-preserving case, it is  possible to define a notion of minimal coupling with mass-changing ~\cite{Aoude:2023fdm}. In fact, it turns out that when one of the incoming particles is a scalar the amplitude is unique and agrees with a general definition of the minimal coupling, regardless of the second particle's spinning nature. Then, the on-shell spinor-helicity version of the three point amplitude can be written as
\begin{align}
q^2 \mathcal{A}_{2,\rma\rmb}^{(0)} (p\to p+q; s,\{ c\}) &= g_{\rma\rmb}(w)\mathcal{D}_{\{ c\}}^{\mu_1\cdots \mu_s}\tilde J_{\mu_1\cdots \mu_s}\\
&= g_{\rma\rmb}(w) M_\rma^{1-2s}\braket{(p+q)^{c} q}^{\odot(s-h)} [(p+q)^{c} q]^{\odot(s+h)}\hat{\delta}(q\cdot V).\nonumber
\end{align}
using the massive spinors  introduced in~\cite{Arkani-Hamed:2017jhn} and $h$ is the helicity of the exchanged messenger.

Let us move on to describe the leading effects of mass changing vertices on diagonal channels. This happens at one loop. We can write down the amplitude  in the  ``$\rma\rma$" channel corresponding to the diagram
\[\label{oneloopmassch}
\begin{tikzpicture}[x=0.8pt,y=0.8pt,yscale=-1,xscale=1]

% Pattern Info
 
\tikzset{
pattern size/.store in=\mcSize, 
pattern size = 5pt,
pattern thickness/.store in=\mcThickness, 
pattern thickness = 0.3pt,
pattern radius/.store in=\mcRadius, 
pattern radius = 1pt}
\makeatletter
\pgfutil@ifundefined{pgf@pattern@name@_8mvdolf4v}{
\pgfdeclarepatternformonly[\mcThickness,\mcSize]{_8mvdolf4v}
{\pgfqpoint{0pt}{0pt}}
{\pgfpoint{\mcSize+\mcThickness}{\mcSize+\mcThickness}}
{\pgfpoint{\mcSize}{\mcSize}}
{
\pgfsetcolor{\tikz@pattern@color}
\pgfsetlinewidth{\mcThickness}
\pgfpathmoveto{\pgfqpoint{0pt}{0pt}}
\pgfpathlineto{\pgfpoint{\mcSize+\mcThickness}{\mcSize+\mcThickness}}
\pgfusepath{stroke}
}}
\makeatother

% Pattern Info
 
\tikzset{
pattern size/.store in=\mcSize, 
pattern size = 5pt,
pattern thickness/.store in=\mcThickness, 
pattern thickness = 0.3pt,
pattern radius/.store in=\mcRadius, 
pattern radius = 1pt}
\makeatletter
\pgfutil@ifundefined{pgf@pattern@name@_hbbjwmrci}{
\pgfdeclarepatternformonly[\mcThickness,\mcSize]{_hbbjwmrci}
{\pgfqpoint{0pt}{0pt}}
{\pgfpoint{\mcSize+\mcThickness}{\mcSize+\mcThickness}}
{\pgfpoint{\mcSize}{\mcSize}}
{
\pgfsetcolor{\tikz@pattern@color}
\pgfsetlinewidth{\mcThickness}
\pgfpathmoveto{\pgfqpoint{0pt}{0pt}}
\pgfpathlineto{\pgfpoint{\mcSize+\mcThickness}{\mcSize+\mcThickness}}
\pgfusepath{stroke}
}}
\makeatother
\tikzset{every picture/.style={line width=0.75pt}} %set default line width to 0.75pt        

%uncomment if require: \path (0,16989); %set diagram left start at 0, and has height of 16989

%Straight Lines [id:da2130646982883273] 
\draw    (123.33,13776.83) -- (79.16,13777.16) ;
\draw [shift={(101.25,13777)}, rotate = 179.57] [fill={rgb, 255:red, 0; green, 0; blue, 0 }  ][line width=0.08]  [draw opacity=0] (5.36,-2.57) -- (0,0) -- (5.36,2.57) -- cycle    ;
%Straight Lines [id:da5502739318571641] 
\draw    (123.33,13776.83) .. controls (125.01,13778.49) and (125.02,13780.16) .. (123.36,13781.83) .. controls (121.71,13783.5) and (121.72,13785.17) .. (123.39,13786.83) .. controls (125.06,13788.49) and (125.07,13790.16) .. (123.42,13791.83) .. controls (121.76,13793.5) and (121.77,13795.17) .. (123.44,13796.83) .. controls (125.11,13798.49) and (125.12,13800.16) .. (123.47,13801.83) .. controls (121.82,13803.5) and (121.83,13805.17) .. (123.5,13806.83) .. controls (125.17,13808.49) and (125.18,13810.16) .. (123.53,13811.83) .. controls (121.87,13813.5) and (121.88,13815.17) .. (123.55,13816.83) .. controls (125.22,13818.49) and (125.23,13820.16) .. (123.58,13821.83) .. controls (121.93,13823.5) and (121.94,13825.17) .. (123.61,13826.83) .. controls (125.28,13828.49) and (125.29,13830.16) .. (123.64,13831.83) .. controls (121.98,13833.5) and (121.99,13835.17) .. (123.66,13836.83) -- (123.66,13837.16) -- (123.66,13837.16) ;
%Straight Lines [id:da36937175917194165] 
\draw    (210.33,13778.33) -- (123.33,13778.33)(210.33,13775.33) -- (123.33,13775.33) ;
\draw [shift={(166.83,13776.83)}, rotate = 180] [fill={rgb, 255:red, 0; green, 0; blue, 0 }  ][line width=0.08]  [draw opacity=0] (7.14,-3.43) -- (0,0) -- (7.14,3.43) -- cycle    ;
%Shape: Ellipse [id:dp7902048938473338] 
\draw  [fill={rgb, 255:red, 255; green, 255; blue, 255 }  ,fill opacity=1 ] (123.34,13831.55) .. controls (119.48,13831.55) and (116.34,13834.82) .. (116.34,13838.86) .. controls (116.34,13842.89) and (119.48,13846.16) .. (123.34,13846.16) .. controls (127.21,13846.16) and (130.34,13842.89) .. (130.34,13838.86) .. controls (130.34,13834.82) and (127.21,13831.55) .. (123.34,13831.55) -- cycle ;
%Shape: Ellipse [id:dp6359216226527717] 
\draw  [pattern=_8mvdolf4v,pattern size=3.2249999999999996pt,pattern thickness=0.75pt,pattern radius=0pt, pattern color={rgb, 255:red, 0; green, 0; blue, 0}] (123.19,13831.12) .. controls (127.04,13831.15) and (130.15,13834.52) .. (130.12,13838.66) .. controls (130.1,13842.81) and (126.96,13846.15) .. (123.11,13846.12) .. controls (119.25,13846.1) and (116.15,13842.73) .. (116.17,13838.58) .. controls (116.19,13834.44) and (119.33,13831.1) .. (123.19,13831.12) -- cycle ;
%Straight Lines [id:da874439339431293] 
\draw    (210.33,13778.83) .. controls (212.01,13780.49) and (212.02,13782.16) .. (210.36,13783.83) .. controls (208.71,13785.5) and (208.72,13787.17) .. (210.39,13788.83) .. controls (212.06,13790.49) and (212.07,13792.16) .. (210.42,13793.83) .. controls (208.76,13795.5) and (208.77,13797.17) .. (210.44,13798.83) .. controls (212.11,13800.49) and (212.12,13802.16) .. (210.47,13803.83) .. controls (208.82,13805.5) and (208.83,13807.17) .. (210.5,13808.83) .. controls (212.17,13810.49) and (212.18,13812.16) .. (210.53,13813.83) .. controls (208.87,13815.5) and (208.88,13817.17) .. (210.55,13818.83) .. controls (212.22,13820.49) and (212.23,13822.16) .. (210.58,13823.83) .. controls (208.93,13825.5) and (208.94,13827.17) .. (210.61,13828.83) .. controls (212.28,13830.49) and (212.29,13832.16) .. (210.64,13833.83) .. controls (208.98,13835.5) and (208.99,13837.17) .. (210.66,13838.83) -- (210.66,13839.16) -- (210.66,13839.16) ;
%Straight Lines [id:da21162442712663831] 
\draw    (254.5,13776.5) -- (210.33,13776.83) ;
\draw [shift={(232.42,13776.66)}, rotate = 179.57] [fill={rgb, 255:red, 0; green, 0; blue, 0 }  ][line width=0.08]  [draw opacity=0] (5.36,-2.57) -- (0,0) -- (5.36,2.57) -- cycle    ;
%Shape: Ellipse [id:dp13080618183211667] 
\draw  [fill={rgb, 255:red, 255; green, 255; blue, 255 }  ,fill opacity=1 ] (210.34,13831.55) .. controls (206.48,13831.55) and (203.34,13834.82) .. (203.34,13838.86) .. controls (203.34,13842.89) and (206.48,13846.16) .. (210.34,13846.16) .. controls (214.21,13846.16) and (217.34,13842.89) .. (217.34,13838.86) .. controls (217.34,13834.82) and (214.21,13831.55) .. (210.34,13831.55) -- cycle ;
%Shape: Ellipse [id:dp301870030392523] 
\draw  [pattern=_hbbjwmrci,pattern size=3.2249999999999996pt,pattern thickness=0.75pt,pattern radius=0pt, pattern color={rgb, 255:red, 0; green, 0; blue, 0}] (210.19,13831.12) .. controls (214.04,13831.15) and (217.15,13834.52) .. (217.12,13838.66) .. controls (217.1,13842.81) and (213.96,13846.15) .. (210.11,13846.12) .. controls (206.25,13846.1) and (203.15,13842.73) .. (203.17,13838.58) .. controls (203.19,13834.44) and (206.33,13831.1) .. (210.19,13831.12) -- cycle ;
%Shape: Ellipse [id:dp9219476585989548] 
\draw  [color={rgb, 255:red, 0; green, 0; blue, 0 }  ,draw opacity=1 ][fill={rgb, 255:red, 74; green, 144; blue, 226 }  ,fill opacity=1 ] (210.36,13781.33) .. controls (212.83,13781.32) and (214.83,13779.29) .. (214.82,13776.81) .. controls (214.8,13774.32) and (212.78,13772.32) .. (210.31,13772.33) .. controls (207.83,13772.35) and (205.84,13774.37) .. (205.85,13776.86) .. controls (205.87,13779.34) and (207.88,13781.34) .. (210.36,13781.33) -- cycle ;
%Shape: Ellipse [id:dp23601942787885455] 
\draw  [color={rgb, 255:red, 0; green, 0; blue, 0 }  ,draw opacity=1 ][fill={rgb, 255:red, 74; green, 144; blue, 226 }  ,fill opacity=1 ] (123.36,13781.33) .. controls (125.83,13781.32) and (127.83,13779.29) .. (127.82,13776.81) .. controls (127.8,13774.32) and (125.78,13772.32) .. (123.31,13772.33) .. controls (120.83,13772.35) and (118.84,13774.37) .. (118.85,13776.86) .. controls (118.87,13779.34) and (120.88,13781.34) .. (123.36,13781.33) -- cycle ;

% Text Node
\draw (59.16,13772.16) node [anchor=north west][inner sep=0.75pt]    {$p$};
% Text Node
\draw (266.16,13770.16) node [anchor=north west][inner sep=0.75pt]    {$p+q$};
% Text Node
\draw (130.16,13750.16) node [anchor=north west][inner sep=0.75pt]    {$p+\ell _{1},s,\{c\}$};
% Text Node
\draw (134.16,13797.16) node [anchor=north west][inner sep=0.75pt]    {$\ell _{1}$};
% Text Node
\draw (219.16,13797.16) node [anchor=north west][inner sep=0.75pt]    {$\ell _{2}$};

\end{tikzpicture}
\]
using the vertex in Eq. \eqref{vertexino}, this is
\[\label{ampli1loop}
\mathcal{A}_{2;\rma\rma}^{(1)}(p \rightarrow p+q)= 2M_\rma\int_{0}^\infty &\dd w\,|g_{\rma\rmb}(w)|^2\rho(w) 
\\&\times
 \int\frac{\hat{\dd}^4 \ell_1  }{\ell^2_1  }\frac{ \hat \dd^4 \ell_2}{  \ell^2_2}\,\frac{ \text{N}(\ell_1,\ell_2)}{2p_\rma\cdot \ell_1-2M_\rma w}\hat{\delta}^{4}(\ell_1+\ell_2-q)
.
\]
where we remind the reader that we have made a change of integration variables $\dd M_{\rmb}^2 = 2M_{\rma}\dd w$ for the internal mass. Note that the integral $\dd w$ introduces an additional power of $\hbar$ which ensures that this contribution has classical scaling at leading order.  
Above we have defined the theory-dependent numerator by the following piece of notation 
\[\label{notat}
 \text{N}(\ell_1, \ell_2)&\equiv \sum_{\{c\}} \tilde J_{\mu_1\cdots \mu_s} (\ell_1; V) \tilde J_{\nu_1\cdots \nu_s}^* (\ell_2; V) \mathcal{D}^{\mu_1\cdots \mu_s}(p, \ell_1; \varepsilon^s_{\{ c\}})\mathcal{D}^{\nu_1\cdots \nu_s}(p, \ell_2; \varepsilon^{*s}_{\{ c\}})
 \\&
=\tilde J_{\mu_1\cdots \mu_s} (\ell_1; V) \tilde J_{\nu_1\cdots \nu_s}^* (\ell_2; V) P^{\mu_1\cdots \mu_s, \nu_1\cdots \nu_s},
\]
where
\begin{equation}
    P^{\mu_1\cdots \mu_s, \nu_1\cdots \nu_s} \equiv  \sum_{\{c\}} \mathcal{D}^{\mu_1\cdots \mu_s}(p, \ell_1; \varepsilon^s_{\{ c\}})\mathcal{D}^{\nu_1\cdots \nu_s}(p, \ell_2; \varepsilon^{*s}_{\{ c\}}),
\end{equation}
is the generalised sum over intermediate states which acts as a metric for the sources $\tilde{J}$. Note that the internal massive propagator in \eqref{ampli1loop} comes from an additional $\dd w$ integral over the intermediate mass, whereas we hid the sum over the exchanged spin d.o.f. inside the definition of $\text{N}(\ell_1, \ell_2).$
Following  \cite{Jones:2023ugm},  we can argue next that the real part of the propagator is odd under  $\ell_1\to \ell_2$ so that its contribution to the integral vanishes. Thus, under the integral, we can make the replacement
\begin{equation}
\frac{1}{2p_\rma\cdot \ell_1-2M_\rma w+i\epsilon} \rightarrow \text{Im}\bigg(\, \frac{1}{2p_\rma\cdot \ell_1-2M_\rma w+i\epsilon}\bigg)=-\frac{i}{4M_\rma}{\hat{\delta}(u\cdot \ell-w)},
\end{equation}
holds true.
After integrating over $w$ with the Dirac delta  we are left with the following weighted scalar  integral
\[
\mathcal{A}_{2;\rma\rma}^{(1)}(p \rightarrow p+q)&=
\frac{-i g_{\rma\rma}}{2}
\int\frac{\hat{\dd}^4 \ell_1  }{\ell^2_1  }\frac{\hat \dd^4 \ell_2}{  \ell^2_2}\,\rho^{(+)}({u\cdot \ell_1 })  \text{N}(\ell_1,\ell_2)\hat{\delta}^{4}(\ell_1+\ell_2-q) 
,
\] 
having used $p^\mu = M_\rma u^\mu$ and defined 
\begin{equation}
    \rho^{(+)}({u\cdot \ell })\equiv \frac{|g_{\rma\rmb}(u\cdot \ell_1)|^2}{g_{\rma\rma}}\rho({u\cdot \ell })\Theta(u\cdot \ell).
\end{equation}
In the equation above   we observed that the combination of effective coupling and spectral function $ 
|g_{\rma\rmb}(w)|^2 \rho(w)$  can be formally matched as done in \cite{Aoude:2023fdm}. To this end we have also rescaled the one loop amplitude by an overall factor of $g_{\rma\rma}$ noting from \cite{Aoude:2023fdm} that the combination of interest always scales, at least, as $ |g_{\rma\rmb}(w)|^2 \rho(w)\sim \mathcal{O}(g_{\rma\rma}),$\footnote{See equation 5.8 there for instance, $g_{0,0,s_2}$ translates to our $g_{\rma\rmb}$ and $r_s \sim g_{\rma\rma}$ (for the gravitational case).} so no singular behaviour can arise.

Following this, we compute the on-shell Fourier transform in impact parameter space. Integrating against the support given in Eq. \ref{eq:on-shell-FT} yields
\begin{equation}\label{eta1l}
\mathcal{A}_{2;\rma\rma}^{(1)}({b})=
\frac{-ig_{\rma\rma}}{4}
\int\frac{\hat{\dd}^4 \ell_1  }{\ell^2_1  }\frac{ \hat\dd^4 \ell_2}{  \ell^2_2}\,\rho^{(+)}({u\cdot \ell_1 })  \text{N}(\ell_1,\ell_2) \hat{\delta}(u\cdot \ell_{12})e^{-ib\cdot(\ell_1+\ell_2 )}.
\end{equation}
It is natural at this stage, to wonder if this one loop process itself controls a resummation of given topologies with mass-changing effects, in the similar way as a tree level graviton exchange, controls the sum of ladder and cross-ladder diagrams. To see if this is the case, let's consider the following process at three loops which can be interpreted as the iterated topology of the one-loop process already studied:
\[\label{diagram3l}
\begin{tikzpicture}[x=0.8pt,y=0.8pt,yscale=-1,xscale=1]

% Pattern Info
 
\tikzset{
pattern size/.store in=\mcSize, 
pattern size = 5pt,
pattern thickness/.store in=\mcThickness, 
pattern thickness = 0.3pt,
pattern radius/.store in=\mcRadius, 
pattern radius = 1pt}
\makeatletter
\pgfutil@ifundefined{pgf@pattern@name@_1o0fwmz0c}{
\pgfdeclarepatternformonly[\mcThickness,\mcSize]{_1o0fwmz0c}
{\pgfqpoint{0pt}{0pt}}
{\pgfpoint{\mcSize+\mcThickness}{\mcSize+\mcThickness}}
{\pgfpoint{\mcSize}{\mcSize}}
{
\pgfsetcolor{\tikz@pattern@color}
\pgfsetlinewidth{\mcThickness}
\pgfpathmoveto{\pgfqpoint{0pt}{0pt}}
\pgfpathlineto{\pgfpoint{\mcSize+\mcThickness}{\mcSize+\mcThickness}}
\pgfusepath{stroke}
}}
\makeatother

% Pattern Info
 
\tikzset{
pattern size/.store in=\mcSize, 
pattern size = 5pt,
pattern thickness/.store in=\mcThickness, 
pattern thickness = 0.3pt,
pattern radius/.store in=\mcRadius, 
pattern radius = 1pt}
\makeatletter
\pgfutil@ifundefined{pgf@pattern@name@_skhrst558}{
\pgfdeclarepatternformonly[\mcThickness,\mcSize]{_skhrst558}
{\pgfqpoint{0pt}{0pt}}
{\pgfpoint{\mcSize+\mcThickness}{\mcSize+\mcThickness}}
{\pgfpoint{\mcSize}{\mcSize}}
{
\pgfsetcolor{\tikz@pattern@color}
\pgfsetlinewidth{\mcThickness}
\pgfpathmoveto{\pgfqpoint{0pt}{0pt}}
\pgfpathlineto{\pgfpoint{\mcSize+\mcThickness}{\mcSize+\mcThickness}}
\pgfusepath{stroke}
}}
\makeatother

% Pattern Info
 
\tikzset{
pattern size/.store in=\mcSize, 
pattern size = 5pt,
pattern thickness/.store in=\mcThickness, 
pattern thickness = 0.3pt,
pattern radius/.store in=\mcRadius, 
pattern radius = 1pt}
\makeatletter
\pgfutil@ifundefined{pgf@pattern@name@_lq9pjbbq4}{
\pgfdeclarepatternformonly[\mcThickness,\mcSize]{_lq9pjbbq4}
{\pgfqpoint{0pt}{0pt}}
{\pgfpoint{\mcSize+\mcThickness}{\mcSize+\mcThickness}}
{\pgfpoint{\mcSize}{\mcSize}}
{
\pgfsetcolor{\tikz@pattern@color}
\pgfsetlinewidth{\mcThickness}
\pgfpathmoveto{\pgfqpoint{0pt}{0pt}}
\pgfpathlineto{\pgfpoint{\mcSize+\mcThickness}{\mcSize+\mcThickness}}
\pgfusepath{stroke}
}}
\makeatother

% Pattern Info
 
\tikzset{
pattern size/.store in=\mcSize, 
pattern size = 5pt,
pattern thickness/.store in=\mcThickness, 
pattern thickness = 0.3pt,
pattern radius/.store in=\mcRadius, 
pattern radius = 1pt}
\makeatletter
\pgfutil@ifundefined{pgf@pattern@name@_9e02hcvrn}{
\pgfdeclarepatternformonly[\mcThickness,\mcSize]{_9e02hcvrn}
{\pgfqpoint{0pt}{0pt}}
{\pgfpoint{\mcSize+\mcThickness}{\mcSize+\mcThickness}}
{\pgfpoint{\mcSize}{\mcSize}}
{
\pgfsetcolor{\tikz@pattern@color}
\pgfsetlinewidth{\mcThickness}
\pgfpathmoveto{\pgfqpoint{0pt}{0pt}}
\pgfpathlineto{\pgfpoint{\mcSize+\mcThickness}{\mcSize+\mcThickness}}
\pgfusepath{stroke}
}}
\makeatother
\tikzset{every picture/.style={line width=0.75pt}} %set default line width to 0.75pt        

%uncomment if require: \path (0,16989); %set diagram left start at 0, and has height of 16989

%Straight Lines [id:da23490118009749494] 
\draw    (190.67,13920.17) -- (146.5,13920.5) ;
\draw [shift={(168.58,13920.33)}, rotate = 179.57] [fill={rgb, 255:red, 0; green, 0; blue, 0 }  ][line width=0.08]  [draw opacity=0] (5.36,-2.57) -- (0,0) -- (5.36,2.57) -- cycle    ;
%Straight Lines [id:da7293604095358261] 
\draw    (191,13919.83) .. controls (192.67,13921.49) and (192.68,13923.16) .. (191.03,13924.83) .. controls (189.37,13926.5) and (189.38,13928.17) .. (191.05,13929.83) .. controls (192.72,13931.49) and (192.73,13933.16) .. (191.08,13934.83) .. controls (189.43,13936.5) and (189.44,13938.17) .. (191.11,13939.83) .. controls (192.78,13941.49) and (192.79,13943.16) .. (191.14,13944.83) .. controls (189.48,13946.5) and (189.49,13948.17) .. (191.16,13949.83) .. controls (192.83,13951.49) and (192.84,13953.16) .. (191.19,13954.83) .. controls (189.54,13956.5) and (189.55,13958.17) .. (191.22,13959.83) .. controls (192.89,13961.49) and (192.9,13963.16) .. (191.25,13964.83) .. controls (189.59,13966.5) and (189.6,13968.17) .. (191.27,13969.83) .. controls (192.94,13971.49) and (192.95,13973.16) .. (191.3,13974.83) .. controls (189.65,13976.5) and (189.66,13978.17) .. (191.33,13979.83) -- (191.33,13980.17) -- (191.33,13980.17) ;
%Straight Lines [id:da14724584070156377] 
\draw    (251.01,13921.33) -- (190.68,13921.67)(250.99,13918.33) -- (190.66,13918.67) ;
\draw [shift={(220.83,13920)}, rotate = 179.68] [fill={rgb, 255:red, 0; green, 0; blue, 0 }  ][line width=0.08]  [draw opacity=0] (7.14,-3.43) -- (0,0) -- (7.14,3.43) -- cycle    ;
%Shape: Ellipse [id:dp049449751004948306] 
\draw  [fill={rgb, 255:red, 255; green, 255; blue, 255 }  ,fill opacity=1 ] (191.01,13974.55) .. controls (187.14,13974.55) and (184.01,13977.83) .. (184.01,13981.86) .. controls (184.01,13985.9) and (187.14,13989.17) .. (191.01,13989.17) .. controls (194.87,13989.17) and (198,13985.9) .. (198,13981.86) .. controls (198,13977.83) and (194.87,13974.55) .. (191.01,13974.55) -- cycle ;
%Shape: Ellipse [id:dp31524902287746737] 
\draw  [pattern=_1o0fwmz0c,pattern size=3.2249999999999996pt,pattern thickness=0.75pt,pattern radius=0pt, pattern color={rgb, 255:red, 0; green, 0; blue, 0}] (190.85,13974.13) .. controls (194.71,13974.15) and (197.81,13977.53) .. (197.79,13981.67) .. controls (197.77,13985.81) and (194.62,13989.15) .. (190.77,13989.13) .. controls (186.92,13989.1) and (183.81,13985.73) .. (183.83,13981.59) .. controls (183.86,13977.44) and (187,13974.11) .. (190.85,13974.13) -- cycle ;
%Straight Lines [id:da5678541424028389] 
\draw    (251,13921.83) .. controls (252.67,13923.49) and (252.68,13925.16) .. (251.03,13926.83) .. controls (249.37,13928.5) and (249.38,13930.17) .. (251.05,13931.83) .. controls (252.72,13933.49) and (252.73,13935.16) .. (251.08,13936.83) .. controls (249.43,13938.5) and (249.44,13940.17) .. (251.11,13941.83) .. controls (252.78,13943.49) and (252.79,13945.16) .. (251.14,13946.83) .. controls (249.48,13948.5) and (249.49,13950.17) .. (251.16,13951.83) .. controls (252.83,13953.49) and (252.84,13955.16) .. (251.19,13956.83) .. controls (249.54,13958.5) and (249.55,13960.17) .. (251.22,13961.83) .. controls (252.89,13963.49) and (252.9,13965.16) .. (251.25,13966.83) .. controls (249.59,13968.5) and (249.6,13970.17) .. (251.27,13971.83) .. controls (252.94,13973.49) and (252.95,13975.16) .. (251.3,13976.83) .. controls (249.65,13978.5) and (249.66,13980.17) .. (251.33,13981.83) -- (251.33,13982.17) -- (251.33,13982.17) ;
%Shape: Ellipse [id:dp973555499374356] 
\draw  [fill={rgb, 255:red, 255; green, 255; blue, 255 }  ,fill opacity=1 ] (251.01,13974.55) .. controls (247.14,13974.55) and (244.01,13977.83) .. (244.01,13981.86) .. controls (244.01,13985.9) and (247.14,13989.17) .. (251.01,13989.17) .. controls (254.87,13989.17) and (258,13985.9) .. (258,13981.86) .. controls (258,13977.83) and (254.87,13974.55) .. (251.01,13974.55) -- cycle ;
%Shape: Ellipse [id:dp7468393283154797] 
\draw  [pattern=_skhrst558,pattern size=3.2249999999999996pt,pattern thickness=0.75pt,pattern radius=0pt, pattern color={rgb, 255:red, 0; green, 0; blue, 0}] (250.85,13974.13) .. controls (254.71,13974.15) and (257.81,13977.53) .. (257.79,13981.67) .. controls (257.77,13985.81) and (254.62,13989.15) .. (250.77,13989.13) .. controls (246.92,13989.1) and (243.81,13985.73) .. (243.83,13981.59) .. controls (243.86,13977.44) and (247,13974.11) .. (250.85,13974.13) -- cycle ;
%Straight Lines [id:da0023059291820244354] 
\draw    (315,13919.83) -- (251,13919.83) ;
\draw [shift={(283,13919.83)}, rotate = 180] [fill={rgb, 255:red, 0; green, 0; blue, 0 }  ][line width=0.08]  [draw opacity=0] (5.36,-2.57) -- (0,0) -- (5.36,2.57) -- cycle    ;
%Straight Lines [id:da45498401249581] 
\draw    (315,13919.83) .. controls (316.67,13921.49) and (316.68,13923.16) .. (315.03,13924.83) .. controls (313.37,13926.5) and (313.38,13928.17) .. (315.05,13929.83) .. controls (316.72,13931.49) and (316.73,13933.16) .. (315.08,13934.83) .. controls (313.43,13936.5) and (313.44,13938.17) .. (315.11,13939.83) .. controls (316.78,13941.49) and (316.79,13943.16) .. (315.14,13944.83) .. controls (313.48,13946.5) and (313.49,13948.17) .. (315.16,13949.83) .. controls (316.83,13951.49) and (316.84,13953.16) .. (315.19,13954.83) .. controls (313.54,13956.5) and (313.55,13958.17) .. (315.22,13959.83) .. controls (316.89,13961.49) and (316.9,13963.16) .. (315.25,13964.83) .. controls (313.59,13966.5) and (313.6,13968.17) .. (315.27,13969.83) .. controls (316.94,13971.49) and (316.95,13973.16) .. (315.3,13974.83) .. controls (313.65,13976.5) and (313.66,13978.17) .. (315.33,13979.83) -- (315.33,13980.17) -- (315.33,13980.17) ;
%Straight Lines [id:da659954411960298] 
\draw    (375,13921.33) -- (315,13921.33)(375,13918.33) -- (315,13918.33) ;
\draw [shift={(345,13919.83)}, rotate = 180] [fill={rgb, 255:red, 0; green, 0; blue, 0 }  ][line width=0.08]  [draw opacity=0] (7.14,-3.43) -- (0,0) -- (7.14,3.43) -- cycle    ;
%Shape: Ellipse [id:dp9876105936117667] 
\draw  [fill={rgb, 255:red, 255; green, 255; blue, 255 }  ,fill opacity=1 ] (315.01,13974.55) .. controls (311.14,13974.55) and (308.01,13977.83) .. (308.01,13981.86) .. controls (308.01,13985.9) and (311.14,13989.17) .. (315.01,13989.17) .. controls (318.87,13989.17) and (322,13985.9) .. (322,13981.86) .. controls (322,13977.83) and (318.87,13974.55) .. (315.01,13974.55) -- cycle ;
%Shape: Ellipse [id:dp28752614452551417] 
\draw  [pattern=_lq9pjbbq4,pattern size=3.2249999999999996pt,pattern thickness=0.75pt,pattern radius=0pt, pattern color={rgb, 255:red, 0; green, 0; blue, 0}] (314.85,13974.13) .. controls (318.71,13974.15) and (321.81,13977.53) .. (321.79,13981.67) .. controls (321.77,13985.81) and (318.62,13989.15) .. (314.77,13989.13) .. controls (310.92,13989.1) and (307.81,13985.73) .. (307.83,13981.59) .. controls (307.86,13977.44) and (311,13974.11) .. (314.85,13974.13) -- cycle ;
%Straight Lines [id:da6850451352367555] 
\draw    (375,13921.83) .. controls (376.67,13923.49) and (376.68,13925.16) .. (375.03,13926.83) .. controls (373.37,13928.5) and (373.38,13930.17) .. (375.05,13931.83) .. controls (376.72,13933.49) and (376.73,13935.16) .. (375.08,13936.83) .. controls (373.43,13938.5) and (373.44,13940.17) .. (375.11,13941.83) .. controls (376.78,13943.49) and (376.79,13945.16) .. (375.14,13946.83) .. controls (373.48,13948.5) and (373.49,13950.17) .. (375.16,13951.83) .. controls (376.83,13953.49) and (376.84,13955.16) .. (375.19,13956.83) .. controls (373.54,13958.5) and (373.55,13960.17) .. (375.22,13961.83) .. controls (376.89,13963.49) and (376.9,13965.16) .. (375.25,13966.83) .. controls (373.59,13968.5) and (373.6,13970.17) .. (375.27,13971.83) .. controls (376.94,13973.49) and (376.95,13975.16) .. (375.3,13976.83) .. controls (373.65,13978.5) and (373.66,13980.17) .. (375.33,13981.83) -- (375.33,13982.17) -- (375.33,13982.17) ;
%Straight Lines [id:da9934980983704906] 
\draw    (419.17,13919.5) -- (375,13919.83) ;
\draw [shift={(397.08,13919.67)}, rotate = 179.57] [fill={rgb, 255:red, 0; green, 0; blue, 0 }  ][line width=0.08]  [draw opacity=0] (5.36,-2.57) -- (0,0) -- (5.36,2.57) -- cycle    ;
%Shape: Ellipse [id:dp45622636645852976] 
\draw  [fill={rgb, 255:red, 255; green, 255; blue, 255 }  ,fill opacity=1 ] (375.01,13974.55) .. controls (371.14,13974.55) and (368.01,13977.83) .. (368.01,13981.86) .. controls (368.01,13985.9) and (371.14,13989.17) .. (375.01,13989.17) .. controls (378.87,13989.17) and (382,13985.9) .. (382,13981.86) .. controls (382,13977.83) and (378.87,13974.55) .. (375.01,13974.55) -- cycle ;
%Shape: Ellipse [id:dp7809338930340579] 
\draw  [pattern=_9e02hcvrn,pattern size=3.2249999999999996pt,pattern thickness=0.75pt,pattern radius=0pt, pattern color={rgb, 255:red, 0; green, 0; blue, 0}] (374.85,13974.13) .. controls (378.71,13974.15) and (381.81,13977.53) .. (381.79,13981.67) .. controls (381.77,13985.81) and (378.62,13989.15) .. (374.77,13989.13) .. controls (370.92,13989.1) and (367.81,13985.73) .. (367.83,13981.59) .. controls (367.86,13977.44) and (371,13974.11) .. (374.85,13974.13) -- cycle ;
%Shape: Ellipse [id:dp6109486901571308] 
\draw  [fill={rgb, 255:red, 255; green, 255; blue, 255 }  ,fill opacity=1 ] (250.51,13924) .. controls (248.03,13924) and (246.01,13922.04) .. (246.01,13919.62) .. controls (246.01,13917.19) and (248.03,13915.23) .. (250.51,13915.23) .. controls (252.99,13915.23) and (255,13917.19) .. (255,13919.62) .. controls (255,13922.04) and (252.99,13924) .. (250.51,13924) -- cycle ;
%Shape: Ellipse [id:dp5555802267273335] 
\draw  [color={rgb, 255:red, 0; green, 0; blue, 0 }  ,draw opacity=1 ][fill={rgb, 255:red, 74; green, 144; blue, 226 }  ,fill opacity=1 ] (250.51,13924) .. controls (252.98,13923.99) and (254.98,13921.96) .. (254.96,13919.48) .. controls (254.95,13916.99) and (252.93,13914.99) .. (250.46,13915) .. controls (247.98,13915.01) and (245.99,13917.04) .. (246,13919.52) .. controls (246.01,13922.01) and (248.03,13924.01) .. (250.51,13924) -- cycle ;
%Shape: Ellipse [id:dp4994057726793333] 
\draw  [color={rgb, 255:red, 0; green, 0; blue, 0 }  ,draw opacity=1 ][fill={rgb, 255:red, 74; green, 144; blue, 226 }  ,fill opacity=1 ] (190.69,13924.67) .. controls (193.17,13924.65) and (195.16,13922.63) .. (195.15,13920.14) .. controls (195.14,13917.66) and (193.12,13915.65) .. (190.64,13915.67) .. controls (188.17,13915.68) and (186.17,13917.71) .. (186.19,13920.19) .. controls (186.2,13922.68) and (188.22,13924.68) .. (190.69,13924.67) -- cycle ;
%Shape: Ellipse [id:dp26502982514068973] 
\draw  [color={rgb, 255:red, 0; green, 0; blue, 0 }  ,draw opacity=1 ][fill={rgb, 255:red, 74; green, 144; blue, 226 }  ,fill opacity=1 ] (315.02,13924.33) .. controls (317.5,13924.32) and (319.5,13922.3) .. (319.48,13919.81) .. controls (319.47,13917.33) and (317.45,13915.32) .. (314.97,13915.33) .. controls (312.5,13915.35) and (310.5,13917.37) .. (310.52,13919.86) .. controls (310.53,13922.34) and (312.55,13924.35) .. (315.02,13924.33) -- cycle ;
%Shape: Ellipse [id:dp9839606120245497] 
\draw  [color={rgb, 255:red, 0; green, 0; blue, 0 }  ,draw opacity=1 ][fill={rgb, 255:red, 74; green, 144; blue, 226 }  ,fill opacity=1 ] (375.02,13924.33) .. controls (377.5,13924.32) and (379.5,13922.3) .. (379.48,13919.81) .. controls (379.47,13917.33) and (377.45,13915.32) .. (374.97,13915.33) .. controls (372.5,13915.35) and (370.5,13917.37) .. (370.52,13919.86) .. controls (370.53,13922.34) and (372.55,13924.35) .. (375.02,13924.33) -- cycle ;
\draw (196.16,13947.16) node [anchor=north west][inner sep=0.75pt]    {$\ell _{1}$};
\draw (256.16,13947.16) node [anchor=north west][inner sep=0.75pt]    {$\ell _{2}$};
\draw (320.16,13947.16) node [anchor=north west][inner sep=0.75pt]    {$\ell _{3}$};
\draw (380.16,13947.16) node [anchor=north west][inner sep=0.75pt]    {$\ell _{4}$};

\end{tikzpicture}
\]
Defining $  \ell_{i\cdots j}^\mu=\sum_{k=i}^j \ell^\mu_k$, we have that this process is given in momentum space as 
\[\mathcal{A}_{2;\rma\rma}^{(3)}(p \rightarrow p+q)= \frac{g^2_{\rma\rma} }{16}\int &\prod_{i=1}^4\frac{\hat{\dd}^4 \ell_i}{\ell_i^2}\, i\rho({u\cdot \ell_1})\text{N}(\ell_1, \ell_2) \hat{\delta}(u\cdot \ell_{12}) \\ & \quad \times  i\rho({u\cdot \ell_{123}}) \text{N}(\ell_3, \ell_4) {\hat{\delta}^{(4)}(q-\ell_{1234})}.
\]
In impact parameter space, we obtain \eqref{eq:AmplitudePotential}
\begin{align} 
\mathcal{A}_{2;\rma\rma}^{(3)}( b)
&= \frac{g^2_{\rma\rma}}{32}
\int\prod_{i=1}^4\frac{\hat{\dd}^4 \ell_i}{\ell_i^2}e^{-ib\cdot \ell_i}\, 
i\rho({u\cdot \ell_1})\text{N}(\ell_1, \ell_2) 
\hat{\delta}(u\cdot \ell_{12})i\rho({u\cdot \ell_{3}}) \text{N}(\ell_3, \ell_4)  {\hat{\delta} (u\cdot \ell_{34})}
\nonumber\\
&= \frac{1}{2}
\left(\frac{-ig_{\rma\rma}}{4}\int \frac{\hat{\dd}^4 \ell_1 }{\ell_1^2 } \frac{ \hat{\dd}^4 \ell_2}{ \ell_2^2}  e^{-ib\cdot(\ell_1+\ell_2)}\, 
\rho^{(+)}({u\cdot \ell_1}){\text{N}(\ell_1, \ell_2)} {\hat{\delta}(u\cdot \ell_{12})}\right)^2
%=\bigg(\mathcal{A}_{2;\rma\rma}^{(1)}( b)\bigg)^2.
\end{align}
This results in the following relation between the three-loop and five-loop amplitude 
\begin{align}
\label{finall}
&\mathcal{A}_{2;\rma\rma}^{(3)}( b)=\frac{1}{2}\bigg(\mathcal{A}_{2;\rma\rma}^{(1)}( b)\bigg)^2.
\end{align}

Despite the factor of $1/2$ in the equation above\footnote{We thank Katsuki Aoki for pointing out a typo in the original equation.} it is clear that an eventual resummation of the $L$-loop comb diagrams of the form   \eqref{diagram3l} does not actually resemble an exponential progression. Naively proceeding to five loops by following the same steps as above we would find immediately the following
\begin{align} 
&\mathcal{A}_{2;\rma\rma}^{(5)}( b)=\frac{1}{2^{2}}\bigg(\mathcal{A}_{2;\rma\rma}^{(1)}( b)\bigg)^3,
\end{align}
which suggests a geometric progression of the type: 
\begin{align} 
&\mathcal{A}_{2;\rma\rma}^{(L=2k+1)}( b)=\frac{1}{2^{k}}\bigg(\mathcal{A}_{2;\rma\rma}^{(1)}( b)\bigg)^{k+1}, \,\,\,\,\, k=1,2,3\cdots.
\end{align}
We indeed expect this pattern to hold at all orders. To support our insight, in appendix \ref{MCExp} we show that this indeed holds true for a certain class of toy models, where the geometric series resummation can be  proved to all loops. While we leave the general case for future work, it is easy to understand why the typical eikonal exponential does not appear here: it is due to the lack of complete permutational symmetry of the interaction ``blobs" seen for instance  figure \eqref{diagram3l}. There, we cannot average over the 1/4! exchanges of the four sources, each injecting momentum $\ell_i$, due to the mass changing propagators alternating in the comb diagram. Now the symmetry is only pairwise: for instance one can average over the pairs $(12)(34)$ but not $(13)(24)$. It is for this simple reason that the resummation inherently resembles a geometrical one,\footnote{Something similar also  happens in hadronic physics, see for instance \cite{Oller:1997ti,AITCHISON1972417}.} with argument given by \eqref{eta1l}. 

Importantly, we believe it is this type of resummation which is at play in classical mass-changing effects: even though at the diagrammatic level the difference in masses is quantum $w\sim \mathcal{O}(\hbar)$ -- see \eqref{MassChangeHbar} -- the suggested resummation will still add up all small contributions to leave a classically relevant effect.  In fact, in the next chapter we will assume the amplitude \eqref{oneloopmassch} to resum beyond the probe limit into an inelasticity function $\eta_{\rma\rma}$ characterizing  the final state. There, using a stationary phase argument similar to the one in \cite{Cristofoli:2021jas} we will indeed be able to compute a finite contribution to the classical mass-change observables. This story is to be compared to what happens in the usual eikonal framework of \cite{Cristofoli:2021jas}: the momentum mismatch $q=p'-p$ is quantum at the level of diagrams, yet it is resummed by eikonalization into a finite conservative contribution of the classical deflection.

\section{Inelasticity effects in the KMOC formalism}
\label{sec:FinalStatesObs}

We have studied how to incorporate a novel notion of inelasticity using channel analysis, and gave a model to describe it in terms of mass- and spin-changing amplitudes. To profit from this understanding we are going to show in this section how to add these effects in the KMOC formalism \cite{Kosower:2018adc}, using directly the final state and the eikonal phase. As shown in \cite{Cristofoli:2021jas}, it is possible to highlight all orders of effects in the final state of a scattering process in terms of on-shell quantities only. Instead of directly computing an observable perturbatively from amplitudes as in \cite{Kosower:2018adc}, one can then derive quantities such as the impulse and the spin kick in a binary scattering using stationary phase methods. The advantage of this method is that it offers an alternative understanding of the perturbative cancellation of box and cross diagrams in $D=4$, providing formulae for classical observables in terms of the eikonal phase only \cite{Cristofoli:2021jas, Aoude:2021oqj,Luna:2023uwd}, as well as pointing out an infinity of relations among on-shell quantities, also known as amplitudes fragments. It is then natural to ask how inelastic effects would enter such a formalism and how they would alter formulae for the impulse or the unitarity of the final state. There are two simple scattering scenarios were such effects can be included, namely the scattering of two point particles and that of a wave scattering on a heavy source. Let's look at each of them in detail.

\subsection{Particle-particle final state}

Consider the final state describing the scattering of two point particles and the emission of radiation without absorption effects within the KMOC formalism \cite{Cristofoli:2021jas}. How can we include inelastic effects in such picture? It is simpler to start by considering the final state without radiation.
For example, considering the scattering without radiation and focusing on the changes of \emph{only} particle one, we have a similar equation to
(\ref{eq:ChannelSpaceMatrix}) but only $\ket{\varphi_{1\rma}}$ changing
\begin{align}
\label{eq:SmatrixChannel}
S|\psi\rangle = 
\begin{bmatrix}
S_{\rma \rma} & S_{\rma \rmb} \\
S_{\rmb \rma} & S_{\rmb \rmb} 
\end{bmatrix}
\begin{bmatrix}
|\varphi_{1\rma};\varphi_2\rangle\\
|\varphi_{1\rmb};\varphi_2\rangle\\ 
\end{bmatrix} \, .
\end{align}
For both diagonal and off-diagonal elements, we are going to use an expression in momentum space which is motivated by a partial wave decomposition of a four-point amplitude near the classical limit. Let's first look at the diagonal case
\begin{equation}
\begin{aligned}
\label{eq:toward-fin}
\left(S_{\rma \rma}-\mathds{1}\right)|\varphi_{1\rma};\varphi_2\rangle 
&= i \int \mathrm{d}\Phi\left(p^{\prime}_{1\rma}, p_2^{\prime}, p_{1\rma}, p_2\right) 
\varphi_b\left(p_{1\rma}, p_2\right) \\
&\quad \times 
\sum_{L=0}^{\infty}\mathcal{A}^{L}_{4;\rma \rma}(p_{1\rma} , p_{2} \rightarrow p_{1\rma}' p_{2}')
\hat{\delta}^{(4)}\left(p_{1\rma} {+} p_2 {-} p^{\prime}_{1\rma} {-} p_2^{\prime}\right) \left|p^{\prime}_{1\rma}, p_2^{\prime}\right\rangle \, .
\end{aligned}
\end{equation}
Note that in this chapter we use the letter $s$ both to indicate the energy dependence $s=(p_{1\rma}+p_2)^2$ in eikonal functions: $\chi(\ell, s), \eta(\ell, s)$ and the sum over spin $\sum_s$, hoping no confusion arises. The four-point amplitude entering in this expression is a sum to all loop orders. Most importantly, it can include intermediate spin and mass change interactions which are not usually considered in the standard elastic eikonal scenario. 
To rewrite the final state in an eikonal fashion, thus highlighting the various differences between elastic and inelastic contributions, we will adopt an ansatz motivated by the partial wave expansion for the four point amplitudes. Defining the transfer momenta as $q_{1,\rma \rma}=p_{1\rma} - p^{\prime}_{1\rma}$ and $q_{2}=p_2- p_2^{\prime}$, we  integrate over the momentum-conservation delta that sets $q_{1,aa} = q_{2}=q$ and expand the amplitude as follows:
\begin{equation}
\begin{aligned}
\label{eq:partialwave}
    \sum_{L=0}^{\infty}\mathcal{A}^{L}_{4;\rma \rma}(p_{1\rma} , p_{2} \rightarrow p_{1\rma}-q , p_{2}+q)&= -i\sum_{\ell=0}^{\infty} (2l+1) P_{\ell}(\cos(\theta)) \bigg[ \tilde{\mathcal{A}}_{\rma \rma}(\ell,s)-1\bigg].  
   % &\equiv-i\sum_{\ell=0}^{\infty} (2l+1) P_{\ell}(\cos(\theta)) \bigg[e^{i \chi_{\rma \rma}(\ell,s){}} \eta_{\rma \rma}(\ell,s)-1\bigg] \, , 
\end{aligned}
\end{equation}
An important step is then to consider 
the complex partial wave coefficients $\tilde{\mathcal{A}}_{\rma \rma}(\ell,s)$ written 
in a polar form,
%\textit{i.e} the radius $\eta_{\rma \rma}$ and the phase $\chi_{\rma \rma}(\ell,s)$
\begin{align}
\tilde{\mathcal{A}}_{\rma \rma}(\ell,s) \equiv\eta_{\rma \rma}(\ell,s) e^{i \chi_{\rma \rma}(\ell,s){}} 
\end{align}
which defines the inelasticity parameter in terms of all-order scattering amplitudes.
The final state can then be written as
\begin{align}
    &S_{\rma \rma}|\varphi_{1\rma};\varphi_2\rangle  =
    i \int\!\dd\Phi\left(p_{1\rma}, p_2\right) 
    \varphi_{\rmb}\left(p_{1\rma}, p_2\right)    \\
    &\quad\times\int \hat{\dd}^4q \: 
    \hat{\delta} (2 \tilde{p}_{1\rma}\cdot q)\hat{\delta} (2 \tilde{p}_2\cdot q) \sum_{\ell=0}^{\infty} (2\ell+1) P_{\ell}(\cos(\theta))  e^{i \chi_{\rma \rma}(\ell,s){}} \eta_{\rma \rma}(\ell,s)  \left|p_{1\rma}, p_2\right\rangle\notag \, ,
\end{align}
where we have introduced $\tilde{p}_{1\rma}=p_{1\rma}-q/2$ and $\tilde{p}_2=p_2+q/2$.
In considering the classical limit of this expression, we can take the continuum limit of the partial wave sum. As a result, we can replace the integrand using the identity 
\begin{align}
\label{eq:identity-cla-eik}
  \hat{\delta} (2 \tilde{p}_{1\rma}\cdot q)\hat{\delta} (2 \tilde{p}_2\cdot q) &\sum_{\ell=0}^{\infty} (2l+1) P_{\ell}(\cos(\theta))  e^{i \chi_{\rma \rma}(\ell,s){}} \eta_{\rma \rma}(\ell,s)  \notag\\
  &=\int \mathrm{d}^4 x\, e^{i q \cdot x }\left\{e^{i \chi_{\rma \rma}\left(x_{\perp} ; s\right) / \hbar} \eta_{\rma \rma}(x_{\perp},s)-1\right\} \, .
\end{align}   
Following Section 4 of \cite{Cristofoli:2021jas}, we have introduced a $\perp$ subscript to highlight that the dependence in $x$ of the eikonal phase is via $x_{\perp}$, which is perpendicular to $\tilde{p}_i$, rather than to $p_i$, so that
\begin{equation}
\label{eq:orthogonal-relations}
    \tilde{p}_{1\rma} \cdot x_{\perp}=0=\tilde{p}_2 \cdot x_{\perp}    \, .
\end{equation}
As a result, $x_{\perp}$ depends on $q$ meaning that in going from the left-hand side to the right-hand side of (\ref{eq:identity-cla-eik}) we have made the eikonal phase a function of $q$ as well.  This leads to the following expression for the final state in the diagonal channel
\begin{align}
\label{eq:DiagonalFinalStateAnsatz}
S_{\rma \rma}|\varphi_{1\rma};\varphi_2\rangle 
{=}\!\int\!\dd\Phi(p_{1\rma},p_2)
&\int\!\hat{\dd}^4\,q\,\dd^4x\,e^{i q \cdot x{}}
\varphi_{\rm b}(p_{1\rma}{-}q,p_2{+}q) \\
&\quad
\times\text{exp}\left[{i \chi_{\rma \rma}(x_\perp;s)}{}\right]\,{\eta}_{\rma \rma}(x_{\perp},s)|p_{1\rma},p_2\rangle. \nonumber 
\end{align} 
The off-diagonal part of the final state follows similar arguments, but it contains an important subtlety that will be relevant in deriving the classical impulse. Following the previous discussion, one gets that one of the two Dirac deltas defining a plane of motion has an additional term
\begin{align}
S_{\rmb \rma}|\varphi_{1\rma};\varphi_2\rangle  
&= i \int_{M_\rma^2}^{\infty} \dd M^2_\rmb \rho( M^2_\rmb) \int \mathrm{d}\Phi\left(p_{1\rmb}, p_2\right) 
\varphi_b\left(p_{1\rmb}, p_2\right)  \int \hat{\dd}^4q \: \hat{\delta} (2 \tilde{p}_{1\rmb}\cdot q{+} M^2_\rmb- M^2_\rma)  \notag\\
&\times \hat{\delta} (2 \tilde{p}_2\cdot q)  \bigg[ \sum_{\ell=0}^{\infty} (2\ell+1) P_{\ell}(\cos(\theta)) e^{i \chi_{\rmb \rma}(\ell,s,M_\rmb^2){}} \eta_{\rmb \rma}(\ell,s, M^2_\rmb) \bigg] \left|p_{1\rmb}, p_2\right\rangle \, .
\end{align}
One interesting aspect of these off-diagonal contributions is that the Dirac delta defining a plane of motion is now a function of the mass change parameter.
It is useful to define the mass difference entering the final state as in~\cite{Aoude:2023fdm} and shift the transfer momentum by
\begin{equation}
w\equiv(M^2_\rmb-M^2_\rma)/2M_\rma
\qquad \qquad
q^\mu \rightarrow q^\mu -  w\, \check{u}^\mu_1 \, ,
\end{equation}
Thanks to this shift, we can remove from one of the Dirac deltas the additional contribution of $w$ where we have introduced the dual velocities
\[
 \check{u}^\mu_1 = \frac{u_1^\mu - \gamma\, u_2^\mu}{1 - \gamma^2}, \quad
  \check{u}^\mu_2 =  \frac{u_2^\mu - \gamma\, u_1^\mu}{1 - \gamma^2} \, ,
\]
satisfying $\check{u}_i\cdot u_j=\delta_{ij}$. We have also defined 
\begin{equation}
    \gamma \equiv u_1\cdot u_2.
\end{equation}
By neglecting $q^2$ terms, the final state in the off diagonal part looks as follows 
\begin{align}
\label{eq:OffDiagonalFinalStateAnsatz}
S_{\rmb \rma}|\varphi_{1\rma};\varphi_2\rangle {=} &\int_{0}^{\infty} \dd w \rho(w) \!\int\!\dd\Phi(p_{1\rmb},p_2)
\int\!\hat{\dd}^4\,q\,\dd^4x\, e^{i q \cdot x} \text{exp}\left[{i \chi_{\rmb \rma}(x_\perp,s,w)} \right]
 \notag\\
&\qquad\qquad
\times \varphi_{\rm b}(p_{1\rmb}{-}q{+}\check{u}_1 w,p_2{+}q-\check{u}_1 w) \,{\eta}_{\rmb \rma}(x_{\perp},s,w)|p_{1\rmb},p_2\rangle .
\end{align}
The various quantities entering the different channels of the final state can be easily expressed in terms of on-shell quantities by perturbatively inverting the partial wave expansion in the diagonal channel (\ref{eq:partialwave}) and its analogue in the off-diagonal one. For example, at leading order one has the following partial wave coefficients
\begin{equation}
\begin{gathered}
\label{eq:tabe1}
    \chi_{\rma \rma}(x,s)=\int \hd^4q \: e^{i q \cdot x} \hat{\delta}(2 p_{1\rma}\cdot q) \hat{\delta}(2 p_{2}\cdot q) \Re \mathcal{A}^{0}_{4;\rma \rma} \, , \\
    {\eta}_{\rma \rma}(x,s)=1-\int \hd^4q \: e^{i q \cdot x} \hat{\delta}(2 p_{1\rma}\cdot q) \hat{\delta}(2 p_{2}\cdot q) \Im \mathcal{A}^{1}_{4;\rma \rma} \ ,
    \end{gathered}
\end{equation}
for the diagonal case (note the correspondence of the lower equation above to the probe limit one loop amplitude in \eqref{oneloopmassch}) and 
\begin{equation}
\begin{gathered}
\label{eq:tabe2}
    \chi_{\rmb \rma}(x,s,w)=\int \hd^4q \: e^{i q \cdot x} \hat{\delta}(2 p_{1\rmb}\cdot q) \hat{\delta}(2 p_{2}\cdot q) \Re \mathcal{A}^{0}_{4;\rmb \rma} \, , \\
    {\eta}_{\rmb \rma}(x,s,w)=-\int \hd^4q \: e^{i q \cdot x} \hat{\delta}(2 p_{1\rmb}\cdot q) \hat{\delta}(2 p_{2}\cdot q) \Im \mathcal{A}^{0}_{4;\rmb \rma} \ ,
    \end{gathered}
\end{equation}
for the off-diagonal one. In the standard PM scattering scenario (purely elastic scattering), all quantities related to the off-diagonal channels are absent, while the quantity $\tilde{\eta}_{\rma \rma}(x_{\perp},s)$ is equal to one, only allowing for potential quantum contributions through the so-called quantum reminder. However, in full generality, inelastic effects alter the structure of the diagonal and off-diagonal final state. It is interesting to understand how this can impact well-known observables, such as the impulse in a $2\rightarrow 2$ scattering process, and how unitarity is still preserved in this context.

Before obtaining the observables, two comments are in order. First, in Eqs. \ref{eq:DiagonalFinalStateAnsatz} and \ref{eq:OffDiagonalFinalStateAnsatz} we have written the inelasticity as a real function in front of the eikonal exponential. Note that without loss of generality, we could have written this product as a complex eikonal phase, for example $e^{iz}$, where its real
value is the usual conservative eikonal and the imaginary part becomes the inelasticity. In a schematic way
\begin{align}
\text{for}\quad z\in \mathds{C}:\qquad
e^{i z} =e^{i (\text{Re}[\chi] + i \text{Im}[\chi])} = e^{-\text{Im}[\chi]}e^{i\text{Re}[\chi]} =:  \eta\, e^{i\text{Re}[\chi]}.
\end{align}
Note that this $\text{Im}[\chi]$ already appears at two loops in gravitational scattering due to radiation reaction~\cite{DiVecchia:2021bdo}. Here we are lifting the imaginary part to include also coupled-channel interactions but both effects lead to the same structure, an inelasticity parameter. This can be both realized as $\eta$ or as $\text{Im}[\chi]$. Note that in writing $\text{Im}[\chi]$ we are simply using a trivial rewriting of the real number $\eta$. We are not imposing that $\text{Im}[\chi]$ truncates at a finite order in perturbation theory, which would be a statement on the exponentiation of perturbative diagrams.  

Secondly, it is interesting to note that our ansatz for the final state could also be turn into an operational ansatz  for the $S_{\rma\rma}$ matrix. In~\cite{Damgaard:2021ipf}, an exponential representation of the $S$-matrix was suggested introducing a proper $\hat{N}$ operator as $S = \exp[i\hat{N}]$. Adding the inelasticity can then be achieved in two ways, either by having a non-hermitian $\hat{N}$ (which is the same argument as the complex eikonal $e^{iz}$) or by adding a $\hat{\eta}$ matrix in front such that 
\begin{align}
S_{\rma\rma} =  \bigg[\hat{\eta}\exp[i\hat{N}]\bigg]_{\rma \rma} \qquad
S_{\rmb\rma} =  \bigg[\hat{\eta}\exp[i\hat{N}]\bigg]_{\rmb \rma} \, .
\end{align} 
where each $\hat{\eta}$ and $\hat{N}$ are operators as in the original proposal~\cite{Damgaard:2021ipf} and not scalar functions.
Relations between these matrix are guaranteed by the unitarity of the full $S$-matrix as in following section. 

\subsubsection{Inelastic effects and unitarity}

Unitarity should be preserved in our approach. If it is not, some degrees of freedom are not being tracked and the system evolves as an open quantum system. Hence, we will demand unitarity of the full $S$-matrix in the left-hand-side of (\ref{eq:SmatrixChannel}). This naturally implies a set of relations for the reduced $S_{ij}$ matrices, which are not unitary by themselves. The matrix equation $S^\dagger S = \mathds{1}$ reads
\begin{subequations} 
\begin{align} \label{eq:unitarity-channels}
S_{\rma \rma}^\dagger S_{\rma \rma} + S_{\rmb \rma}^\dagger S_{\rmb \rma} &= 1 = S_{\rma \rmb}^\dagger S_{\rma \rmb} + S_{\rmb \rmb}^\dagger S_{\rmb \rmb},\\  S_{\rma \rma}^\dagger S_{\rma \rmb} + S_{\rmb \rma}^\dagger S_{\rmb \rmb} &= 0 = S_{\rma \rmb}^\dagger S_{\rma \rma} + S_{\rmb \rmb}^\dagger S_{\rmb \rma}  \, .
\end{align} 
\end{subequations}
By averaging, it is trivial to realize that the norm of the various channels contributing to the final state are related. The first line for example will give us a precise relation which we are now going to look at in detail. First of all, let's notice that the diagonal channel part of the final state can be evaluated via a saddle point approximation when\footnote{We remind that in fact there is an $\hbar$ in the exponents, for instance in  $e^{i\chi/\hbar}$.} $\hbar \rightarrow 0$. Within this limit, the saddles are
\begin{equation}
\label{eq:saddle-points}
    q_{*,\rma \rma}^{\mu}=-\partial^\mu \chi_{\rma \rma}\left(x_{\perp}, s\right), \quad 
    x^{\mu}_{*,\rma \rma}=b^\mu-\frac{\partial}{\partial q_\mu} \chi_{\rma \rma}\left(x_{*\perp,\rma \rma}, s\right) .
\end{equation}
Notice that the derivative in $q$ is non-vanishing as a consequence of (\ref{eq:orthogonal-relations}) meaning that the saddle point in $x$ is not the impact parameter $b^{\mu}$ but the so called eikonal impact parameter 
\cite{Amati:1990xe,Bern:2020gjj,Cristofoli:2021jas}. As a result, we obtain 
\begin{equation}
\label{eq:Diagonal_Saddle}
\begin{aligned}
S_{\rma \rma}\left|\varphi_{1\rma} ; \varphi_2\right\rangle=\int \mathrm{d} \Phi\left(p_{1\rma}, p_2\right) 
e^{i q_*(s) \cdot x_*(s)+ i \chi_{\rma \rma}(x_{*, \perp,\rma \rma}(s) ; s)}  \\ \times \varphi_{\mathrm{b}}\left(p_1{+}q_{*,\rma \rma}, p_2{-}q_{*,\rma \rma}\right) \eta_{\rma \rma}
\left(x_{*\perp,\rma \rma},s\right)\left|p_{1\rma}, p_2\right\rangle \, .
\end{aligned}
\end{equation}
This representation for the final state has a simple interpretation as a free wavepacket where the momenta, instead of being localized at a momenta $p_0$, it is localized on a shifted momenta $p_0+q$. The discussion is unchanged for the off-diagonal part of the final state apart from the fact that it is present an additional shift in the off-diagonal part of the final state , which we need to take into account. Doing so, we obtain
\begin{equation}
    q_{*\rm \rmb \rma}^{\mu}=-\partial^\mu \chi_{\rmb \rma}\left(x_{*\perp,ba}, s,w\right), \quad \quad 
    x^{\mu}_{*\rm,\rmb \rma}=b^\mu-\frac{\partial}{\partial q_\mu} \chi_{\rmb \rma}(x_{*\perp,\rmb \rma}, s) .
\end{equation}
which are different from the saddle points of the diagonal channel
\begin{equation}
\label{eq:OffDiagonal_saddle}
\begin{aligned}
S_{\rmb \rma}\left|\varphi_{1\rma} ; \varphi_2\right\rangle= \int_{0}^{\infty} \dd w \rho(w) \: \int \mathrm{d} \Phi\left(p_{1\rmb}, p_2\right) 
e^{i q_{*,\rmb \rma} \cdot x_{*,\rmb \rma}+ i \chi_{\rmb \rma}(x_{*, \perp,\rmb \rma},s,w)}  \\ \times \varphi_{\mathrm{b}}\left(p_1{+}q_{*,\rmb \rma}{-}w\check{u}_1, p_2{-}q_{*,\rmb \rma}{+}w\check{u}_1\right) \eta_{\rmb \rma}
\left(x_{*\perp,\rmb \rma},s,w\right)\left|p_{1\rmb}, p_2\right\rangle \, .
\end{aligned}
\end{equation}
We can now easily check that the final state including diagonal and off diagonals channels is unitary. Using (\ref{eq:OffDiagonal_saddle}), the norms in the various channels are
\begin{equation}
\label{eq:Norm_SmatricesChannel}
\|S_{\rma \rma}\left|\varphi_{1\rma} ; \varphi_2\right\rangle \|^2
=\int\! \mathrm{d} \Phi\left(p_{1\rma}, p_2\right)\left|\varphi_{\mathrm{b}}\left(p_1{-}q_{*,\rma \rma}, p_2{+}q_{*,\rma \rma}\right)\right|^2\left|\eta_{\rma \rma}
\left( x_{*\perp,\rma \rma},s\right)\right|^2 ,
\end{equation}
\begin{equation}
\begin{aligned}
\|S_{\rmb \rma}\left|\varphi_{1\rma} ; \varphi_2\right\rangle \|^2
&=\int_{0}^{\infty}\dd w \rho(w) \int \mathrm{d} \Phi\left(p_{1\rmb}, p_2\right)
\left|\varphi_{\mathrm{b}}\left(p_1{-}q_{*,\rmb \rma}^{\rm }{+}w\check{u}_1 , p_2{+}q_{*,\rmb \rma}{-}w\check{u}_1 \right)\right|^2  \\ 
&\qquad\qquad\qquad\qquad
\times 
\left[\eta_{\rmb \rma}^*\left(s, x_{*\perp,\rmb \rma},w\right) 
\cdot 
\eta_{\rma \rmb} \left(s, x^{\rm}_{*\perp,\rmb \rma},w\right) \right] \, .
\end{aligned}
\end{equation}
Hence, we can obtain the relation between the classical inelasticity parameters from unitarity relations
\begin{equation}
\label{eq:unit-constraint-eta}
 \lexp |\eta_{\rma \rma}(x_{* \perp,\rma \rma})|^2 \rexp +   \int_{0}^{\infty} \dd w \rho(w) \lexp \eta_{\rma \rmb}(x_{* \perp,\rma \rmb},w) \cdot \eta_{\rmb \rma}(x_{* \perp,\rmb \rma},w)^* \rexp =1.
\end{equation}
It is important to remember here  that $\eta_{\rmb \rma}^* \cdot \eta_{\rma \rmb}$  has an implicitly inner product in the Hilbert space of $\rmb $, \text{i.e.} there is a summation as well as in integral over the possible mass variations. As pointed out in~\cite{Aoude:2023fdm,Jones:2023ugm}, we expect small mass deviations from the original mass (they differ by $\hbar$ at quantum amplitude level), in other words, near-threshold light modes are only relevant. We now turn to using such relations in the context of computing classical observables. This will be useful to understand how inelastic effects affects classical observable as seen from on-shell amplitudes. 

\subsubsection{The impulse with inelastic effects} 
The linear impulse in KMOC is given as
\begin{align}
\Delta p_1^\mu = 
\langle {\rm in} | S^\dagger \mathds{P}_1^\mu S | {\rm in} \rangle -
\langle {\rm in} | \mathds{P}_1^\mu | {\rm in} \rangle \, .
\end{align}
Starting with the two particle state $|\varphi_{1\rma};\varphi_2\rangle$ and inserting an identity~\footnote{Formally, this is a two-particle identity where only particle one is allowed to change.} using Eq. \eqref{eq:identity} in the first term, we have
\begin{align}
\Delta p_1^\mu = 
\langle \varphi_{1\rma};\varphi_2| S^\dagger_{\rma \rma} \mathds{P}_1^\mu S_{\rma \rma} | \varphi_{1\rma};\varphi_2 \rangle +
\langle\varphi_{1\rma};\varphi_2 | S^\dagger_{\rma \rmb} &\mathds{P}_1^\mu S_{\rmb \rma} | \varphi_{1\rma};\varphi_2 \rangle\\
& -
\langle \varphi_{1\rma};\varphi_2 | \mathds{P}_1^\mu | \varphi_{1a};\varphi_2\rangle \notag \, .
\end{align}
Since $S_{\rma \rma}$ keeps its channel, it has an identity part and can also be written as
${S_{\rma \rma} = \mathds{1} + i T_{\rma \rma}}$, but $S_{\rmb \rma}$ is off-diagonal and does not have the identity part, just the scattering matrix. We can proceed to calculating it by order-by-order perturbation in the $T$-matrix or by using directly the final state ansatz\"e and using the saddle. We are going follow the latter approach but one should bear in mind that knowledge about the inelasticity in obtained through a matching calculation which in currently performed perturbatively~\cite{Aoude:2023fdm}. See also Ref.~\cite{Jones:2023ugm} for a related analysis of the KMOC with absorption. 

Using the saddle point expression for the final states in the diagonal and off-diagonal channel respectively (\ref{eq:Diagonal_Saddle}-\ref{eq:OffDiagonal_saddle}), it is straightforward to derive an expression for the impulse only in terms of partial wave coefficients defining the outgoing state
\begin{align}
    \Delta p_{1}^{\mu} &= \partial^\mu \chi_{\rma \rma}\left(x_{* \perp,\rma \rma}, s\right) |\eta_{\rma \rma}(x_{* \perp,\rma \rma},s)|^2\\
    &+ \int_{0}^{+\infty} \dd w \,\rho(w) \, w \,  \big[\eta_{\rmb \rma}^{*}(x^{\rm}_{*\perp,\rmb \rma},w)^{*} \cdot \eta_{\rma \rmb}
     (x^{\rm}_{*\perp,\rmb \rma},w)\big]_{\rm}
    (\check{u}_1^\mu + \partial^\mu \chi_{\rma \rmb}\left(x_{*\perp,\rma \rmb}, s,w\right))\nonumber \, .
\end{align}

At LO, the first line contains the standard transverse impulse. The second line contains the desired longitudinal impulse $\Delta p_{1,{\rm L}}^\mu$,  which at leading order reads
\[  \label{dpmc}
    \Delta p_{1,{\rm L}}^\mu = \int_{0}^{\infty} \dd w \,\rho(w) \, w \,   \big[\eta_{\rmb \rma}^{*}(x^{\rm}_{*\perp,\rmb \rma},w) \cdot \eta_{\rma \rmb}
     (x^{\rm}_{*\perp,\rmb \rma},w)\big]_{\rm}   \, \check{u}_1^\mu \, ,
\]
and, as a cross-check, we can also extract  the mass change as 
\begin{align}\label{dmmc}
    \Delta M_1^2 = 2M_1\int_{0}^{\infty} \dd w \,\rho(w) \, w \,      \big[\eta_{\rmb \rma}^{*}( x_{*\perp,\rma \rmb},w) \cdot \eta_{\rma \rmb}
     (x_{*\perp,\rmb \rma},w)\big] \, \, .
\end{align}
At this point we can note a few things. First that both \eqref{dpmc} and \eqref{dmmc} can be seen to  agree with the results  of \cite{Goldberger:2020wbx,Jones:2023ugm} once the matching to a specific $\rho(w)$ is done and   $\eta_{\rma\rmb}$ is  identified with the mass/spin-changing tree-level interaction. Secondly, we observe how the kernel of both equations has the structure of the one-loop integral we computed earlier in chapter \ref{sec:ICCE_QFT}, equation \eqref{eta1l}. This can be  seen by recognizing that  $\eta_{\rma\rmb}$  is nothing but the mass/spin-changing tree-level interaction of Eq. \eqref{vertexino}, at least at leading order. This means that one can indeed interpret mass-changing amplitudes as momentum space representations of 
the inelasticity parameter $\eta$.
Finally we note that  $\Delta M_1^2>0$  as  the integrand of \eqref{dmmc} is an absolute square:   the  mass increases  as expected by Hawking's area theorem.

\subsection{Wave-particle final state}
\begin{figure}[!htb]
% trim={<left> <lower> <right> <upper>}
    \centering
    \begin{subfigure}{0.44\textwidth}
     \includegraphics[trim={12cm 0 12cm 0},clip,width=1.0\textwidth]
     {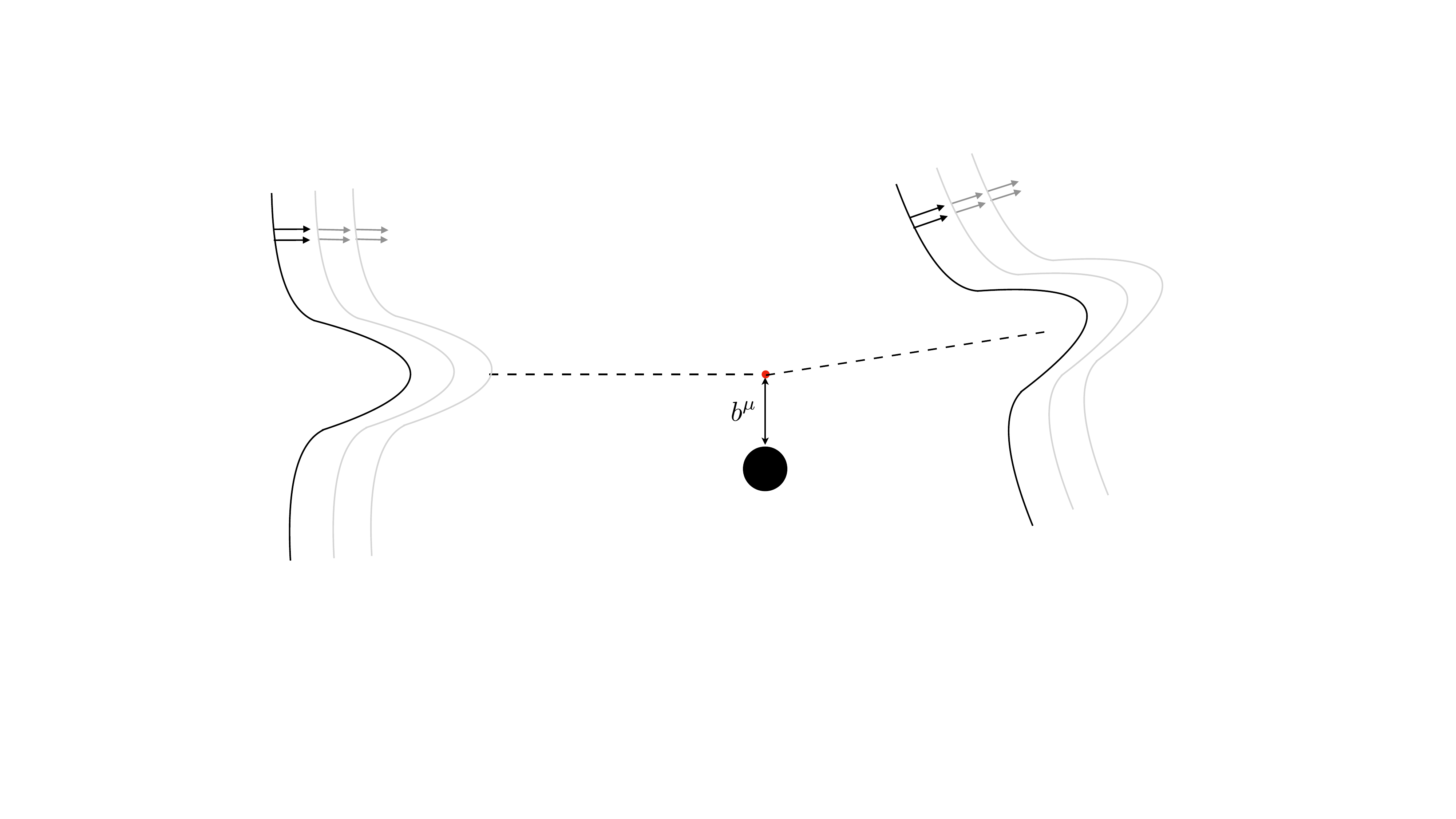}
     \caption{'Laser'-like wave scattering}
     \label{fig:sub1}
     \end{subfigure}
     \begin{subfigure}{0.55\textwidth}
     \includegraphics[trim={12cm 0 2cm 0},clip,width=1.2\textwidth]{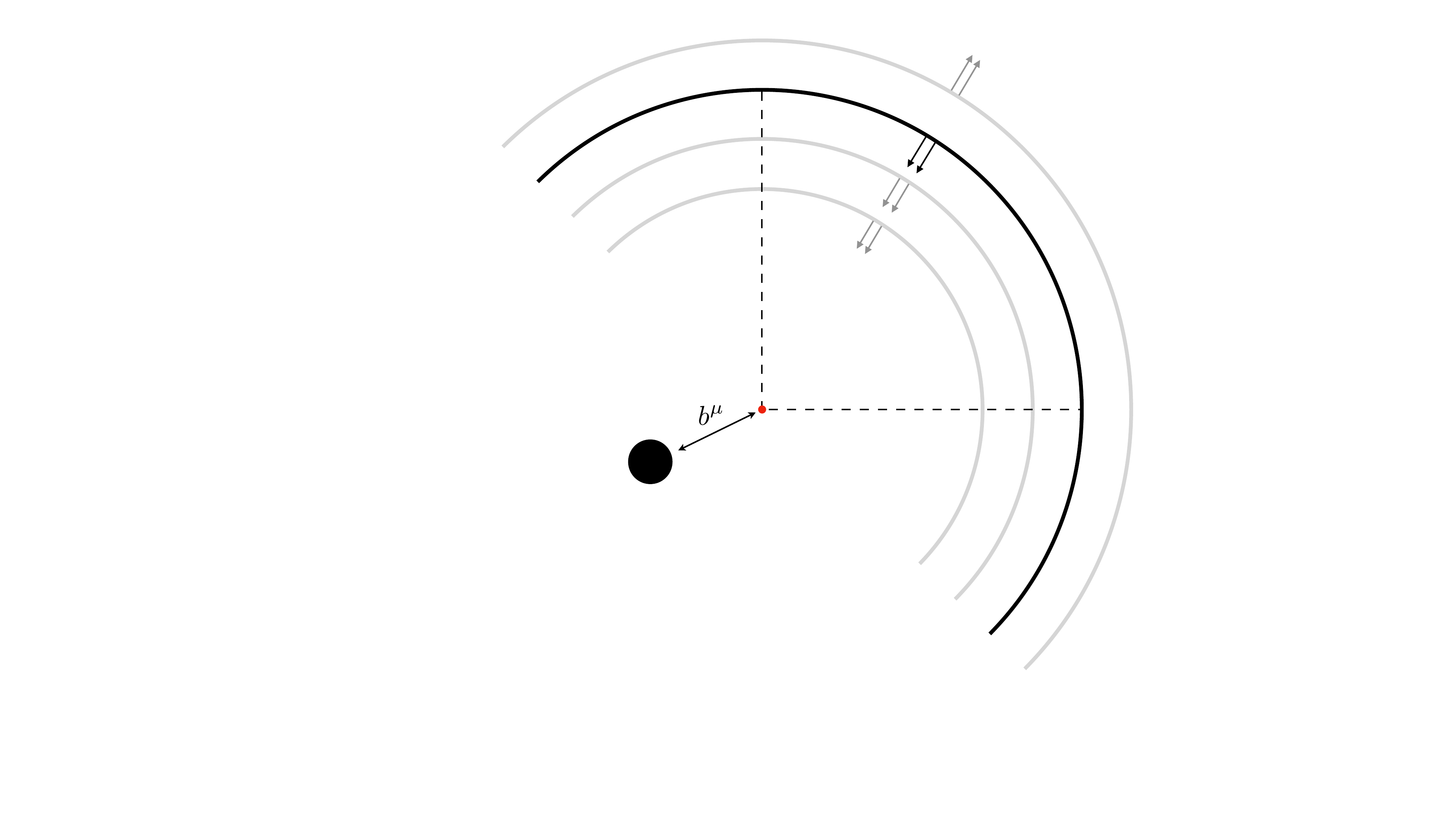}
     \caption{Delocalized spherical wave scattering.}
     \label{fig:sub2}
     \end{subfigure}
    \caption{(Left) Setup of a 'laser'-like coherent wave scattering~\cite{Cristofoli:2021vyo}. (Right) Setup of a spherical coherent wave impinging on a black hole~\cite{Aoude:2023dui}. We leave the impact parameter, the distance between the particle's rest frame position and the centre of the wave, explicit in our calculation but in principle we have $b = 0$. Note that the wave is delocalized in space. Here we have depicted the $(j,m)=(0,0)$ (spherically symmetric) wave although our setup also includes higher $(j,m)$ modes. 
    }
    \label{figure2}
\end{figure}

Having explored how inelastic effects are incorporated into the final state for particle-particle scattering, we now turn our attention to the case where a classical wave scatters off a massive particle. Specifically, we consider an incoming classical wave represented by a coherent state $\ket{\alpha_h}$ approaching a massive particle $p^\mu$ taken to be the heavy particle in the probe limit. We represent this by the initial state in the $\rma$-channel: 
=\[
\ket{\varphi_\rma, \alpha} &= \int \dd \Phi(p_{\rma}) \varphi_{\rma}(p)\, e^{ib\cdot p} |p_\rma , \alpha_h \rangle = \int \dd \Phi(p_{\rma}) \varphi_{\rma}(p)\, e^{ib\cdot p}\, \mathbb{D}[\alpha_h]|p_\rma \rangle,
\]
where $h$ is a helicity index and 
\[
&\mathbb{D}[\alpha_h] 
= \mathcal{N}_{\alpha} \exp\left( \int \dd \Phi(k) \alpha(k)  a^\dagger_h(k) \right), \quad \mathcal{N}_{\alpha} = \exp \left( - \frac{1}{2}\int \dd \Phi(k) |\alpha(k)|^2 \right).
\]
We want to write down an ansatz\"e for the final state, taking into account the scattering of the wave $\ket{\alpha_h}$ and neglecting the recoil of the massive particle. First, considering the case where there is no absorption (\textit{i.e.} no channel-transitions), we write
\[\label{eq:WaveParticleAnsatz}
S_{\rma \rma} \ket{\varphi_\rma, \alpha} =\int \dd \Phi(p_{\rma}) \varphi_{\rma}(p)\, e^{ib\cdot p}\,\mathbb{D}[\beta(p)]|p_\rma \rangle,
\]
where $\beta$ is the waveshape of the final state. Note that this ansatz  implies that $\beta(p)$ admits an expansion in powers of $\alpha$:
\[\label{eq:BetaExpansion}
\beta(k,p) = \sum_{n} \beta^{(n)}(k,p),
\]
where $\beta^{(n)}(k,p)$ denotes the $\mathcal{O}(\alpha^n)$ contribution. It is nevertheless possible to determine $\beta(p)$ to all-orders in $\alpha$ by  projecting \eqref{eq:WaveParticleAnsatz} into the state $\bra{p'_\rma,k'}$ to get
\[\label{eq: BetaAlpha}
 \mathcal{N}_\beta\,  \varphi_\rma(p')e^{i b \cdot p'} \beta(k',p') &= \int \dd \Phi(p_{\rma}) \varphi_{\rma}(p)\, e^{ib\cdot p}\,\mathcal{N}_{\alpha} \\
 \times & \bra{p'_\rma,k'} S \exp\left[ \int \dd \Phi(k) \alpha(k)  a^\dagger_h(k) \right] \ket{p_\rma}. 
\]
Now, squaring both sides and integrating with respect to the measure $\dd \Phi(k')$ we end up with the equation
\[\label{eq: LambertEq}
e^{-\mathcal{I}_{\beta}}\mathcal{I}_{\beta}= z(\alpha), 
\]
where
\[
\mathcal{I}(\beta) \equiv \int \dd\Phi(k) |\beta(k,p)|^2,
\]
represents the intensity of the outgoing wave and 
\[
z(\alpha)\equiv \int \dd\Phi(k')\bigg|\int \dd \Phi(p_{\rma}) \varphi_{\rma}(p)\, e^{ib\cdot p}\,\mathcal{N}_{\alpha} \bra{p_\rma',k'} S_{\rma \rma} \exp\left[ \int \dd \Phi(k) \alpha(k)  a^\dagger_h(k) \right] \ket{p_\rma}\bigg|^2. 
\]
We can now solve eq. \eqref{eq: LambertEq} for $\mathcal{I}(\beta)$ in terms of the Lambert-W function, so that
\[
\mathcal{I}(\beta) = W_{n}(-z(\alpha)).
\]
where the subscript $n$ distinguishes between the branches of the W-function. In our case  the argument $(-z(\alpha))$, being real and negative, picks out the branch denoted by $n=-1$. Now the solution to eq. \eqref{eq: LambertEq} reads:
\[
\varphi_{\rma}(p') \beta(k',p') = e^{W_{-1}(-z(\alpha))/2} \int \dd \Phi(p_{\rma}) \varphi_{\rma}(p)\, e^{ib\cdot (p-p')}\,  \bra{p_\rma',k'} S_{\rma \rma}  \ket{p_\rma,\alpha}. 
\]
Alternatively, we can solve eq.\eqref{eq: BetaAlpha} perturbatively in $\alpha$. At leading order in $\alpha$, we can neglect the normalisations $\mathcal{N}_\beta$ and $\mathcal{N}_\alpha$ ,which enter at higher orders in this expansion, to get
\[
e^{i b \cdot p'} \varphi_\rma(p') \beta^{(1)}(k',p') =  e^{i b \cdot p'} \varphi_\rma(p') \alpha(k') + \int\! \dd \Phi(p_{\rma},k) \varphi_\rma(p) e^{i b \cdot p}\alpha(k) \bra{p'_\rma,k'}iT_{\rma \rma}\ket{p_\rma,k}, \\
 =  e^{i b \cdot p'} \varphi_\rma(p') \alpha(k') + \int \!\dd\Phi(k) \, \hdelta(2p'\cdot(k'{-}k)) \,\alpha(k) \, \varphi_\rma(p'+k'-k)e^{i (p'+ k'-k) \cdot b} \, i \mathcal{A}_{\rma \rma}(k,k').
\]
Approximating $\varphi(p'+k'-k) \simeq \varphi(p')$ we find that 
\[
\beta^{(1)}(k',p') =  \alpha(k') + \int \dd\Phi(k) \, \hdelta(2p'\cdot(k'-k)) \,\alpha(k) \, e^{i b \cdot (k'-k)}\, i \mathcal{A}_{\rma \rma}(p,k \rightarrow p',k').
\]
Notice that the first term corresponds to the forward scattering ($S=1$) part of the $S$-matrix. It is possible to proceed in this manner to obtain $\beta(k,p)$ to any order in $\alpha(k)$. Here, we will not be concerned with the explicit form of $\beta(k,p)$, focusing instead on how to incorporate inelastic effects into the final state ansatz. \eqref{eq:WaveParticleAnsatz}. To this end, let us introduce an additional channel $\rmb$, defining the state $\ket{\varphi_\rmb}$ as in Eq. \eqref{KMOCbeta}. Next, we introduce channel-couplings to  the final state by defining the operators $\eta_{\rma\rmb}$ and $\eta_{aa}$ such that our final state becomes
\[
&S_{\rma \rma} \ket{\varphi_\rma, \alpha}  =  \int \dd \Phi(p_{\rma}) \varphi_{\rma}(p)\, e^{ib\cdot p} \, \eta_{\rma \rma}  | p_\rma , \beta\rangle\\
&S_{\rmb \rma} \ket{\varphi_\rma, \alpha}  =  \int \dd \Phi(p_{\rma}) \varphi_{\rma}(p)\, e^{ib\cdot p} \, \eta_{\rma \rmb}  | p_\rma , \beta\rangle.
\]
We will not restrict the form of this operator at this stage, for now we simply note that the diagonal $S$-matrix ceases to be unitary. Instead, the unitarity condition is now
\[
||S_{ \rma \rma} \ket{\varphi_\rma}||^2 + ||S_{ \rmb \rma} \ket{\varphi_\rma}||^2 =1. 
\]
In particular, we interpret the off-diagonal contribution
\[
||S_{ \rmb \rma} \ket{\varphi_\rma}||^2 =  \int \dd \Phi(p,p') \varphi_b(p,p') \bra{p_\rmb',\beta} \eta_{\rmb \rma}^\dagger \eta_{\rma \rmb} \ket{p_\rma, \beta},
\]
as the total inelasticity. Note that it receives contributions from the standard perturbative expansion of $\beta$ to all orders. In addition, it contains terms arising from insertions of the operator $\eta_{\rma \rmb}$. The nature of this operator is clarified by matching this ansatz to classical observables such as the absorption cross section. To this end, let us slightly modify our ansatz to account for an incoming partial waves with angular momenta $(j,m)$. We define the incoming coherent state by
\[
\ket{\alpha_{j,m}^h} = \mathcal{N}_{\alpha} \exp \left(  \int \hd \omega \, \alpha(\omega) \, a^\dagger_{j,m,h} (\omega)\right)
\]
Here we have restricted the incoming wave to only contain modes of specific angular momentum numbers $(j,m)$, ie 
\[
 a_{j',m',h'}(\omega) \ket{\alpha_{j,m}^h}  = \delta_{j,j'} \delta_{m,m'} \delta_{h,h'} \,\alpha(\omega)\ket{\alpha_{j,m}^h},
\]
whereas a generic wave is composed of a superposition of such modes. We now define the initial state in the $\rma$ channel by 
\[
&\ket{\varphi_\rma, \alpha_{j,m}^h} = \int \dd\Phi(p_{\rma})\, \varphi_{\rma}(p)\, e^{ib\cdot p} \ket{p_{\rma},\alpha_{j,m}^h},
\]
in which case the final state ansatz  becomes
\[
S_{ \rmb \rma} \ket{\varphi_\rma,\alpha_{j,m}^h}  &= \int \dd\Phi(p_{\rma})\, \varphi_{\rma}(p)\, e^{ib\cdot p}  \, \eta_{\rmb \rma} \,\mathcal{N}_{\beta} \exp \left(  \int \hd \omega \, \beta^h_{jm}(\omega) \, a^\dagger_{j,m,h} (\omega)\right) | p_\rma \rangle, \\
&= \int \dd\Phi(p_{\rma})\, \varphi_{\rma}(p)\, e^{ib\cdot p} \, \eta_{ \rmb \rma} \ket{p_\rma,\beta_{j,m}^h}.
\] 
 Having defined this setup, let us proceed to make contact with observables. To start, let us assume that our incoming partial wave profile $\alpha(\omega)$ is peaked around a frequency which we denote $\omega_{\rm cl.}$. In this sense, it plays an analogous role to the KMOC wavepacket $\varphi(p)$ but only smeared in frequency. Note  that the classical limit requires that the wavelength of the classical wave, $\lambda_{\rm cl.}$, is much larger than the Compton wavelength, $\lambda_C$, of the massive particle so that $\omega \ll M$. However, unlike in the scattering of point masses, we do not need to impose a similar inequality on the momentum transfer $q$, since we are no longer restricted to the large impact parameter regime. This large $b$ regime is approached in the geometric optics limit considered in \cite{Cristofoli:2021vyo}. In our case, we consider partial wave scattering with small impact parameter $b \sim 0$, see figure \ref{figure2}. In practice this implies that amplitudes are no longer dominated by the $t$-channel (ladder) diagrams and we should keep all classical contributions including contact terms. 
 
We now apply our ansatz  to calculate the absorptive part of the cross section, incorporating mass and spin-changing effects. We can probe this by inserting the off-diagonal part of the completeness relation 

\[
\mathds{1}_{\Delta \rm MS} &= \sum_{s>0}^\infty \sum_{\{c\}}\int_{M_\rma^2}^\infty \dd M_\rmb^2 \rho(M_\rmb^2)\int \dd\Phi(p_{\rmb}) |p_{\rmb},s,\{c\}\rangle\langle p_{\rmb},s,\{c\}|
\]
so that the absorptive cross section is given 
by
\[
\sigma_{\text{abs}}\left(\omega_{\mathrm{cl}}, j, m, h\right)=\frac{\pi}{\omega_{\mathrm{cl}}^2} P_{\Delta \rm MS}\left(\omega_{\mathrm{cl}}, j, m, h\right) ,
\]
where
\[
P_{\Delta \rm MS}\left(\omega_{\mathrm{cl}}, j, m, h\right) = \sum_{s>0}^\infty \sum_{\{c\}} \int \dd \Phi(p'_{\rmb}) \int_{M_\rma^2}^\infty \dd M_b^2 \, \rho(M_b^2) \, |\bra{p'_{\rmb},\{c\}} S_{ \rmb \rma} \ket{\varphi_\rma} |^2. 
\]
That is, the absorptive cross section is determined by summing over final in channel $(\rmb)$. Let us calculate this by considering the quantity  
\[
\bra{p'_{\rmb},\{c\}}  S_{ \rmb \rma} \ket{\varphi_\rma,\alpha_{j,m}^h} = \int \dd\Phi(p_{\rma} )\,  \varphi_{\rma}(p)  e^{i b \cdot p} \bra{p'_{\rmb},\{c\}} \eta_{\rmb \rma} \ket{p_\rma,\beta_{j,m}^h},
\]
and expanding it to linear order in $\beta$ to obtain  
\[
\bra{p'_{\rmb},\{c\}} \eta_{\rmb \rma} \ket{p_\rma,\beta_{j,m}^h} = \bra{p'_{\rmb},\{c\}} \eta_{\rmb \rma} \ket{p_\rma}  +  \int \hd \omega \, \beta(\omega) \bra{p'_{\rmb},\{c\}} &\eta_{\rmb \rma} \ket{p_\rma,\omega,j,m,h}  \\&\qquad\qquad+\mathcal{O}(\beta^2) .  
\]
Demanding that the operator $ \eta_{\rmb \rma}$ conserves angular momentum, we note that the first term above must vanish since it represents the transition of a spinless state to a spinning state with no additional graviton contributions. 
Turning to the second term, and  recalling that $\beta$ admits a series expansion in powers of $\alpha$ such that $\beta^{(0)}(\omega) = \alpha(\omega)$, we have
\[
\bra{p'_{\rmb},\{c\}} \eta_{\rmb \rma} \ket{p_\rma,\beta_{j,m}^h} =   \int \hd \omega \, \alpha(\omega) \bra{p'_{\rmb},\{c\}} &\eta_{\rmb \rma} \ket{p_\rma,\omega,j,m,h} + \mathcal{O}(\alpha^2) .
\]
Expanding the state $\ket{\omega,j,m,h} $ in a plane wave basis $\ket{k^h}$, we obtain
\[
 \bra{p'_{\rmb},\{c\}} \eta_{\rmb \rma} \ket{p_\rma,\omega,j,m,h} =\frac{2\pi}{\omega}  \int \dd \Phi(k) \hdelta(k \cdot u - \omega)\, Y_{j,m}^h(k;u,n) \bra{p'_{\rmb},\{c\}} \eta_{\rmb \rma} \ket{p,k^h},
\]
where we have introduced the helicity-weighted spherical harmonics $Y_{j,m}^h$ and adopted the covariantized representation of \cite{Aoude:2023fdm}  and expressed $p^\mu_\rma = M_\rma u^\mu$. Using this, we find that at leading order in $\alpha$, the off-diagonal matrix element $ \bra{p'_{\rmb},\{c\}} S_{ \rmb \rma} \ket{\varphi_\rma,\alpha_{j,m}^h}$ is 
\[
 \bra{p'_{\rmb},\{c\}} S_{ \rmb \rma} \ket{\varphi_\rma,\alpha_{j,m}^h} = \int \dd\Phi(p_{\rma},k ) \varphi_{\rmb}(p)\,  \varphi_{\rma}(p) \alpha(\omega)  e^{i b \cdot (p' -p)} \\
\times \, \frac{2\pi}{\omega}  \, Y_{j,m}^h(k;u,n) \bra{p'_{\rmb},\{c\}} \eta_{\rmb \rma} \ket{p_\rma,k^h}. 
\]
Expressing things in this notation, it is easy to see that the quantity $\bra{p'_{\rmb},\{c\}} \eta_{\rmb \rma} \ket{k^h}$ can be related (to this order) to the mass-changing amplitudes of \cite{Aoude:2023fdm}. Indeed, comparing this expression to Eq. (2.18) in that reference, we identify 
\[\label{eta3pt}
\bra{p'_{\rmb},\{c\}} \eta_{\rmb \rma} \ket{p_\rma,k^h} = i \hdelta^4(p' - p - k) \mathcal{A}_{\rmb \rma}(p,k \rightarrow p'),
\]
where $\mathcal{A}_{\rmb \rma}(p,k \rightarrow p')$ is the mass-changing three-point amplitude of \cite{Aoude:2023fdm}. For our purposes, we will not explicitly construct this amplitude. Instead, we will match directly to the  absorptive cross-section. At leading order, our ansatz  gives 
\[\label{eq:LOmatching}
&\sigma_{\text{abs}}\left(\omega_{\mathrm{cl}}, j, m, h\right)=\frac{\pi}{\omega_{\mathrm{cl}}^2} \sum_{s>0}^\infty \sum_{\{c\}} \int \dd \Phi(p'_{\rmb}) \int_{M_\rma^2}^\infty \dd  M^2_\rmb\rho( M^2_\rmb) \\
 &\times \left| \frac{2\pi}{\omega}   \int \dd\Phi(p_{\rma},k ) \varphi_{\rmb}(p)\,  \varphi_{\rma}(p) \alpha(\omega)  e^{i b \cdot (p' -p)} \, Y_{j,m}^h(k;u,n) \bra{p'_{\rmb},\{c\}} \eta_{\rmb \rma} \ket{p_\rma,k^h}\right|^2, 
\]
which is then matched to the part of the absorptive cross section \cite{Chia:2020yla,Ivanov:2022qqt}
\[\label{AbsCrossSection}
\sigma_{\text{abs}}(\omega, j, m, h) = 4 \pi (GM)^{2j+2}(2\omega)^{2j} \left[ \frac{(j+h)!(j-h)!}{(2 j)!(2 j+1)!}\right]^2 \prod_{l=1}^j\left[l^2+\left(2 GM\omega\right)^2\right],
\]
to leading order in $G$. 
Equating eqs. \eqref{eq:LOmatching} and \eqref{AbsCrossSection} fixes the matrix element in eq. \eqref{eta3pt} to $\mathcal{O}(G^{j+1})$ with the identification of $M\equiv M_\rma$. 

Notice that our ansatz has allowed us to explicitly isolate the absorptive part of the $S$-matrix at this order. It is interesting to explore whether this ansatz simplifies the matching at higher orders. Although we will not do this explicitly, let us outline how the matching calculation is organised at higher orders. To this end, consider the matrix element
\[
\bra{p'_{\rmb},\{c\}}  S_{ \rmb \rma} \ket{p_\rma,\beta_{j,m}^h} = \bra{p'_{\rmb},\{c\}} \eta_{\rmb \rma} \ket{p_\rma,\beta_{j,m}^h},
\]
and note that there are two sources of higher order corrections. First, we have at linear order in $\beta$:
\[
\bra{p'_{\rmb},\{c\}} \eta_{\rmb \rma} \ket{p_\rma,\beta_{j,m}^h}= \int \hd \omega \, \beta(\omega) \bra{p'_{\rmb},\{c\}} \eta_{\rmb \rma} \ket{p_\rma,\omega,j,m,h} + \mathcal{O}(\beta^2).
\]
Since $\beta$ admits an expansion in powers of $\alpha$ (and the coupling), this contains terms to all orders in $G$. However, since $\bra{p'_{\rmb},\{c\}} \eta_{\rmb \rma} \ket{p_\rma,\omega,j,m,h} $ is already determined by the LO matching in \eqref{eq:LOmatching}, this quantity  is fully determined to all orders from standard perturbation theory. Moving on to contributions quadratic in $\beta$, we encounter the term
\[
\int \hd \omega \, \hd \omega' \, \beta(\omega) \beta(\omega') \bra{p'_{\rmb},\{c\}} \eta_{\rmb \rma} \ket{p,\omega,\omega'},
\]
where we have suppressed angular momentum indices for brevity. At this order, we encounter the new quantity $\bra{p'_{\rmb},\{c\}} \eta_{\rmb \rma} \ket{p_\rma,\omega,\omega'}$ which is not fixed by the LO matching.  We can interpret this as a higher order contact term representing the absorption of two gravitons. This term is fixed by comparing to the NLO part of the cross section. Proceeding in this manner, we see that higher order corrections can come from either standard perturbative vertices (originating from $\beta)$ or higher order absorptive vertices from $\eta_{\rmb \rma}$. This separation between absorptive and perturbative scattering contributions makes this ansatz particularly suitable for performing matching calculations with black hole perturbation theory, where a similar factorization
is realized in the solution of the wave equation for Schwarzchild and Kerr metrics.  It is interesting to explore whether this approach leads to an efficient way to match the absorptive cross section at higher orders.

\section{Conclusions and outlook}

In this work, we have formulated the ICCE, or \emph{Inelastic Coupled-Channel Eikonal}, with the aim of studying inelastic effects in black hole dynamics using on-shell amplitudes. In Section 2, we introduced this formalism, highlighting how the language of hadron physics and quantum information provides a natural interpretation of inelastic scattering dynamics in terms of an original channel analysis. We have introduced the concept of channel space, including diagonal and off-diagonal channels, and related these to the insertion of a completeness relation with a spectral density that accounts for inelastic effects. In Section 3, following a review of the eikonal formalism in the probe limit in QFT, we focused on describing inelastic effects in a potential theory. We considered a toy model on a  background, where one particle can change its mass and spin multiple times before returning to its original Casimir labels. Naturally expressed in terms of channel analysis, we have identified an important feature that distinguishes this eikonal analysis from the standard elastic case. Specifically, while in $2 \rightarrow 2$ scattering, boxes and crossed-boxes resum into an exponential, in this context, mass change effects can alter the iteration of a certain topology, reducing exchange symmetry and suggesting instead a novel geometric progression. 

Section 3 provides a simple yet interesting model to illustrate these effects and the absence of the usual $1/n!$ terms: the resummation scheme resembles a geometric progression now. It is this new type of resummation which now yields a classically finite mass changing effect in physical observables, despite the mass-mismatch $w$ being quantum in each single diagrammatic exchange. This result is clearly non trivial from a pure KMOC perspective of the calculation. In the Appendix \ref{MCExp}, we also presented another example of this resummation inspired by previous analyses by Harrington and Rudin \cite{Rudin:1970jv, Harrington:1971ax, Harrington:1968zza}. 

Section 4 relaxes the probe limit assumption  and instead focuses on the arbitrary $2 \rightarrow 2$ scattering. There, we focuses on the extraction of classical observables from the KMOC formalism using a channel analysis. We have examined two physical scenarios where a channel analysis of inelastic effects is relevant in KMOC: the inelastic scattering of two black holes and the scattering of a wave off a heavy source. In the first case, we find that inelastic effects introduce off-diagonal contributions, providing additional contributions beyond those studied previously. When radiation is neglected, the final state is no longer described solely by an eikonal phase but also includes an inelastic function that governs both diagonal and off-diagonal contributions. We have also conjectured an all order structure for such state, at the same time noting that unitarity imposes non-trivial relations among these inelastic functions (\ref{eq:unit-constraint-eta}). Then,  we have applied our framework to the computation of the linear impulse in  $2 \rightarrow 2$ gravitational scattering; this analysis naturally yields an additional inelastic contribution arising from the presence of event horizons through a saddle point approximation. The ICCE framework accounts for these inelastic contributions systematically.
Our results are consistent with those first presented in \cite{Goldberger:2020wbx} and more recently rederived in \cite{Jones:2023ugm}, and further provide a non trivial consistency check that our ansatz\"e are physically sound. 

{Following the same strategy, we then examined the structure of the final state for the case of a wave scattering inelastically off a heavy background. We provided a new representation that is non-perturbative in the coherent waveshape of the final state, which, interestingly, can be expressed in terms of the Lambert-W function. By then expanding this} and matching it with the absorption cross section as in \cite{Aoude:2023fdm}, this has allowed us to determine a $3$-point for inelastic mass change effects to $\mathcal{O}(G^{j+1})$ where $j$ denotes the angular momenta entering a partial wave expansion of the incoming wave. 

Looking into future directions we can already outline a few promising ones. It would be interesting to extend this analysis to cases where complete absorption of one of the massive particles occurs, thereby expanding the channel analysis to include complete capture scenarios, as recently studied in~\cite{Aoki:2024boe}. A prominent feature of the final state ansatz of \cite{Cristofoli:2021jas, Monteiro:2020plf} was the inclusion of dynamically produced radiation in terms of coherent states. There, the coherent parameter is essentially the five point scattering amplitude. A straightforward extension of our model is to then consider emitted radiation in the scattered state, and to understand how this interplays with inelasticity. Finally, we believe another fascinating direction is offered by the  investigation of black hole laws of thermodynamics using amplitudes. This is where we believe the channel language of the ICCE framework will be particularly effective. Since our formalism naturally retains both aspects of information theory and scattering amplitudes, it appears  the natural choice for the study of black hole laws in an on-shell manner. 
We leave these and other exciting directions for upcoming future works.

\label{sec:Conclusion}

\acknowledgments
We are sincerely grateful to Katsuki Aoki, Alex Ochirov and Donal O'Connell for valuable conversations and useful feedbacks to earlier versions of the draft. We also thank Fabian Bautista and Max Hansen for useful conversation. A.C. and R.A. thank David Kosower and  the financial support from  the Grant [ERC–AdG–885414] during their visits to the IPhT during the spring of 2024. A.C. was supported by JSPS KAKENHI Grant No. [JP24KF0153]. M.S. has been supported by the European Research Council under Advanced Investigator Grant [ERC–AdG–885414]. 
R.A. is supported by UK Research and Innovation (UKRI) under the UK
government’s Horizon Europe Marie Skłodowska-Curie funding guarantee grant [EP/Z000947/1] and by the STFC grant “Particle Theory at the Higgs Centre”.

\newpage
\appendix

\section{Multi-Channel Exponentiation}\label{MCExp}
In this appendix we give a brief overview of a model where eikonal resummation is realised in the multi-channel scattering. 
Here we will do this following a class of models introduced by Harrington and Rudin \cite{Rudin:1970jv, Harrington:1971ax, Harrington:1968zza}, where it is possible to obtain all-orders expressions in the coupling between different channels. 

Consider a scattering setup involving two channels $\rma$ and $\rmb$. Involving initial states of the form
\[
|\varphi_\rma\rangle = \int \dd\Phi(p_{1\rma},p_2)\, \varphi_{\rma}(p_1) \varphi(p_2)\, e^{ib\cdot p_1}|p_{1\rma},p_2\rangle,
\]
and likewise for $|\varphi_\rmb\rangle$. Here we consider particle $p_2$ to be the heavy particle in the probe limit, which is taken to always remain in the same channel. We take our initial particle to be in channel $\rma$, and we allow for an arbitrary number of channel transitions between $\rma$ and $\rmb$.
 For example, the amplitude $\langle \varphi_{\rma}|S_{\rma \rma} |\varphi_\rma\rangle$ may include any (even) number of intermediate $\rma \rightarrow \rmb$ and $\rmb \rightarrow \rma$ transitions between channels. It is therefore helpful to introduce the notation
\[
\langle \varphi_{\rmb}|S^{(n)}_{\rmb \rma } |\varphi_\rma\rangle = \int \dd\Phi(p_{1\rma}, p'_{1\rmb},p_2,p_2')\, \varphi^*_{\rmb}(p_1')\varphi_{\rma}(p_1) \varphi^*(p_2')\varphi(p_2') \, e^{ib\cdot (p_1 - p_1')}  \\
\times \left( \delta_{\rma\rmb} \,\hdelta_{\Phi} ( p_1' - p_1 )\hdelta_{\Phi} ( p_2' - p_2 ) + i \hdelta^4 \left(p_1 + p_2 - p_1' -p_2' \right)\mathcal{A}_{\rmb \rma }^{(n)}(p_1,p_2\rightarrow p_1',p_2') \right),
\]
where $ \mathcal{A}_{\rmb \rma}^{(n)}$ is the amplitude with initial channel $\rma$ and final channel $\rmb$, with $n$ intermediate channel transitions. We have also introduced a Kronecker delta $\delta_{\rma\rmb}$ to indicate that there is no forward scattering term between different channels.  Note that the coupling between different channels need not be the same as the couplings between the same channel. It is therefore useful to denote the coupling between channels $\rma$ and $\rmb$ by $g_{\rma \rmb}$ so that $ \mathcal{A}_{\rmb \rma}^{(n)}$ is $\mathcal{O}(g^{n/2}_{\rma \rmb} g^{n/2}_{\rmb \rma})$. 

With this notation in place, let us proceed to define our eikonal functions $\chi_{\rma \rma}$ and $\chi_{\rmb \rma}$. To do so, we define the amplitude in impact parameter space using the on-shell Fourier transform:
\[
\mathcal{A}_{\rmb \rma}^{(n)}(b)= \mathcal{F}_{\text{o.s.}}\left[ \mathcal{A}_{\rmb \rma}^{(n)}(p_1 \rightarrow p_1+q) \right] \equiv \int \dd^4 q \, \hdelta(2p_1\cdot q) \hdelta(2 p_2 \cdot q) \, e^{i q \cdot b}  \, \mathcal{A}_{\rmb \rma}^{(n)}(p_1 \rightarrow p_1+q),
\]
where we have defined $q=p_1'-p_1$ and abbreviated 
\[
\mathcal{A}_{\rmb \rma}^{(n)}(p_1 \rightarrow p_1+q) \equiv \mathcal{A}_{\rmb \rma}^{(n)}(p_1,p_2 \rightarrow p_1+q,p_2-q)\]
by suppressing its dependence on the heavy particle.
To start, consider the amplitude $\mathcal{A}_{\rma \rma}^{(0)}(b)$, representing the diagonal amplitude with no intermediate state changes. Here the amplitude is related to the eikonal phase by the standard formula
\[
i \mathcal{A}_{\rma \rma}^{(0)}(b) = e^{i \chi_{\rma \rma}(b)} -1,
\]
which we take as a given, ie we assume that the purely diagonal amplitude in channel $\rma$ eikonalizes in the usual manner. We want to understand whether this exponentiation persists when channel ($\rmb$) becomes accessible. To this end we consider the off-diagonal amplitudes, adopting the following relation due to Rudin \cite{Rudin:1970jv}:
\[\label{eq:ODeikonalAnsatz}
i \mathcal{A}_{\rmb \rma}^{(1)}(b) = e^{i \chi_{\rma \rma}(b)} \, \chi_{\rmb \rma}(b).
\]
Let us emphasise that the above equation is taken to define the off-diagonal eikonal phase $i \chi_{\rmb \rma}$. 
In other words, we will seek out conditions on the amplitude $\mathcal{A}_{\rmb \rma}^{(1)}$ such that eq. \eqref{eq:ODeikonalAnsatz} is realised for an appropriate choice of $\chi_{\rmb \rma}$ . To get a better sense of this ansatz, let us invert the above equation to get
\[\label{eq:OffD1}
\chi_{\rmb \rma}(b) = i \, e^{-i \chi_{\rma \rma} ( b)}\mathcal{F}_{\text{o.s.}}\left[    \mathcal{A}_{\rmb \rma}^{(1)}(p \rightarrow p+q) \right]
\]
The amplitude $\mathcal{A}_{\rmb \rma}^{(1)}(p \rightarrow p+q)$ is obtained by summing all diagrams starting from channel $\rma$ and ending at channel $\rmb$, with only one intermediate $\rma \rightarrow \rmb$ transition. Considering the diagram at $\mathcal{O}(g_\rmb^m g_{\rmb \rma} g_\rma^n)$, the leading order contribution in the eikonal approximation takes the form
\[\label{eq:FactorisedA}
\int_{ {\ell}_{n+m+1}}   \frac{\mathcal{A}^{(0)}_{\rma \rma}(\ell_{1 \cdots n}) i \Pi(n,n+1) \tilde{\mathcal{A}}^{(1)}_{\rmb \rma}(\ell_{1 \cdots n+m+1})}{2 p_1 \cdot\ell_{1 \cdots n} + i \epsilon}
  \hat{\delta}(2p_2\cdot \ell_{1 \cdots n}) \hat{\delta}^4(\ell_{1 \cdots n+m+1} - q).
 \]
Here we have defined the truncated amplitude ${\tilde{\mathcal{A}}}^{(1)}_{\rmb \rma}$ containing no iterations in the initial channel ($\rma$), ie the $\mathcal{O}(g_\rma^0)$ part of the amplitude $\mathcal{A}^{(1)}_{\rmb \rma}$. We have also adopted the following abbreviations
\[
&\mathcal{A}^{(0)}_{\rma \rma}(\ell_{1 \cdots n}) \equiv \mathcal{A}^{(0)}_{\rma \rma}(p_1 \rightarrow p_1+\ell_{1\cdots n}) ,\\
&\tilde{\mathcal{A}}^{(1)}_{\rmb \rma}(\ell_{1 \cdots n+m+1}) \equiv \tilde{\mathcal{A}}^{(1)}_{\rmb \rma}(p_1+ \ell_{1 \cdots n} \rightarrow p_1+ \ell_{1 \cdots n+m+1}),
\]
where $\ell_{ij} \equiv \ell_i + \ell_{i+1}+ \cdots + \ell_j$. In addition, we defined
\[
 \int_{ \{\ell\}_n} \equiv \int \hd^4 \ell_1 \cdots \int \hd^4 \ell_{n}, \quad
 \int_{ \{\ell\}_n^m} \equiv \int \hd^4 \ell_n \cdots \int \hd^4 \ell_{m},
\]
 and used $ \Pi(n,n+1) $ to denote the projector over states flowing through the cut. We will leave this projector implicit in what follows, noting that it is nontrivial in the case where the states $p_{1\rmb}$ carry spin. 
Note that the form \eqref{eq:OffD1} is merely a rewriting of the amplitude $\mathcal{A}^{(1)}_{\rmb \rma}$ by explicitly isolating iterations in the initial $\rma$ channel. We have done so by exposing the $n$'th linearised propagator $1/(2 p_1 \cdot 
 \ell_{1 \cdots n} + i \epsilon)$ and noting that the numerator is fixed (up to terms subleading in the eikonal approximation) by requiring the correct factorisation when this propagator is on-shell. Note that $\hdelta(2p_2\cdot \ell_{1 \cdots n})$ is simply the exposed propagator of the heavy particle $p_2$,which always appears as a delta function in the probe limit. 

So far, we have placed no restrictions on the off-diagonal amplitude $\mathcal{A}_{\rmb\rma}$. Now, we introduce an additional assumption which is that the principle-valued part of the propagator in eq.\eqref{eq:FactorisedA} vanishes, so that we may replace it by a cut propagator. This constitutes a restriction on the interaction potential connecting the two channels. See \cite{Rudin:1970jv, Harrington:1971ax, Harrington:1968zza} for examples of such potentials.  With this assumption we have

% Pattern Info
 
\tikzset{
pattern size/.store in=\mcSize, 
pattern size = 5pt,
pattern thickness/.store in=\mcThickness, 
pattern thickness = 0.3pt,
pattern radius/.store in=\mcRadius, 
pattern radius = 1pt}
\makeatletter
\pgfutil@ifundefined{pgf@pattern@name@_hrh0z2leu}{
\pgfdeclarepatternformonly[\mcThickness,\mcSize]{_hrh0z2leu}
{\pgfqpoint{0pt}{0pt}}
{\pgfpoint{\mcSize+\mcThickness}{\mcSize+\mcThickness}}
{\pgfpoint{\mcSize}{\mcSize}}
{
\pgfsetcolor{\tikz@pattern@color}
\pgfsetlinewidth{\mcThickness}
\pgfpathmoveto{\pgfqpoint{0pt}{0pt}}
\pgfpathlineto{\pgfpoint{\mcSize+\mcThickness}{\mcSize+\mcThickness}}
\pgfusepath{stroke}
}}
\makeatother

% Pattern Info
 
\tikzset{
pattern size/.store in=\mcSize, 
pattern size = 5pt,
pattern thickness/.store in=\mcThickness, 
pattern thickness = 0.3pt,
pattern radius/.store in=\mcRadius, 
pattern radius = 1pt}
\makeatletter
\pgfutil@ifundefined{pgf@pattern@name@_s402rxgj2}{
\pgfdeclarepatternformonly[\mcThickness,\mcSize]{_s402rxgj2}
{\pgfqpoint{0pt}{0pt}}
{\pgfpoint{\mcSize+\mcThickness}{\mcSize+\mcThickness}}
{\pgfpoint{\mcSize}{\mcSize}}
{
\pgfsetcolor{\tikz@pattern@color}
\pgfsetlinewidth{\mcThickness}
\pgfpathmoveto{\pgfqpoint{0pt}{0pt}}
\pgfpathlineto{\pgfpoint{\mcSize+\mcThickness}{\mcSize+\mcThickness}}
\pgfusepath{stroke}
}}
\makeatother

% Pattern Info
 
\tikzset{
pattern size/.store in=\mcSize, 
pattern size = 5pt,
pattern thickness/.store in=\mcThickness, 
pattern thickness = 0.3pt,
pattern radius/.store in=\mcRadius, 
pattern radius = 1pt}
\makeatletter
\pgfutil@ifundefined{pgf@pattern@name@_mmdhrf110}{
\pgfdeclarepatternformonly[\mcThickness,\mcSize]{_mmdhrf110}
{\pgfqpoint{0pt}{0pt}}
{\pgfpoint{\mcSize+\mcThickness}{\mcSize+\mcThickness}}
{\pgfpoint{\mcSize}{\mcSize}}
{
\pgfsetcolor{\tikz@pattern@color}
\pgfsetlinewidth{\mcThickness}
\pgfpathmoveto{\pgfqpoint{0pt}{0pt}}
\pgfpathlineto{\pgfpoint{\mcSize+\mcThickness}{\mcSize+\mcThickness}}
\pgfusepath{stroke}
}}
\makeatother

% Pattern Info
 
\tikzset{
pattern size/.store in=\mcSize, 
pattern size = 5pt,
pattern thickness/.store in=\mcThickness, 
pattern thickness = 0.3pt,
pattern radius/.store in=\mcRadius, 
pattern radius = 1pt}
\makeatletter
\pgfutil@ifundefined{pgf@pattern@name@_c5t4209r2}{
\pgfdeclarepatternformonly[\mcThickness,\mcSize]{_c5t4209r2}
{\pgfqpoint{0pt}{0pt}}
{\pgfpoint{\mcSize+\mcThickness}{\mcSize+\mcThickness}}
{\pgfpoint{\mcSize}{\mcSize}}
{
\pgfsetcolor{\tikz@pattern@color}
\pgfsetlinewidth{\mcThickness}
\pgfpathmoveto{\pgfqpoint{0pt}{0pt}}
\pgfpathlineto{\pgfpoint{\mcSize+\mcThickness}{\mcSize+\mcThickness}}
\pgfusepath{stroke}
}}
\makeatother
\tikzset{every picture/.style={line width=0.75pt}} %set default line width to 0.75pt        

\[
\mathcal{A}_{\rmb \rma}^{(1)}(p\rightarrow p+q) = \sum_{n,m} \left(\vcenter{\hbox{ \begin{tikzpicture}[x=0.75pt,y=0.75pt,yscale=-1,xscale=1]
%uncomment if require: \path (0,148); %set diagram left start at 0, and has height of 148

%Straight Lines [id:da3477566671280202] 
\draw [color={rgb, 255:red, 0; green, 0; blue, 0 }  ,draw opacity=1 ][line width=1.5]    (131,38.37) -- (278.45,38.37) ;
%Shape: Ellipse [id:dp10776038437869861] 
\draw  [pattern=_hrh0z2leu,pattern size=3.2249999999999996pt,pattern thickness=0.75pt,pattern radius=0pt, pattern color={rgb, 255:red, 0; green, 0; blue, 0}] (218.78,98.7) .. controls (223.06,98.72) and (226.51,102.18) .. (226.48,106.42) .. controls (226.46,110.65) and (222.98,114.07) .. (218.7,114.04) .. controls (214.42,114.02) and (210.98,110.56) .. (211,106.33) .. controls (211.02,102.09) and (214.51,98.68) .. (218.78,98.7) -- cycle ;
%Straight Lines [id:da019282458363405475] 
\draw    (218.45,38.37) .. controls (220.13,40.02) and (220.14,41.69) .. (218.48,43.37) .. controls (216.83,45.04) and (216.84,46.71) .. (218.51,48.37) .. controls (220.18,50.03) and (220.19,51.7) .. (218.53,53.37) .. controls (216.88,55.04) and (216.89,56.71) .. (218.56,58.37) .. controls (220.23,60.03) and (220.24,61.7) .. (218.59,63.37) .. controls (216.94,65.04) and (216.95,66.71) .. (218.62,68.37) .. controls (220.29,70.03) and (220.3,71.7) .. (218.64,73.37) .. controls (216.99,75.04) and (217,76.71) .. (218.67,78.37) .. controls (220.34,80.03) and (220.35,81.7) .. (218.7,83.37) .. controls (217.05,85.04) and (217.06,86.71) .. (218.73,88.37) .. controls (220.4,90.03) and (220.41,91.7) .. (218.75,93.37) .. controls (217.1,95.04) and (217.11,96.71) .. (218.78,98.37) -- (218.78,98.7) -- (218.78,98.7) ;
%Shape: Ellipse [id:dp4550089951910685] 
\draw  [pattern=_s402rxgj2,pattern size=3.2249999999999996pt,pattern thickness=0.75pt,pattern radius=0pt, pattern color={rgb, 255:red, 0; green, 0; blue, 0}] (278.78,98.7) .. controls (283.06,98.72) and (286.51,102.18) .. (286.48,106.42) .. controls (286.46,110.65) and (282.98,114.07) .. (278.7,114.04) .. controls (274.42,114.02) and (270.98,110.56) .. (271,106.33) .. controls (271.02,102.09) and (274.51,98.68) .. (278.78,98.7) -- cycle ;
%Straight Lines [id:da7088381614826345] 
\draw    (278.45,38.37) .. controls (280.13,40.02) and (280.14,41.69) .. (278.48,43.37) .. controls (276.83,45.04) and (276.84,46.71) .. (278.51,48.37) .. controls (280.18,50.03) and (280.19,51.7) .. (278.53,53.37) .. controls (276.88,55.04) and (276.89,56.71) .. (278.56,58.37) .. controls (280.23,60.03) and (280.24,61.7) .. (278.59,63.37) .. controls (276.94,65.04) and (276.95,66.71) .. (278.62,68.37) .. controls (280.29,70.03) and (280.3,71.7) .. (278.64,73.37) .. controls (276.99,75.04) and (277,76.71) .. (278.67,78.37) .. controls (280.34,80.03) and (280.35,81.7) .. (278.7,83.37) .. controls (277.05,85.04) and (277.06,86.71) .. (278.73,88.37) .. controls (280.4,90.03) and (280.41,91.7) .. (278.75,93.37) .. controls (277.1,95.04) and (277.11,96.71) .. (278.78,98.37) -- (278.78,98.7) -- (278.78,98.7) ;
%Shape: Ellipse [id:dp024096887377100917] 
\draw  [color={rgb, 255:red, 0; green, 0; blue, 0 }  ,draw opacity=1 ][fill={rgb, 255:red, 74; green, 144; blue, 226 }  ,fill opacity=1 ] (278.5,46.34) .. controls (282.77,46.31) and (286.21,42.73) .. (286.19,38.33) .. controls (286.16,33.92) and (282.68,30.37) .. (278.41,30.4) .. controls (274.14,30.42) and (270.69,34.01) .. (270.72,38.41) .. controls (270.74,42.81) and (274.23,46.36) .. (278.5,46.34) -- cycle ;
%Straight Lines [id:da8383978856560432] 
\draw [color={rgb, 255:red, 0; green, 0; blue, 0 }  ,draw opacity=1 ][line width=0.75]    (286.19,36.83) -- (401,36.87)(286.18,39.83) -- (401,39.87) ;
%Shape: Ellipse [id:dp3456976381905048] 
\draw  [pattern=_mmdhrf110,pattern size=3.2249999999999996pt,pattern thickness=0.75pt,pattern radius=0pt, pattern color={rgb, 255:red, 0; green, 0; blue, 0}] (373.3,98.7) .. controls (377.58,98.72) and (381.02,102.18) .. (381,106.42) .. controls (380.98,110.65) and (377.49,114.07) .. (373.22,114.04) .. controls (368.94,114.02) and (365.49,110.56) .. (365.52,106.33) .. controls (365.54,102.09) and (369.02,98.68) .. (373.3,98.7) -- cycle ;
%Straight Lines [id:da6539928863061485] 
\draw    (372.97,38.37) .. controls (374.64,40.02) and (374.65,41.69) .. (373,43.37) .. controls (371.34,45.04) and (371.35,46.71) .. (373.02,48.37) .. controls (374.69,50.03) and (374.7,51.7) .. (373.05,53.37) .. controls (371.4,55.04) and (371.41,56.71) .. (373.08,58.37) .. controls (374.75,60.03) and (374.76,61.7) .. (373.11,63.37) .. controls (371.45,65.04) and (371.46,66.71) .. (373.13,68.37) .. controls (374.8,70.03) and (374.81,71.7) .. (373.16,73.37) .. controls (371.51,75.04) and (371.52,76.71) .. (373.19,78.37) .. controls (374.86,80.03) and (374.87,81.7) .. (373.22,83.37) .. controls (371.56,85.04) and (371.57,86.71) .. (373.24,88.37) .. controls (374.91,90.03) and (374.92,91.7) .. (373.27,93.37) .. controls (371.62,95.04) and (371.63,96.71) .. (373.3,98.37) -- (373.3,98.7) -- (373.3,98.7) ;
%Shape: Ellipse [id:dp21572799326523873] 
\draw  [pattern=_c5t4209r2,pattern size=3.2249999999999996pt,pattern thickness=0.75pt,pattern radius=0pt, pattern color={rgb, 255:red, 0; green, 0; blue, 0}] (153.3,98.7) .. controls (157.58,98.72) and (161.02,102.18) .. (161,106.42) .. controls (160.98,110.65) and (157.49,114.07) .. (153.22,114.04) .. controls (148.94,114.02) and (145.49,110.56) .. (145.52,106.33) .. controls (145.54,102.09) and (149.02,98.68) .. (153.3,98.7) -- cycle ;
%Straight Lines [id:da5421108314098562] 
\draw    (152.97,38.37) .. controls (154.64,40.02) and (154.65,41.69) .. (153,43.37) .. controls (151.34,45.04) and (151.35,46.71) .. (153.02,48.37) .. controls (154.69,50.03) and (154.7,51.7) .. (153.05,53.37) .. controls (151.4,55.04) and (151.41,56.71) .. (153.08,58.37) .. controls (154.75,60.03) and (154.76,61.7) .. (153.11,63.37) .. controls (151.45,65.04) and (151.46,66.71) .. (153.13,68.37) .. controls (154.8,70.03) and (154.81,71.7) .. (153.16,73.37) .. controls (151.51,75.04) and (151.52,76.71) .. (153.19,78.37) .. controls (154.86,80.03) and (154.87,81.7) .. (153.22,83.37) .. controls (151.56,85.04) and (151.57,86.71) .. (153.24,88.37) .. controls (154.91,90.03) and (154.92,91.7) .. (153.27,93.37) .. controls (151.62,95.04) and (151.63,96.71) .. (153.3,98.37) -- (153.3,98.7) -- (153.3,98.7) ;
%Straight Lines [id:da04134879368686184] 
\draw [color={rgb, 255:red, 208; green, 2; blue, 27 }  ,draw opacity=1 ][line width=1.5]  [dash pattern={on 1.69pt off 2.76pt}]  (248,28.91) -- (248,76.83) -- (248,109.37) ;

% Text Node
\draw (112,30.77) node [anchor=north west][inner sep=0.75pt]    {$\rma$};
% Text Node
\draw (412,30.77) node [anchor=north west][inner sep=0.75pt]    {$\rmb$};
% Text Node
\draw (132,70.77) node [anchor=north west][inner sep=0.75pt]    {$\ell_{1}$};
% Text Node
\draw (195,70.77) node [anchor=north west][inner sep=0.75pt]    {$\ell_{n}$};
% Text Node
\draw (168,70.77) node [anchor=north west][inner sep=0.75pt]    {$\cdots $};
% Text Node
\draw (289,70.77) node [anchor=north west][inner sep=0.75pt]    {$\cdots $};
% Text Node
\draw (314,70.77) node [anchor=north west][inner sep=0.75pt]    {$\ell_{n+m+1}$};

\end{tikzpicture}

 }} \right)
\] 

In this case the off-diagonal eikonal function becomes
\[\label{eq:ODCut}
\chi_{\rmb \rma}(b) = i \sum_{n,m}   e^{-i \chi_{\rma \rma} ( b)}\,\mathcal{F}_{\text{o.s.}} \, \bigg[  \int_{ {\ell}_{n+m+1}}   & i \mathcal{A}^{(0)}_{\rma \rma}(\ell_{1 \cdots n})\; 
\hdelta(2 p_1 \cdot 
 \ell_{1 \cdots n}) \hdelta(2 p_2 \cdot 
 \ell_{1 \cdots n}) \\
& \times \tilde{\mathcal{A}}^{(1)}_{\rmb \rma}(\ell_{1 \cdots n+m+1}) 
\hdelta^4(\ell_{1 \cdots n+m+1} - q) \bigg].
\]
The effect of assuming the cut condition is that it allows the loop integrations to factorise so that
\[\label{eq:ODfactorisation}
\chi_{\rmb \rma}(b) = e^{-i \chi_{\rma \rma}(b)} \sum_{n,m} \mathcal{I}_{\rma \rma}(n) \, \tilde{\mathcal{I}}_{\rmb \rma}(m)
\]
where
\[
&\mathcal{I}_{\rma \rma}( n)  \equiv  i \int_{ \{\ell\}_{n}} \hdelta(2 p_1 \cdot 
\ell_{1 \cdots n}) \hdelta(2 p_2 \cdot 
 \ell_{1 \cdots n})   \mathcal{A}^{(0)}_{\rma \rma}(\ell_{1 \cdots n})e^{i b \cdot \ell_{1 \cdots n}},  \\
&\tilde{\mathcal{I}}_{\rmb \rma}(m) \equiv \int_{ \{\ell\}_{n+1}^{n+m+1}} \hdelta(2 p_1 \cdot \ell_{n+1 \cdots n+m+1}) \hdelta(2 p_2 \cdot \ell_{n+1 \cdots n+m+1}) e^{i b \cdot \ell_{n+1 \cdots n+m+1}} \tilde{\mathcal{A}}^{(1)}_{\rmb \rma}(\ell_{1 \cdots n+m+1}).
\]
This factorization is realized provided that the amplitudes $\mathcal{A}^{(0)}_{\rma \rma}$ and $\tilde{\mathcal{A}}^{(1)}_{\rmb \rma}$ depend on different subsets of the loop momenta. In this case, it is straightforward to verify that
\[
\sum_{n} \mathcal{I}_{\rma \rma}(n) = e^{i \chi_{\rma \rma}(b)}.
\]
In other words, the diagonal iterations in the initial channel ($\rma$) exponentiate separately. Returning to $\chi_{\rmb \rma}(b)$, we see that the prefactor $e^{-i \chi_{\rma \rma}}$ removes  diagonal iterations in the $(\rma)$ channel, so that the off-diagonal eikonal function is defined by the Fourier transform of the truncated amplitude
\[
\chi_{\rma \rma}(b) = \mathcal{F}_{\text{o.s.}}\left[\tilde{\mathcal{A}}^{(1)}_{\rmb \rma}(\ell_{1 \cdots n+m+1}) \right].
\]
This gives a simple interpretation to the off-diagonal eikonal function $\chi_{\rmb \rma}$: It is obtained by truncating diagonal iterations in the initial channel $(\rma)$ and taking the Fourier transform to impact parameter space. Note that $\chi_{\rmb \rma}$ still contains an arbitrary number of iterations in the final channel $\rmb$, ie we can expand in powers of $g_\rmb$: 
\[
\chi_{\rmb \rma}(b) = g_{\rmb \rma}\sum_i g_\rmb^i \,\chi^{(i)}_{\rmb \rma}(b). 
\]
This structure becomes useful when considering higher order diagrams.
To see this, let us return to the diagonal amplitude, this time allowing intermediate channel transitions. 
The leading contribution is $\mathcal{A}^{(2)}_{\rma \rma}$ obtained by inserting two off-diagonal interactions. A generic diagram contributing to this amplitude at $\mathcal{O}(g_\rma^n g_\rmb^m g_{\rmb \rma}^2)$ is obtained by inserting $n_1$ diagonal vertices in the initial $(\rma)$ channel, $n_2 = n - n_1$  iterations in the final channel $(\rma)$, and $m$ diagonal iterations in the intermediate channel $(\rmb)$.  
The full amplitude is then obtained by summing over all values of the $m$, $n_1$ and $n$. Diagrammatically, we have

% Pattern Info
 
\tikzset{
pattern size/.store in=\mcSize, 
pattern size = 5pt,
pattern thickness/.store in=\mcThickness, 
pattern thickness = 0.3pt,
pattern radius/.store in=\mcRadius, 
pattern radius = 1pt}
\makeatletter
\pgfutil@ifundefined{pgf@pattern@name@_wlsxkbwul}{
\pgfdeclarepatternformonly[\mcThickness,\mcSize]{_wlsxkbwul}
{\pgfqpoint{0pt}{0pt}}
{\pgfpoint{\mcSize+\mcThickness}{\mcSize+\mcThickness}}
{\pgfpoint{\mcSize}{\mcSize}}
{
\pgfsetcolor{\tikz@pattern@color}
\pgfsetlinewidth{\mcThickness}
\pgfpathmoveto{\pgfqpoint{0pt}{0pt}}
\pgfpathlineto{\pgfpoint{\mcSize+\mcThickness}{\mcSize+\mcThickness}}
\pgfusepath{stroke}
}}
\makeatother

% Pattern Info
 
\tikzset{
pattern size/.store in=\mcSize, 
pattern size = 5pt,
pattern thickness/.store in=\mcThickness, 
pattern thickness = 0.3pt,
pattern radius/.store in=\mcRadius, 
pattern radius = 1pt}
\makeatletter
\pgfutil@ifundefined{pgf@pattern@name@_6dguie6jz}{
\pgfdeclarepatternformonly[\mcThickness,\mcSize]{_6dguie6jz}
{\pgfqpoint{0pt}{0pt}}
{\pgfpoint{\mcSize+\mcThickness}{\mcSize+\mcThickness}}
{\pgfpoint{\mcSize}{\mcSize}}
{
\pgfsetcolor{\tikz@pattern@color}
\pgfsetlinewidth{\mcThickness}
\pgfpathmoveto{\pgfqpoint{0pt}{0pt}}
\pgfpathlineto{\pgfpoint{\mcSize+\mcThickness}{\mcSize+\mcThickness}}
\pgfusepath{stroke}
}}
\makeatother

% Pattern Info
 
\tikzset{
pattern size/.store in=\mcSize, 
pattern size = 5pt,
pattern thickness/.store in=\mcThickness, 
pattern thickness = 0.3pt,
pattern radius/.store in=\mcRadius, 
pattern radius = 1pt}
\makeatletter
\pgfutil@ifundefined{pgf@pattern@name@_jbhv1n2v6}{
\pgfdeclarepatternformonly[\mcThickness,\mcSize]{_jbhv1n2v6}
{\pgfqpoint{0pt}{0pt}}
{\pgfpoint{\mcSize+\mcThickness}{\mcSize+\mcThickness}}
{\pgfpoint{\mcSize}{\mcSize}}
{
\pgfsetcolor{\tikz@pattern@color}
\pgfsetlinewidth{\mcThickness}
\pgfpathmoveto{\pgfqpoint{0pt}{0pt}}
\pgfpathlineto{\pgfpoint{\mcSize+\mcThickness}{\mcSize+\mcThickness}}
\pgfusepath{stroke}
}}
\makeatother

% Pattern Info
 
\tikzset{
pattern size/.store in=\mcSize, 
pattern size = 5pt,
pattern thickness/.store in=\mcThickness, 
pattern thickness = 0.3pt,
pattern radius/.store in=\mcRadius, 
pattern radius = 1pt}
\makeatletter
\pgfutil@ifundefined{pgf@pattern@name@_wlgh5lik4}{
\pgfdeclarepatternformonly[\mcThickness,\mcSize]{_wlgh5lik4}
{\pgfqpoint{0pt}{0pt}}
{\pgfpoint{\mcSize+\mcThickness}{\mcSize+\mcThickness}}
{\pgfpoint{\mcSize}{\mcSize}}
{
\pgfsetcolor{\tikz@pattern@color}
\pgfsetlinewidth{\mcThickness}
\pgfpathmoveto{\pgfqpoint{0pt}{0pt}}
\pgfpathlineto{\pgfpoint{\mcSize+\mcThickness}{\mcSize+\mcThickness}}
\pgfusepath{stroke}
}}
\makeatother

% Pattern Info
 
\tikzset{
pattern size/.store in=\mcSize, 
pattern size = 5pt,
pattern thickness/.store in=\mcThickness, 
pattern thickness = 0.3pt,
pattern radius/.store in=\mcRadius, 
pattern radius = 1pt}
\makeatletter
\pgfutil@ifundefined{pgf@pattern@name@_ccyvpvib1}{
\pgfdeclarepatternformonly[\mcThickness,\mcSize]{_ccyvpvib1}
{\pgfqpoint{0pt}{0pt}}
{\pgfpoint{\mcSize+\mcThickness}{\mcSize+\mcThickness}}
{\pgfpoint{\mcSize}{\mcSize}}
{
\pgfsetcolor{\tikz@pattern@color}
\pgfsetlinewidth{\mcThickness}
\pgfpathmoveto{\pgfqpoint{0pt}{0pt}}
\pgfpathlineto{\pgfpoint{\mcSize+\mcThickness}{\mcSize+\mcThickness}}
\pgfusepath{stroke}
}}
\makeatother

% Pattern Info
 
\tikzset{
pattern size/.store in=\mcSize, 
pattern size = 5pt,
pattern thickness/.store in=\mcThickness, 
pattern thickness = 0.3pt,
pattern radius/.store in=\mcRadius, 
pattern radius = 1pt}
\makeatletter
\pgfutil@ifundefined{pgf@pattern@name@_hqcbax7rl}{
\pgfdeclarepatternformonly[\mcThickness,\mcSize]{_hqcbax7rl}
{\pgfqpoint{0pt}{0pt}}
{\pgfpoint{\mcSize+\mcThickness}{\mcSize+\mcThickness}}
{\pgfpoint{\mcSize}{\mcSize}}
{
\pgfsetcolor{\tikz@pattern@color}
\pgfsetlinewidth{\mcThickness}
\pgfpathmoveto{\pgfqpoint{0pt}{0pt}}
\pgfpathlineto{\pgfpoint{\mcSize+\mcThickness}{\mcSize+\mcThickness}}
\pgfusepath{stroke}
}}
\makeatother
\tikzset{every picture/.style={line width=0.75pt}} %set default line width to 0.75pt        

\[\label{Diagrm2}
\mathcal{A}^{(2)}_{\rma \rma} = \sum_{n_1+n_2=n}\left(\vcenter{\hbox{

\begin{tikzpicture}[x=0.73pt,y=0.73pt,yscale=-1,xscale=1]
%uncomment if require: \path (0,130); %set diagram left start at 0, and has height of 130

%Shape: Ellipse [id:dp006434411813384444] 
\draw  [pattern=_wlsxkbwul,pattern size=3.2249999999999996pt,pattern thickness=0.75pt,pattern radius=0pt, pattern color={rgb, 255:red, 0; green, 0; blue, 0}] (91.02,78.86) .. controls (94.91,78.88) and (98.05,81.95) .. (98.03,85.72) .. controls (98.01,89.49) and (94.84,92.53) .. (90.95,92.51) .. controls (87.06,92.48) and (83.92,89.41) .. (83.94,85.64) .. controls (83.96,81.87) and (87.13,78.84) .. (91.02,78.86) -- cycle ;
%Straight Lines [id:da2641468142648167] 
\draw    (90.72,25.2) .. controls (92.4,26.85) and (92.41,28.52) .. (90.75,30.2) .. controls (89.1,31.87) and (89.11,33.54) .. (90.78,35.2) .. controls (92.45,36.86) and (92.46,38.53) .. (90.81,40.2) .. controls (89.15,41.87) and (89.16,43.54) .. (90.83,45.2) .. controls (92.5,46.86) and (92.51,48.53) .. (90.86,50.2) .. controls (89.21,51.87) and (89.22,53.54) .. (90.89,55.2) .. controls (92.56,56.86) and (92.57,58.53) .. (90.92,60.2) .. controls (89.27,61.87) and (89.28,63.54) .. (90.95,65.2) .. controls (92.62,66.86) and (92.63,68.53) .. (90.97,70.2) .. controls (89.32,71.87) and (89.33,73.54) .. (91,75.2) -- (91.02,78.86) -- (91.02,78.86) ;
%Straight Lines [id:da9851104817165511] 
\draw [color={rgb, 255:red, 0; green, 0; blue, 0 }  ,draw opacity=1 ][line width=1.5]    (74.84,25.2) -- (236.32,25.2) ;
%Shape: Ellipse [id:dp6827926481779844] 
\draw  [pattern=_6dguie6jz,pattern size=3.2249999999999996pt,pattern thickness=0.75pt,pattern radius=0pt, pattern color={rgb, 255:red, 0; green, 0; blue, 0}] (150.61,78.86) .. controls (154.5,78.88) and (157.64,81.95) .. (157.62,85.72) .. controls (157.6,89.49) and (154.43,92.53) .. (150.54,92.51) .. controls (146.65,92.48) and (143.51,89.41) .. (143.53,85.64) .. controls (143.55,81.87) and (146.72,78.84) .. (150.61,78.86) -- cycle ;
%Straight Lines [id:da12109506431617045] 
\draw    (150.31,25.2) .. controls (151.99,26.85) and (152,28.52) .. (150.34,30.2) .. controls (148.69,31.87) and (148.7,33.54) .. (150.37,35.2) .. controls (152.04,36.86) and (152.05,38.53) .. (150.39,40.2) .. controls (148.74,41.87) and (148.75,43.54) .. (150.42,45.2) .. controls (152.09,46.86) and (152.1,48.53) .. (150.45,50.2) .. controls (148.8,51.87) and (148.81,53.54) .. (150.48,55.2) .. controls (152.15,56.86) and (152.16,58.53) .. (150.51,60.2) .. controls (148.86,61.87) and (148.87,63.54) .. (150.54,65.2) .. controls (152.21,66.86) and (152.22,68.53) .. (150.56,70.2) .. controls (148.91,71.87) and (148.92,73.54) .. (150.59,75.2) -- (150.61,78.86) -- (150.61,78.86) ;
%Straight Lines [id:da5421615096063696] 
\draw [color={rgb, 255:red, 208; green, 2; blue, 27 }  ,draw opacity=1 ][line width=1.5]  [dash pattern={on 1.69pt off 2.76pt}]  (198.13,18.16) -- (198.13,60.77) -- (198.13,89.71) ;
%Shape: Ellipse [id:dp9226709993353296] 
\draw  [color={rgb, 255:red, 0; green, 0; blue, 0 }  ,draw opacity=1 ][fill={rgb, 255:red, 74; green, 144; blue, 226 }  ,fill opacity=1 ] (236.36,32.28) .. controls (240.24,32.26) and (243.38,29.07) .. (243.35,25.16) .. controls (243.33,21.25) and (240.16,18.09) .. (236.28,18.11) .. controls (232.39,18.13) and (229.26,21.32) .. (229.28,25.24) .. controls (229.3,29.15) and (232.47,32.31) .. (236.36,32.28) -- cycle ;
%Straight Lines [id:da008358152705195465] 
\draw [color={rgb, 255:red, 0; green, 0; blue, 0 }  ,draw opacity=1 ][line width=0.75]    (243.35,23.66) -- (343.72,23.7)(243.35,26.66) -- (343.72,26.7) ;
%Shape: Ellipse [id:dp17692060534294296] 
\draw  [pattern=_jbhv1n2v6,pattern size=3.2249999999999996pt,pattern thickness=0.75pt,pattern radius=0pt, pattern color={rgb, 255:red, 0; green, 0; blue, 0}] (296.21,78.86) .. controls (300.1,78.88) and (303.24,81.95) .. (303.21,85.72) .. controls (303.19,89.49) and (300.02,92.53) .. (296.13,92.51) .. controls (292.24,92.48) and (289.1,89.41) .. (289.12,85.64) .. controls (289.14,81.87) and (292.31,78.84) .. (296.21,78.86) -- cycle ;
%Straight Lines [id:da014524767166518515] 
\draw    (295.91,25.2) .. controls (297.58,26.86) and (297.59,28.53) .. (295.93,30.2) .. controls (294.28,31.87) and (294.29,33.54) .. (295.96,35.2) .. controls (297.63,36.86) and (297.64,38.53) .. (295.99,40.2) .. controls (294.34,41.87) and (294.35,43.54) .. (296.02,45.2) .. controls (297.69,46.86) and (297.7,48.53) .. (296.05,50.2) .. controls (294.39,51.87) and (294.4,53.54) .. (296.07,55.2) .. controls (297.74,56.86) and (297.75,58.53) .. (296.1,60.2) .. controls (294.45,61.87) and (294.46,63.54) .. (296.13,65.2) .. controls (297.8,66.86) and (297.81,68.53) .. (296.16,70.2) .. controls (294.51,71.87) and (294.52,73.54) .. (296.19,75.2) -- (296.21,78.86) -- (296.21,78.86) ;
%Shape: Ellipse [id:dp2722953684165509] 
\draw  [color={rgb, 255:red, 0; green, 0; blue, 0 }  ,draw opacity=1 ][fill={rgb, 255:red, 74; green, 144; blue, 226 }  ,fill opacity=1 ] (350.8,32.25) .. controls (354.69,32.23) and (357.82,29.04) .. (357.8,25.12) .. controls (357.77,21.21) and (354.6,18.05) .. (350.72,18.07) .. controls (346.83,18.09) and (343.7,21.28) .. (343.72,25.2) .. controls (343.75,29.11) and (346.91,32.27) .. (350.8,32.25) -- cycle ;
%Straight Lines [id:da2102766677754483] 
\draw [color={rgb, 255:red, 0; green, 0; blue, 0 }  ,draw opacity=1 ][line width=1.5]    (357.8,25.12) -- (489.32,25.2) ;
%Shape: Ellipse [id:dp6053009834136588] 
\draw  [pattern=_wlgh5lik4,pattern size=3.2249999999999996pt,pattern thickness=0.75pt,pattern radius=0pt, pattern color={rgb, 255:red, 0; green, 0; blue, 0}] (236.62,78.86) .. controls (240.51,78.88) and (243.65,81.95) .. (243.63,85.72) .. controls (243.61,89.49) and (240.44,92.53) .. (236.54,92.51) .. controls (232.65,92.48) and (229.51,89.41) .. (229.53,85.64) .. controls (229.56,81.87) and (232.73,78.84) .. (236.62,78.86) -- cycle ;
%Straight Lines [id:da9501343964752452] 
\draw    (236.36,32.29) -- (236.37,33.86) -- (236.37,33.86) .. controls (238.04,35.51) and (238.05,37.18) .. (236.4,38.86) .. controls (234.75,40.53) and (234.76,42.2) .. (236.43,43.86) .. controls (238.1,45.52) and (238.11,47.19) .. (236.45,48.86) .. controls (234.8,50.53) and (234.81,52.2) .. (236.48,53.86) .. controls (238.15,55.52) and (238.16,57.19) .. (236.51,58.86) .. controls (234.86,60.53) and (234.87,62.2) .. (236.54,63.86) .. controls (238.21,65.52) and (238.22,67.19) .. (236.56,68.86) .. controls (234.91,70.53) and (234.92,72.2) .. (236.59,73.86) .. controls (238.26,75.52) and (238.27,77.19) .. (236.62,78.86) -- (236.63,80.43) -- (236.63,80.43) ;
%Shape: Ellipse [id:dp07880733067372159] 
\draw  [pattern=_ccyvpvib1,pattern size=3.2249999999999996pt,pattern thickness=0.75pt,pattern radius=0pt, pattern color={rgb, 255:red, 0; green, 0; blue, 0}] (350.8,80.67) .. controls (354.7,80.69) and (357.83,83.76) .. (357.81,87.53) .. controls (357.79,91.3) and (354.62,94.34) .. (350.73,94.31) .. controls (346.84,94.29) and (343.7,91.22) .. (343.72,87.45) .. controls (343.74,83.68) and (346.91,80.64) .. (350.8,80.67) -- cycle ;
%Straight Lines [id:da8535603610660709] 
\draw    (350.55,34.09) -- (350.8,32.25) -- (350.8,32.25) .. controls (352.47,33.92) and (352.47,35.58) .. (350.8,37.25) .. controls (349.13,38.92) and (349.13,40.58) .. (350.8,42.25) .. controls (352.47,43.92) and (352.47,45.58) .. (350.81,47.25) .. controls (349.14,48.92) and (349.14,50.58) .. (350.81,52.25) .. controls (352.48,53.92) and (352.48,55.58) .. (350.81,57.25) .. controls (349.14,58.92) and (349.14,60.58) .. (350.81,62.25) .. controls (352.48,63.92) and (352.48,65.58) .. (350.81,67.25) .. controls (349.14,68.92) and (349.14,70.58) .. (350.81,72.25) .. controls (352.48,73.92) and (352.48,75.58) .. (350.81,77.25) -- (350.82,82.24) -- (350.82,82.24) ;
%Shape: Ellipse [id:dp0072262376651854865] 
\draw  [pattern=_hqcbax7rl,pattern size=3.2249999999999996pt,pattern thickness=0.75pt,pattern radius=0pt, pattern color={rgb, 255:red, 0; green, 0; blue, 0}] (469.1,78.86) .. controls (472.99,78.88) and (476.13,81.95) .. (476.11,85.72) .. controls (476.09,89.49) and (472.92,92.53) .. (469.03,92.51) .. controls (465.14,92.48) and (462,89.41) .. (462.02,85.64) .. controls (462.04,81.87) and (465.21,78.84) .. (469.1,78.86) -- cycle ;
%Straight Lines [id:da8689006380700681] 
\draw    (468.8,25.2) .. controls (470.47,26.85) and (470.48,28.52) .. (468.83,30.2) .. controls (467.18,31.87) and (467.19,33.54) .. (468.86,35.2) .. controls (470.53,36.86) and (470.54,38.53) .. (468.88,40.2) .. controls (467.23,41.87) and (467.24,43.54) .. (468.91,45.2) .. controls (470.58,46.86) and (470.59,48.53) .. (468.94,50.2) .. controls (467.29,51.87) and (467.3,53.54) .. (468.97,55.2) .. controls (470.64,56.86) and (470.65,58.53) .. (469,60.2) .. controls (467.34,61.87) and (467.35,63.54) .. (469.02,65.2) .. controls (470.69,66.86) and (470.7,68.53) .. (469.05,70.2) .. controls (467.4,71.87) and (467.41,73.54) .. (469.08,75.2) -- (469.1,78.86) -- (469.1,78.86) ;

% Text Node
\draw (161.74,44.11) node [anchor=north west][inner sep=0.75pt]    {$\ell_{n_{1}}$};
% Text Node
\draw (71.01,44.11) node [anchor=north west][inner sep=0.75pt]    {$\ell_{1}$};
% Text Node
\draw (106.73,44.11) node [anchor=north west][inner sep=0.75pt]    {$\cdots $};
% Text Node
\draw (251.42,49.11) node [anchor=north west][inner sep=0.75pt]    {$\cdots $};
% Text Node
\draw (357.53,49.11) node [anchor=north west][inner sep=0.75pt]    {$\cdots $};
% Text Node
\draw (383.93,44.11) node [anchor=north west][inner sep=0.75pt]    {$\ell_{n_{1} +n_{2} +m+2}$};
% Text Node
\draw (58.36,17.6) node [anchor=north west][inner sep=0.75pt]    {$\rma$};
% Text Node
\draw (498.79,17.6) node [anchor=north west][inner sep=0.75pt]    {$\rma$};

\draw (290.36,7.6) node [anchor=north west][inner sep=0.75pt]    {$\rmb$};

\end{tikzpicture}

}} \right)
\]

Note that we have attached the off-diagonal vertices to cut propagators as before. The cut propagator leads again to a factorised form for the amplitude
\[\label{eq:Diag2}
\mathcal{A}_{\rma \rma}^{(2)}(b) =  \sum_{m} \sum_{n_1+n_2 =n}\mathcal{I}^L_{\rma \rma} (n_1) \, \tilde{\mathcal{I}}^M_{\rmb \rma}(m) \,\tilde{\mathcal{I}}^R_{\rma \rmb}(n-n_1),
\]
The diagonal part exponentiates independently as before.
Furthermore, from our definition of the eikonal $\chi_{\rmb \rma}$ it is straightforward to show that 
\[
 \tilde{\mathcal{I}}^M_{\rmb \rma}(m) = ig_{\rmb}^m \chi^{(m)}_{\rmb \rma} , \quad \tilde{\mathcal{I}}^R_{\rmb \rma}(n-n_1) = i g_{\rma}^{n-n_1} \chi_{\rma \rmb}^{(n-n_1)} .
\]
Using this, we recognise that eq. \eqref{eq:Diag2} is simply the $\mathcal{O}(g_\rma^n g_\rmb^m g_{\rmb \rma}^2)$ term in the expansion of 
\[
i \mathcal{A}_{\rma \rma}^{(2)}(b)  = - e^{i \chi_{\rma \rma}(b)} \langle \chi_{\rma \rmb}(b)\chi_{\rmb \rma} (b) \rangle.
\]
Here we have introduced the bracket notation to emphasise that the product $\chi_{\rmb \rma} \chi_{\rma \rmb}$ may involve a sum over internal degrees of freedom. For example, in the case where channel $(\rmb)$ has a different mass that channel $(\rma)$ we have
\[
 \langle \chi_{\rma \rmb}(b)\chi_{\rmb \rma} (b) \rangle= \int \dd w \,  \rho(w)\,\chi_{\rma \rmb}(b,w)\chi_{\rmb \rma} (b,w).
\]
where $w= (M_b^2 -M_a^2)/2M_a$ is the mass difference between the two channels.
Now, let us observe the general pattern at higher orders: Inserting additional off-diagonal iterations merely adds powers of $ \chi_{\rmb \rma}(b)\chi_{\rma \rmb} (b)$ to the diagonal amplitude, thanks to the factorisation assumption. We now have
\[\label{eq:GeometricSeries}
1 + \sum_{n} i \mathcal{A}_{\rma \rma}^{(n)}(b) &= e^{i \chi_{\rma \rma}(b)} \big[1 -\sum_{n} \langle \chi_{\rmb \rma}(b)\chi_{\rma \rmb} (b) \rangle^n \big],\\
& = \frac{e^{i \chi_{\rma \rma}(b)} }{1 + \langle \chi_{\rmb \rma}(b)\chi_{\rma \rmb} (b) \rangle}. 
\]
Let us reiterate that this form holds provided that the factorisation condition is satisfied. Equation \eqref{eq:GeometricSeries} provides a simple example in which intermediate channel transitions can be resummed to all orders. Here, we have not committed to any particular theory or model. Instead we have imposed certain conditions on the diagonal and off-diagonal amplitudes which in practice may not be satisfied by an arbitrary theory. In this model, we see that the total inelastisity
\begin{align}\label{eq:A-geom}
\text{inelasticity} = \frac{1}{1 + \langle \chi_{\rmb \rma}(b)\chi_{\rma \rmb} (b) \rangle}, 
\end{align}
emerges by an all-order resummation of off-diagonal vertices. In more general models, this elasticity cannot be derived to all orders from perturbation theory, but instead has to be obtained by matching to a classical observable. We conclude by observing how the argument of the geometric series indeed resembles both the one loop diagram already encountered in \eqref{ampli1loop} in the diagonal channel and the integrand of the mass-change computed in   \eqref{dmmc}.

\newpage
\bibliographystyle{JHEP}
\bibliography{InelasticEikonal.bib}

\end{document}